\documentclass[journal,10pt,twocolumn]{IEEEtran} 
\usepackage{balance}
\usepackage{comment}
\usepackage{booktabs}
\usepackage{threeparttable}
\usepackage{array}
\usepackage{multirow}
\usepackage{amsmath}
\usepackage{geometry}
\usepackage{algpseudocode}
\usepackage[caption=false,font=footnotesize]{subfig}
\usepackage[ruled,vlined,linesnumbered,resetcount]{algorithm2e}
\SetKwRepeat{Do}{do}{while}%
\usepackage[linesnumbered,ruled,vlined]{algorithm2e}
\usepackage{cite}
\usepackage{amsmath,amssymb,amsfonts,bm}

\usepackage{textcomp}
\usepackage{soul}
\usepackage{graphics}
\usepackage{graphicx} 
\usepackage{epsfig}
\usepackage{amssymb} 
\usepackage{amsthm}
\usepackage{times} % assumes new font selection scheme installed
\usepackage{amsmath} % assumes amsmath package installed
\usepackage{amssymb}  % assumes amsmath package installed
\usepackage{epstopdf}
\usepackage{array}
\usepackage{color}
\usepackage[numbers]{natbib} 
\usepackage{mathtools}
\usepackage{lipsum}

\def\BibTeX{{\rm B\kern-.05em{\sc i\kern-.025em b}\kern-.08em
    T\kern-.1667em\lower.7ex\hbox{E}\kern-.125emX}}
\usepackage{url}
\usepackage{tikz} 
\usepackage{lipsum}
\usetikzlibrary{shapes,shapes.geometric,arrows,fit,calc,positioning,automata,arrows.meta}
\definecolor{behavioral}{RGB}{255, 193, 71}
\definecolor{internal}{RGB}{102, 204, 204}
\definecolor{hybrid}{RGB}{255, 128, 171}

\renewcommand{\SetKwInOut}[2]{%
	\sbox\algocf@inoutbox{\KwSty{#2}\algocf@typo:}%
	\expandafter\ifx\csname InOutSizeDefined\endcsname\relax% if first time used
	\newcommand\InOutSizeDefined{}%
	\sbox\algocf@inoutbox{\KwSty{#2}\algocf@typo\textbf{:}~}\setlength{\inoutindent}{\wd\algocf@inoutbox}%
	\else% else keep the larger dimension
	\ifdim\wd\algocf@inoutbox>\inoutsize%
	\sbox\algocf@inoutbox{\KwSty{#2}\algocf@typo\textbf{:}~}\setlength{\inoutindent}{\wd\algocf@inoutbox}%
	\fi%
	\fi% the dimension of the box is now defined.
	\algocf@newcommand{#1}[1]{%
		\ifthenelse{\boolean{algocf@inoutnumbered}}{\relax}{\everypar={\relax}}%
		{\let\\\algocf@newinout\hangindent=\inoutindent\hangafter=1\KwSty{#2}\algocf@typo\textbf{:}~##1\par}%
		\algocf@linesnumbered% reset the numbering of the lines
}}
\usepackage{titlesec}

\begin{document}
%\title{A Tutorial on Science-Informed Deep Learning (ScIDL) and its Applications in Wireless Communications}
%\title{Science-Informed Deep Learning, Modeling, and Analysis: A Tutorial}
\title{Science-Informed Design of Deep Learning With Applications to Wireless Systems: A Tutorial}
%\title{A Tutorial on Science-Informed Deep Learning (ScIDL) for Wireless Communications: Concepts, Methods and Applications}

\author{Atefeh Termehchi, Ekram Hossain, Angelo Vera-Rivera, Muhammad Ibrahim,
%\IEEEmembership{Fellow, IEEE}, 
and  Isaac Woungang

%\IEEEmembership{Senior Member, IEEE}
%\affil{Department of Electrical and Computer Engineering at the University of Manitoba, Winnipeg, Canada}
%\affil{\textit{Department of Computer Science, Toronto Metropolitan University, Toronto, Canada}}
\thanks{Atefeh Termehchi, Ekram Hossain, Angelo Vera-Rivera, and Mumhammad Ibrahim are with the Department of Electrical and Computer Engineering at the University of Manitoba, Winnipeg, Canada (emails: \{atefeh.termehchi, ekram.hossain, angelo.verarivera\}@umanitoba.ca, ibrah101@myumanitoba.ca), and Isaac Woungang is with the Department of Computer Science, Toronto Metropolitan University, Toronto, Canada (email: iwoungan@torontomu.ca). \textbf{E. Hossain} is the corresponding author.}}

\maketitle
\begin{abstract}
Recent advances in computational infrastructure and large-scale data processing have accelerated the adoption of data-driven inference methods, particularly deep learning (DL), to solve problems in many scientific and engineering domains. 
In wireless systems, DL has been applied to problems where analytical modeling or optimization is difficult to formulate, relies on oversimplified assumptions, or becomes computationally intractable. However, conventional DL models are often regarded as non-transparent, as their internal reasoning mechanisms are difficult to interpret even when model parameters are fully accessible. This lack of transparency undermines trust and leads to three interrelated challenges: limited interpretability, weak generalization, and the absence of a principled framework for parameter tuning. Science-informed deep learning (ScIDL) has emerged as a promising paradigm to address these limitations by integrating scientific knowledge into deep learning pipelines. This integration enables more precise characterization of model behavior and provides clearer explanations of how and why DL models succeed or fail. Despite growing interest, the existing literature remains fragmented and lacks a unifying taxonomy. This tutorial presents a structured overview of ScIDL methods and their applications in wireless systems. We introduce a structured taxonomy that organizes the ScIDL landscape, present two representative case studies illustrating its use in challenging wireless problems, and discuss key challenges and open research directions. The pedagogical structure guides readers from foundational concepts to advanced applications, making the tutorial accessible to researchers in wireless communications without requiring prior expertise in AI.
\end{abstract}
\section{Introduction: From Data-Driven to Science-Informed Intelligence}
Over recent decades, advances in computational capability and the recent ability to gather, store, and process massive datasets have propelled data-driven methods to the forefront of science and engineering. Among these methods, deep learning (DL), a subset of machine learning (ML), has emerged as particularly powerful. DL employs multi-layer neural networks to learn representations of real-world phenomena directly from data. This enables complex decision-making without relying on explicit, analytically derived mathematical models. Indeed, DL has achieved remarkable progress due to higher-capacity deep neural network (DNN) architectures, the development of efficient training algorithms, the availability of large-scale datasets, and powerful computing infrastructure~\cite{karpatne2017theory}.

These advancements have expanded the applicability of DL across diverse domains, including wireless communications. In wireless systems, DL has been successfully applied to problems where traditional analytical modeling or optimization is difficult to formulate, relies on oversimplified assumptions, or becomes computationally intractable. DL provides a unified framework to perceive environments, infer latent system states, and make autonomous decisions directly from data. Consequently, DL-aided solutions have been developed for tasks such as channel estimation, beamforming design, resource allocation, mobility prediction, interference detection, and traffic forecasting. These methods often achieve performance comparable to or better than simplified model-based approaches, while enabling faster inference and lower online computational complexity. These successes highlight the potential of DL to replace certain traditional scientific approaches based on physical laws, analytical modeling, and conventional optimization.

%However, growing concerns about DL models' non-transparent nature have received increasing attention 
However, growing concerns about DL models have intensified, as they are often treated as non-transparent boxes \cite{suh2025survey,hellstrom2025generalization, zhang2021survey, xu2025interpretability, csahin2025unlocking,li2022interpretable}. Specifically, DL models are often labeled ``non-transparent” because, even with full access to their weights, biases, and activation functions, it remains difficult to determine what the model has actually learned. More formally, non-transparent models refer to systems whose internal mechanisms are largely unknown or difficult to characterize \cite{bunge1963general}. In particular, forward propagation repeatedly applies complex, high-dimensional nonlinear transformations across many layers, obscuring the structure of internal representations. Consequently, it is challenging to identify which features drive the decisions, how specific inputs map to particular outputs, and how intermediate states evolve inside the model.

%This lack of transparency can hinder trust when deploying DL in safety-critical applications, such as medical image analysis, intelligent driving, and security-sensitive systems. In these applications, failures may lead to life-threatening outcomes or substantial financial losses. The need to question, understand, and trust DL-based systems has become more pressing, driven by regulations and legal requirements. For example, U.S. Federal Aviation Administration enforces stringent validation and certification procedures for ML-enabled flight control systems \cite{faa_ml}. In the same spirit, the European Commission’s communication Artificial Intelligence for Europe highlights the importance of research on explainable artificial intelligence (AI) to strengthen public confidence. It notes that ``AI systems should be developed in a manner which allows humans to understand (the basis of) their actions" \cite{eu_ai_act}. In sum, the non-transparent nature of conventional DL and the limited theoretical foundations for its purely data-driven engine pose significant challenges for its reliable deployment, particularly in high-stakes settings.%thereby improving transparency and reducing the risk of bias and error \cite{eu_ai_act}.

%is a primary barrier to deploying these models in safety-critical applications, because it 
This opaque nature contributes to three interrelated challenges: limited interpretability, weak generalization assurances, and the lack of a rigorous, principled framework for parameter tuning. Limited interpretability makes it difficult to explain a model’s reasoning or to verify that its decisions are driven by meaningful, reliable cues rather than spurious correlations. This limitation is problematic, as it can conceal hidden failure modes and hinder timely diagnosis and correction of errors before they escalate. Moreover, weak generalization further undermines trust, since performance can degrade unpredictably under unseen conditions that are not represented in the training data. In addition, non-transparency often forces model design and hyperparameter tuning to rely on trial-and-error, making the development and training of large-scale DL models expensive and time-consuming. Beyond these issues, obtaining large labeled datasets for large-scale DL training is costly and labor-intensive, often requiring extensive measurement campaigns and sophisticated instrumentation.

A promising direction to address these limitations is to embed well-established scientific principles and domain knowledge into DL models, an approach we refer to here as \emph{science-informed deep learning} (ScIDL).
ScIDL (also referred to as ``physics-guided ML,'' ``physics-informed ML,'' ``theory-guided data science,'' ``physics-informed neural networks,'' ``hybrid model-based/data-driven learning,'' or ``physics-aware AI'') is an emerging paradigm that integrates scientific knowledge, including physical laws, classical optimization-based insights, and domain-specific knowledge, into DL models~\cite{baker2019workshop, willard2022integrating, von2021informed, karniadakis2021physics}. In wireless communications, this integration can incorporate components such as Maxwell’s equations, channel propagation models, Shannon capacity constraints, network geometry, and queueing theory into the neural network (NN) architecture, loss function, or optimization procedure. 

\begin{figure*}
    \centering
    \includegraphics[width=0.75\linewidth]{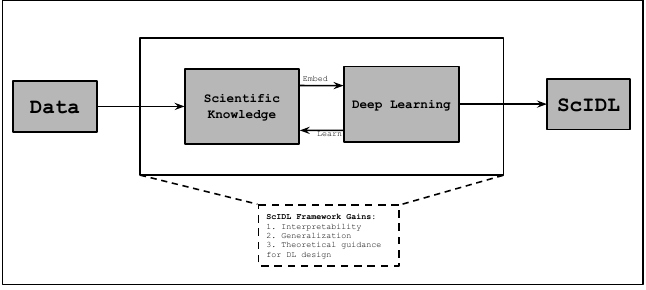}
    \caption{Conceptual overview of ScIDL, showing the integration of raw data, scientific knowledge, and DL architectures. By embedding physics and domain principles into the learning pipeline, DL models improve their interpretability, domain-consistency, and generalization, and enable meta-applications such as scientific discovery and theory-guided DL design.}%Knowledge + DL models = ScIDL models
    \label{fig:ScIDL}
\end{figure*}

By bridging data-driven learning with established scientific knowledge, ScIDL can enhance both generalization and interpretability while reducing computational burden and dependence on large measurement datasets. Embedding science-based principles into otherwise purely data-driven learning pipelines yields physically consistent models with theoretically grounded outcomes. 
%Moreover, incorporating scientific structure into purely data-driven interpretability and generalization analyses enables a more precise characterization of model behavior with clearer explanations of how and why DL models succeed (or fail) under distribution shifts. 
A conceptual overview of ScIDL, showing the integration of data and scientific knowledge into DL model pipelines is shown in Figure~\ref{fig:ScIDL}.
\subsection{Motivation and Contribution}
In recent years, numerous studies in wireless communications have explored strategies for embedding established scientific knowledge into DL models, leading to the emergence of physics-informed and hybrid learning paradigms. For instance, physics-based algebraic or differential equations can be incorporated as constraints within NN loss functions, as in physics-informed NNs (PINNs)~\cite{karpatne2017physics} and their variational extensions (VPINNs)~\cite{kharazmi2019variational}, where the variational form of governing equations is integrated into the training objective. Despite the growing body of research in this area, the literature lacks a unified taxonomy that organizes the  diverse ScIDL approaches and their applications to wireless  problems (e.g., related to communication and sensing). Existing works still remain fragmented, only addressing specific tasks without providing a comprehensive framework that classifies existing methods, outlines their gains, and highlights future research directions.

This tutorial presents a structured overview of ScIDL methods and their applications in the field of wireless communications. It aims to explain, contextualize, and unify the existing ScIDL literature related to wireless systems within a coherent organization framework. 
%Our goal is to guide the readers toward understanding the potential advantages of this emerging DL paradigm for addressing complex challenges in wireless system design, modeling, and optimization. 
Our main contributions are as follows:
\begin{enumerate}
\item A structured taxonomy that defines and organizes ScIDL methods across architecture design, training data and output validation, and loss function formulation and optimization.
\item A comprehensive review of existing ScIDL approaches, methods, and applications in wireless communications.
%\item A taxonomy of science-informed methods for the interpretability and generalization analysis of DL models. 
\item Two representative ScIDL case studies illustrating how science-informed models can address complex wireless communication problems.
\item Identification of key challenges and open research directions for applying ScIDL to wireless systems.
\end{enumerate}
\subsection{Methodology}
This tutorial adopts a structured pedagogical approach~\cite{IEEEComSocTutorials} to organize and synthesize the emerging body of knowledge on ScIDL. We introduce a unified taxonomy that defines the ScIDL design space and serves as the organizing backbone of the tutorial. The reviewed literature is classified according to the proposed ScIDL taxonomy, which identifies where and how scientific knowledge can be incorporated into DL models -- specifically their architecture, initialization, and optimization. 
%It further surveys interpretability and generalization analysis techniques that embed science-based principles into the evaluation process. This moves beyond purely data-driven assessments and provides clearer, more rigorous explanations of DL behavior. 
Following the literature review, we present two representative case studies that demonstrate how ScIDL methods can be applied in key wireless problems, assessing the practicality and impact of ScIDL in real-world scenarios. The discussion concludes with a synthesis of open challenges and future research directions for this emerging field. 

%The pedagogical structure is designed to guide readers from foundational concepts, through the taxonomy of ScIDL methods, toward advanced applications and a forward-looking perspective, ensuring conceptual clarity and facilitating learning for readers familiar with wireless systems but not necessarily experts in AI. 
%
\subsection{Organization}
\begin{figure*}
    \centering
    \includegraphics[width=\linewidth]{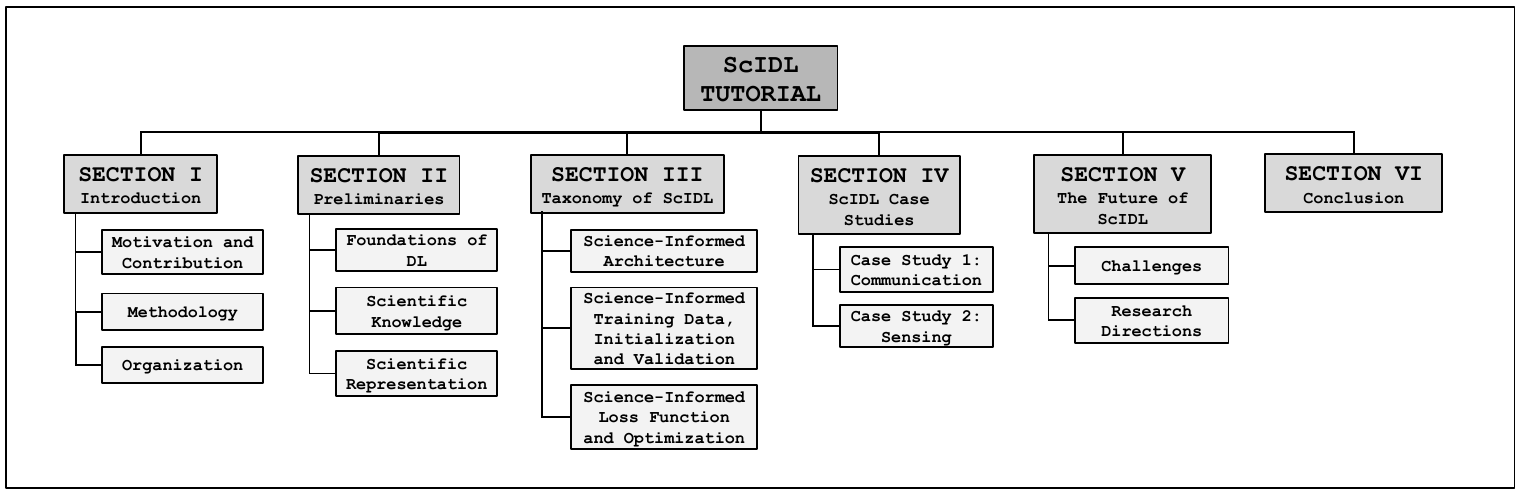}
    \caption{Organizational structure of the ScIDL Tutorial to guide readers through their reading progression. The map outlines the tutorial's core sections: Introduction, Preliminary Discussion, ScIDL Taxonomy,  ScIDL Case Studies, the Future of ScIDL, and Conclusion.}
    \label{fig:organization}
\end{figure*}

The remaining of the tutorial is organized as follows. Section II introduces the background of ScIDL, discussing the foundations of DL, various sources and representations of scientific knowledge. Section IV defines a structured taxonomy for ScIDL, organizing the emerging domain. Section III provides a comprehensive overview of existing ScIDL approaches, methods, and applications in wireless communications. 
%Section IV discusses science-informed methods for interpretability and generalization analysis of DL models. 
Section IV presents representative case studies illustrating how science-informed models can be applied to solve complex wireless problems. Section V outlines key challenges and future research directions.
 %for extending ScIDL to broader challenges in wireless systems. 
 Section VI %summarizes the main insights and lessons learned, followed by Section VII, which concludes the article. 
concludes the paper. A visual map of the article’s organization is shown in Figure~\ref{fig:organization}.
\section{Preliminaries: Foundations of DL, the Nature of Knowledge, and Scientific Representation}
%In the following subsections, we present the fundamental concepts that govern DL models, followed by a discussion on sources and representations of scientific knowledge, followed by a detailed taxonomy for ScIDL, which explains how this knowledge can be integrated into DL models to enhance and extend scientific analysis. 
%
\subsection{Foundations of DL}
%
%\begin{figure*}
%    \centering
%    \includegraphics[width=0.75\linewidth]{Figures/empty.pdf}
%    \caption{Placeholder (Figure: NN fundamental concepts and architecture).}
%    \label{fig:taxonomy}
%\end{figure*}
%
DL is a subset of machine learning that focuses on learning complex representations of data through hierarchical layers of processing. At its core, DL relies on artificial NNs, which are computational models inspired by the human brain, that consist of interconnected neurons grouped by layers capable of mapping inputs into abstract features. A neuron consists of a mathematical function that receives one or more input values, each multiplied by a corresponding weight representing its importance. The neuron then additively combines these weighted inputs, adds a bias term, and applies a non-linear activation function -- such as ${\rm ReLU}$, ${\rm sigmoid}$, or ${\rm tanh}$ -- to determine the output. Mathematically, a neuron computes its output $y$ as:
\begin{equation}
y = \sigma\left(\sum_{i=1}^{n} w_i x_i + b\right),   
\end{equation}
where $x_i$ are the input features, $w_i$ are the weights, $b$ is the bias, and $\sigma(\cdot)$ is the activation function. When the neurons are connected across layers -- input, hidden, and output -- they form an NN capable of learning the hierarchical feature representations from data. NNs form the mathematical foundation of DL and are known as universal function approximators, meaning they can approximate any continuous function given sufficient neurons and appropriate parameters. 

%This property was formally introduced in 1989 by  the Universal Approximation Theorem~\cite{Cybenko1989, Hornik1991}. The theorem states that a feedforward NN $f$ with at least one hidden layer, appropriate weights and biases, and a non-linear activation function can approximate any continuous function $\phi:\mathbb{R}^n \to \mathbb{R}$ to arbitrary precision $\epsilon$. Formally, for any $\epsilon>0$, there exist an NN $f$ such that
%%
%\begin{equation}
%|\phi(\mathbf{x})-f(\mathbf{x})| < \epsilon, \quad \forall \mathbf{x} \in \mathcal{K},    
%\end{equation}
%%
%where $\mathcal{K} \subset \mathbb{R}^n$ is compact. Given its nature, NNs are a flexible tool to model diverse phenomena, making them useful to solve problems in numerous scientific domains.   
%
%\begin{figure}
%    \centering
%    \includegraphics[width=0.8\linewidth]{Figures/NN.pdf}
%    \caption{NNs with at least one hidden layer are universal approximators of any continuous function $f:\mathbb{R}^n \to \mathbb{R}$ on compact sets $\mathcal{K}\in\mathbb{R}^n$ to arbitrary precision $\epsilon$.}
%    \label{fig:placeholder}
%\end{figure}
%
In an NN, information flows in two directions during training: feedforward and backpropagation. Feedforward refers to the process of passing the input data through the network -- layer by layer, from input to output -- to compute a prediction. Backpropagation, on the other hand, is the learning mechanism that follows. After the network produces a prediction, a loss function quantifies the difference between the predicted and true outputs, measuring how far the model deviates from the true target. Common examples include the mean squared error (MSE) for regression and cross-entropy loss for classification. The backpropagation algorithm then computes the gradient of the loss function with respect to each weight, and sends this error signal backward through the network. The gradient indicates how much each weight should be adjusted to reduce future errors. The most basic backpropagation optimizer is gradient descent, which iteratively adjusts weights in the direction that most reduces the loss. More advanced optimizers, such as Adam, RMSProp, or Adagrad, adapt learning rates automatically and accelerate convergence, making them widely used in modern DL models.

DNNs are NNs with multiple hidden layers stacked between the input and output layers. Each layer learns to extract hierarchical features from data -- from simple patterns in early layers to complex structures in deeper ones. This behavior enables DNNs to map highly nonlinear relationships between the inputs and outputs, significantly increasing their capacity to approximate complex functions found in many scientific domains. 

%DL models typically operate in two stages: training and production. During the training stage, the model adjusts its internal parameters through repeated feedforward and backpropagation rounds using the training data, guided by a defined loss function and an optimization algorithm. Once trained, the model transitions to the production stage, where it is deployed to make predictions on new, unseen data.    

A DL architecture can be formally defined as a parameterized composite function $f$ that maps an input vector $\mathbf{x} \in \mathbb{R}^n$ to an output vector $\mathbf{y} \in \mathbb{R}^m$ through a sequence of $L$ transformation layers. Formally,
\begin{equation}
\mathbf{y} = f(\mathbf{x}; \boldsymbol{\theta}) = f^{(L)} \circ f^{(L-1)} \circ \cdots \circ f^{(1)}(\mathbf{x}),  
\label{equation:DLmodel}
\end{equation}
where $f^{(\ell)}(\cdot)$ denotes the transformation performed at the $\ell$-th layer, $\boldsymbol{\theta} = \{ \mathbf{W}^{(1)}, \mathbf{b}^{(1)}, \ldots, \mathbf{W}^{(L)}, \mathbf{b}^{(L)} \}$ represents the set of learnable parameters (weights and biases), and $\circ$ denotes functional composition (i.e., $(f \circ g)(x)=f(g(x))$). Typically, each layer $\ell$ with parameters $\theta^{(\ell)} = \{\mathbf{W}^{(\ell)}, \mathbf{b}^{(\ell)}\}$ takes the output of the previous layer $\mathbf{h}^{(\ell-1)}$, applies the linear transformation $\mathbf{W}^{(\ell)} \mathbf{h}^{(\ell-1)} + \mathbf{b}^{(\ell)}$, and pass it through a non-linear activation function $\sigma$ to produce the output. Mathematically, 
\(
\mathbf{h}^{(\ell)}=f^{(\ell)}(\mathbf{h}^{(\ell-1)}; \theta^{(\ell)})
= \sigma^{(\ell)}\big(\mathbf{W}^{(\ell)} \mathbf{h}^{(\ell-1)} + \mathbf{b}^{(\ell)}\big). 
\)
The goal of the architecture is to find the optimal parameter set $\boldsymbol{\theta}^*$ that minimizes the average of a loss function $\mathcal{L}$ over the training dataset $\mathcal{D} = \{(\mathbf{x}_k, \mathbf{y}_k)\}$, $k=1:N$. That is, 
\begin{equation}
\boldsymbol{\theta}^* = \arg\min_{\boldsymbol{\theta}} \frac{1}{N} \sum_{k=1}^{N} \mathcal{L}(f(\mathbf{x}_k; \boldsymbol{\theta}), \mathbf{y}_k).    
\end{equation}
Different DL architectures have been developed to address the requirements of specific data types and learning objectives. The following discussion introduces widely used DL architectures, highlighting their key characteristics and applications.%The following discussion introduces some of the most widely used DL architectures, highlighting their fundamental characteristics and main applications. 
\subsubsection{Convolutional Neural Networks (CNNs)}
CNNs are a type of DL architectures specifically designed to process data with grid-like topology~\cite{lecun1998gradient}. These include images, time-frequency representations, or spatial channel maps in wireless systems. CNNs replace fully connected mappings with structured convolutional layers to capture spatial or temporal correlations while preserving local grid structure. The term convolutional comes from the mathematical convolution operation. In CNNs, each neuron applies a convolution between a small filter -- also called a kernel -- and a local region of the input data. Formally, let $x_m^{(l-1)}(\cdot,\cdot)$ denote $m$-th input feature map at layer $l-1$ with kernel support $\mathcal{K}=\{0,\dots,U-1\}\times\{0,\dots,V-1\}$ and $\mathcal{M}_k$ a set of input maps connected to output map $k$. A convolutional layer computes the $k$-th output feature map as
\begin{align}
%\scalebox{0.7}{
z_k^{(l)}(i,j) =&   
\sum_{m\in\mathcal{M}_k}\;
\sum_{(u,v)\in\mathcal{K}}  \notag \\
     &  \left(W_{k,m}^{(l)}(u,v)\; x_m^{(l-1)}(i+u,j+v)+b_k^{(l)}\right),
    % },
\end{align}
where $(u,v)$ are the spatial offsets over the kernel support, $W_{k,m}^{(l)}(u,v)$ is the convolution kernel connecting input map $m$ to output map $k$ at layer $l$ and $b_k^{(l)}$ is the bias for output map $k$. The output $y_k^{(l)}(i,j)$ is then
\(y_k^{(l)}(i,j)=\phi\!\big(z_k^{(l)}(i,j)\big),\)
where $\phi(\cdot)$ is a non-linear activation function, applied element-wise. The operation slides the filter across the input grid, multiplying and summing overlapping values to produce an output. The convolution allows CNNs to detect local features regardless of their position in the input, and since the same kernel weights are applied across all positions, the number of network parameters is reduced dramatically, making the model computationally efficient and widely used for spatial and spatio-temporal data processing. 
%
%%\vspace{0.2cm}
\subsubsection{Recurrent Neural Networks (RNNs)}
RNNs are a class of DL architectures designed to model sequential or time-indexed data~\cite{Elman1990Finding}. Unlike feedforward networks, that treat inputs independently, RNNs incorporate feedback connections, allowing information from previous time steps to persist in the network. Given an input sequence $\{\mathbf{x}_t\}$, $t=\{1, \ldots, T\}$, a standard RNN updates its hidden state $\mathbf{h}_t \in \mathbb{R}^d$ as
\(\mathbf{h}_t = \phi\!\left( \mathbf{W}_{xh}\mathbf{x}_t + \mathbf{W}_{hh}\mathbf{h}_{t-1} + \mathbf{b}_h\right),\)
where $\mathbf{W}_{xh}$ and $\mathbf{W}_{hh}$ are input-to-hidden and recurrent weight matrices, $\mathbf{b}_h$ is a bias vector, and $\phi(\cdot)$ is a nonlinear activation function. The output is then computed as
\(\mathbf{y}_t =  \psi\!\left(\mathbf{W}_{hy}\mathbf{h}_t + \mathbf{b}y \right),\) by using the output activation $\psi(\cdot)$. RNNs extend feedforward NNs by introducing temporal recursion, enabling them to capture the temporal dependencies across sequences, making them effective for time-dependent tasks such as speech recognition, natural language processing, and time-series forecasting. During training, the network updates its hidden state at each time step based on both the current input and the previous state, allowing it to memorize patterns over time. Traditional RNNs struggle with long-term dependencies, motivating advanced variants such as Long Short-Term Memory (LSTM) and Gated Recurrent Units (GRUs) that enable stable learning over long sequences.%Traditional RNNs often struggle with long-term sequences, motivating the development of advanced variants such as Long Short-Term Memory (LSTM) and Gated Recurrent Units (GRUs) networks, that maintain a stable learning over long sequences. 
%

%\vspace{0.2cm}
\subsubsection{Deep Reinforcement Learning (DRL)}
DRL combines DNNs with reinforcement learning (RL) to solve sequential decision-making problems~\cite{mnih2015human}. A DRL problem is usually modeled as a Markov decision process (MDP) defined by the tuple $(\mathcal{S},\mathcal{A},P,r,\gamma)$, where $\mathcal{S}$ and $\mathcal{A}$ denote a high-dimensional state and action spaces, $P(s'|s,a)$ is the state-transition probability, $r(s,a)$ is the reward function, and $\gamma \in (0,1)$ is a discount factor. The objective in DRL systems is learning a policy $\pi(a|s)$ that maximizes the expected cumulative return 
\begin{equation}
J(\theta)=\mathbb{E}_{\pi}\!\left[\sum_{t=0}^{\infty}\gamma^t r(s_t,a_t)\right].
\end{equation}
A DRL model approximates the action-value function
\begin{equation}
Q^*(s,a)=\max_\pi \mathbb{E}_\pi\left[\sum_t \gamma^t r_t \mid s_0=s,a_0=a\right],
\end{equation}
using a neural network $Q_\theta(s,a)$. The DNN serves as a function approximator for policies, enabling high-dimensional state and action space learning, common in wireless optimization applications.
%~\cite{xu2025interpretability}
%\addatefeh{Please take a look at reference  to include DBN as well.}
%\addatefeh{Briefly explain DRL as well.}
%
%%\vspace{0.2cm}
\subsubsection{Generative Adversarial Networks (GANs)}
GANs are DL-based architectures designed for data generation through adversarial training~\cite{goodfellow2014generative}. A GAN consists of two NNs: a generator and a discriminator, competing in a zero-sum game. Let $p_{data}(\mathbf{x})$ denote the true data distribution and $p_{\mathbf{z}}(\mathbf{z})$ a random prior (e.g., Gaussian). The generator $G_\theta:\mathbf{z}\mapsto \tilde{\mathbf{x}}$ maps noise samples $\mathbf{z}\sim p_{\mathbf{z}}$ to synthetic data that closely resemble real samples $p_G$. The discriminator 
$D_\phi:\mathbf{x}\mapsto [0,1]$ outputs the probability that a sample is real. During training, the generator learns to produce synthetic data to fool the discriminator, while the discriminator is trained on both real and generated data to improve its ability to distinguish between them. Training solves the following adversarial objective:
\begin{align}
%\scalebox{0.8}{
\min_{\theta}\max_{\phi}\;
\mathbb{E}_{\mathbf{x}\sim p_{data}} 
\big[\log D_\phi(\mathbf{x})\big] \; + & \notag\\
  \mathbb{E}_{\mathbf{z}\sim p_{\mathbf{z}}}
\big[\log(1-D_\phi(G_\theta(\mathbf{z})))\big]. & 
%$}    
\end{align}
For fixed $G_\theta$, the optimal discriminator estimates the density ratio between real and generated data; for fixed $D_\phi$, the generator updates $\theta$ to minimize the discriminator’s ability to distinguish fake samples. This adversarial learning process has been proven effective for generating high-fidelity data, with successful applications in images, audio, and text generation.
\subsubsection{Transformers}
Transformers are a type of neural network designed to process sequences (like text, audio, or time-series) by letting every part of the input {\em pay attention} to every other part at once.
They use a mechanism called {\em self-attention}~\cite{vaswani2017attention} to decide which words or elements are most important when making predictions. This mechanism computes relationships among all input components, allowing the network to learn long-range dependencies and contextual relationships across data points.
%Transformers are DL architectures introduced for modeling sequential data such as, natural language, speech, or time series that replace recurrence and convolution with self-attention mechanisms~\cite{vaswani2017attention}. Transformers rely on an attention mechanism that enables the model to weigh the importance of different elements in an input sequence simultaneously. 
Along with self-attention, the key building blocks for a transformer 
include multi-head attention (to look at the relationships in multiple ways at the same time), positional encoding (to add information about word order), 
%(since attention alone doesn’t know sequence order)
and feed-forward layers (to process the attention results to make predictions). The self-attention produces context-aware representation of the input sequence. Multi-head attention repeats this computation in parallel so the model learns different types of relationships simultaneously. The resulting outputs are then combined into a single representation. Due to their scalability and parallel processing, transformers have become foundational to state-of-the-art architectures such as GPT.
%Process the attention results to make predictionsTransformers are composed of stacked encoder-decoder blocks, each containing self-attention layers and deep feedforward networks.  

%Formally, given an input sequence $\{\mathbf{x}_t\} \in \mathbb{R}^d$, $t=\{1,\ldots,T\}$, transformers first add positional encodings $\mathbf{p}_t$ to preserve order:
%
%\begin{equation}
%\mathbf{h}_t^{(0)} = \mathbf{x}_t + \mathbf{p}_t.    
%\end{equation}
%
%In the following layers, the model computes queries, keys, and values via linear projections $\mathbf{Q}=\mathbf{H}W_Q$, $\mathbf{K}=\mathbf{H}W_K$, and $\mathbf{V}=\mathbf{H}W_V$, where $\mathbf{H}$ stacks all token representations. Then, the similarity score matrix is defined as 
%
%\begin{equation}
%\mathbf{S}=\mathbf{Q}\mathbf{K}^\top/\sqrt{d_k}%\in\mathbb{R}^{T\times T}
%\end{equation}
%
%with entries $S_{tj}=\mathbf{q}_t^\top\mathbf{k}j/\sqrt{d_k}$ measuring how relevant token $j$ is to token $t$. The self-attention produces context-aware representations as 
%
%\begin{equation}
%\mathbf{Z}=\mathrm{Att}(\mathbf{Q},\mathbf{K},\mathbf{V})
%\;=\;
%\mathrm{softmax}\!\left(\mathbf{S}\right)\mathbf{V},
%\end{equation}
%
%where $\mathbf{Z}\in\mathbb{R}^{T\times d_v}$ contains the context aware representations 
%%\vspace{0.2cm}
\subsubsection{Autoencoders}
Autoencoders are DL-based architectures designed for unsupervised learning, where the primary objective is to learn low-dimensional representations of data~\cite{Rumelhart1986Learning}. An autoencoder consists of two main components: an encoder and a decoder. The encoder compresses the input into a compact, low-dimensional space representation---or code. The decoder reconstructs the original input from the compact space. Formally, given an input $\mathbf{x}\in\mathbb{R}^d$ drawn from an unknown data distribution $p_{\text{data}}(\mathbf{x})$, the encoder maps it to a latent variable $\mathbf{z}\in\mathbb{R}^k$, with $k<d$, as
\(\mathbf{z}(\mathbf{x})=\phi_e(\mathbf{W}_e\mathbf{x}+\mathbf{b}_e),\) where $\phi_e(\cdot)$ is typically a non-linear function, $\mathbf{W_e}$ is the encoder weight matrix, and $\mathbf{b}_e$ is the bias vector. The decoder reconstructs the input $\mathbf{x}$ from the latent code as:
\(\hat{\mathbf{x}} = \phi_d\!\left(\mathbf{W}_d \mathbf{z} + \mathbf{b}_d\right),\)
where $\phi_d(\cdot)$ is a non-linear function, $\mathbf{W}_d$ is the decoder weight matrix, and $\mathbf{b}_e$ is the bias vector. The training stage minimizes a reconstruction error as:
\begin{equation}
\mathcal{L}(\theta_e,\theta_d) = \mathbb{E}_{\mathbf{x}}
\Big[\ell\big(\mathbf{x},\hat{\mathbf{x}}\big)\Big],
\end{equation}
where $\theta_e=(\mathbf{W}_e,\mathbf{b}_e)$ and $\theta_d=(\mathbf{W}_d , \mathbf{b}_d)$. Then, the model finds the optimal parameters $(\theta_e,\theta_d)$ that minimize the reconstruction error as
\((\theta_e^\star,\theta_d^\star) = \min_{(\theta_e,\theta_d)} \mathcal{L} (\theta_e,\theta_d).\)
In short, autoencoders are NNs trained to reconstruct inputs $\mathbf{x} \;\mapsto\; \hat{\mathbf{x}} \approx \mathbf{x}$ through a dimensionality reduction process, only capturing meaningful structures in high dimensional data. More advanced variants of this framework include denoising autoencoders, sparse autoencoders, and variational autoencoders, specifically designed for noise reduction, feature extraction, and generative modeling, respectively.
%
%%\vspace{0.2cm}
\subsubsection{Large Language Models (LLMs)}

LLMs (e.g., GPT-4) are neural networks trained to understand and generate human language.
They have millions or billions of parameters (weights) to capture the complex patterns. They are trained using huge amounts of text from books, articles, and the web, and they learn grammar, facts, reasoning patterns, and some context understanding. Given a prompt, they generate text that is statistically likely based on training. LLMs characterize the probability distribution of token sequences~\cite{bengio2003neural}. 

%Given a token (or word) sequence $\{x_t\}$, $t=\{1,\dots,T\}$, LLMs model the autoregressive conditional distribution $p_\theta(x_t \mid x_1, \ldots x_{t-1})$. During training, LLMs maximize the negative log-likelihood loss:
%%
%\begin{eqnarray}
%%\scalebox{0.9}{
%\mathcal{L}(\theta) & = & -\mathbb{E}_{(x_1,\dots,x_T)\sim\mathcal{D}}   \notag \\
% &  &\left[ \sum_{t=1}^{T}
%\log p_\theta(x_t \mid x_1,\dots,x_{t-1})\right],  
%%}    
%\end{eqnarray}
%%
%where $\mathcal{D}$ is the training dataset and $\theta$ are the model parameters. 

LLMs are typically implemented using transformer architectures, where tokens are mapped to vector embeddings and  processed through stacked self-attention layers. These layers compute context-aware representations by weighting the relevance of all tokens in the sequence relevant to specific inputs. Since LLMs are typically trained on very large datasets, they normally acquire broad reasoning capabilities, serving as ``foundation'' models for general-purpose sequence applications such as conversational AI systems. 
\begin{table*}[htbp]
\centering
\caption{Comparison among major forms of scientific knowledge representation and their characteristics in scientific and engineering domains}
\label{tab:interpretability_methods}
\resizebox{\linewidth}{!}{
%\small
\begin{tabular}{p{3.5cm} p{4.5cm} p{4.5cm} p{4.5cm}}
%\begin{tabular}{|p{2cm}|p{5cm}|p{5cm}|p{5cm}|p{5cm}|}
\toprule
\textbf{Representation} & \textbf{Structure} & \textbf{Examples} & \textbf{Relevance for ScIDL} \\
\midrule
\textbf{Algebraic Equations} &
Closed-form relationships between variables from an element set are combined through algebraic operators. Typical algebraic expressions take the form of polynomial relations:
$$
p(x)=a_0+a_1x+\ldots+a_nx^n,
$$ 
for an independent variable x. & ${\rm SINR}$ for user $k$ in a wireless access network is:
$$
\mathrm{SINR}_k = \frac{p_k |h_{kk}|^2}{\sum_{j\neq k} p_j |h_{kj}|^2 + \sigma_k^2},
$$ 
where $p_k$, $p_j$ are transmit powers, $h_{kk}$, $h_kj$ are channel gains, and $\sigma^2_k$ is the noise power.
& Encode static constraints, feasibility regions, and steady-state relationships into DL architectures, loss functions, or optimization algorithm steps. \\
\addlinespace
\midrule
\textbf{Differential Equations} &
Generalized algebraic equations that describe relationships between functional quantities and their rates of change. They can be written as:
$$
F\!\left(x,\,u(x),\,\frac{du}{dx},\,\ldots,\,\frac{d^n u}{dx^n}\right) = 0,
$$
for an independent variable x, functional relationship $u(x)$ and its derivatives, and a non-linear function $F$. &
Electromagnetic-wave propagation equation:
$$
\nabla^2 \mathbf{E}(\mathbf{r},t) - \mu\varepsilon\,\frac{\partial^2 \mathbf{E}(\mathbf{r},t)}{\partial t^2} = \mathbf{0}, 
$$
where $\mathbf{E}(\mathbf{r},t)$ is the electric field at spatial location $\mathbf{r}$, time-stamp $t$. $\mu$ is the magnetic permeability, and $\epsilon$ is the electric permittivity.
& Provide physics residuals for loss functions, constrain learning outputs, and enforce physical consistency.
 \\
\addlinespace
\midrule
\textbf{Logic Rules} &
Propositional logic expressions representing discrete relationships among system variables. Typical logic rules take the form $r(\mathbf{z}) \in \{0,1\}$, where $\mathbf{z}=\{z_1 \ldots z_n\}$ are relevant system variables, and $r$ comprises first-order logical operators AND $(\wedge)$, OR $(\vee)$, NOT $(\neg)$, and IMPLIES $(\Rightarrow)$.
&
Control systems, protocol design, safety specifications, scheduling constraints, etc. &
Enable enforcement of rule-based science regularization, safety specification compliance, and physics constraint agreements for DL models etc. \\
\addlinespace
\midrule
\textbf{Geometric / Topological Properties} &
%Object or function characteristic
Invariances, symmetries, and manifold structures observed in objects or functions within the system’s space. For a transformation group $\mathcal{G}$ acting on an input parameter space $\mathcal{X}$, a function $f$ is invariant if $(f \circ g)(x)=f(x)$, for all $g \in \mathcal{G}$, $x \in \mathcal{X}$.
& Spatial symmetries in antenna arrays, permutation invariance in multi-user networks, and signal manifolds in wireless channels. & Enable symmetry and invariance incorporation in DL architectures through equivariant layers, parameter sharing, and geometry-aware embeddings. \\
\addlinespace
\midrule
\textbf{Graph-Based Models} &
Structured node and edges representation of a system’s entities and their interactions. A graph $\mathcal{G}$ is encoded as $\mathcal{G} = (\mathcal{V},\mathcal{E})$, where $\mathcal{V}$ is a set of nodes and $\mathcal{E}$ denotes their connecting edges. & Network topology representations. & Encode topology, relational structure, and message passing through graph structures. \\
\addlinespace
\midrule
\textbf{Production Rules} &
Control rules in role-based reasoning systems. A set of $M$ production rules define a rule base $\mathcal{R}$ from which production rules $r_j \in \mathcal{R}$ are written as $r_j:\quad \text{IF } c_j(\mathbf{z}) = \text{true} \;\; \text{THEN } a_j(\mathbf{z})$, where $\mathbf{z} \in \mathcal{Z}$ is the system state, $c_j:\mathcal{Z}\rightarrow\{0,1\}$ is the condition, and $a_j:\mathcal{Z}\rightarrow\mathcal{A}$ is an action.  
& Expert systems, protocol control logic, scheduling policies, fault diagnosis, decision-making systems. & Enable the incorporation of rule-based domain knowledge into learning systems. \\
\bottomrule
\end{tabular}
}
\end{table*}
%   
%\end{comment}
%Used to impose equality/inequality constraints, normalize outputs, reduce parameter search spaces, or design specialized layers (e.g., normalization tied to conservation laws).
\subsection{The Nature of Scientific Knowledge}
Scientific knowledge refers to the understanding gained through the systematic study of the structure and dynamics of the natural and physical world. It is primarily developed through observation, measurement, and experimentation, followed by the formulation of formal models and theories that explain and predict observed phenomena. The sources of scientific knowledge can be broadly categorized into physics-based knowledge, empirical world facts, and expert knowledge. A discussion on each category follows. 
%
%%\vspace{0.2cm}
\subsubsection{Physics Knowledge} 
Physics knowledge refers to the fundamental principles, theories, and laws that describe and govern the physical phenomena, typically expressed through mathematical formulations. For instance, in wireless communication, Maxwell’s equations define the generation, propagation, and interaction of electromagnetic waves as follows:
\begin{equation}
\nabla \times \mathbf{E} = -\frac{\partial \mathbf{B}}{\partial t}, \quad \nabla \times \mathbf{H} = \mathbf{J} + \frac{\partial \mathbf{D}}{\partial t},
\end{equation}
where $\mathbf{E}$ and $\mathbf{H}$ represent the electric and magnetic field intensities that carry information, $\mathbf{D}$ is the electric displacement field and $\mathbf{B}$ is the magnetic flux density that characterize how materials respond to electromagnetic fields, and $\mathbf{J}$ denotes the electric current sources that generate electromagnetic radiation. In the information-theoretic domain, the Shannon–Hartley theorem establishes the upper bound of reliable communication data rate over a noisy channel as 
\(C = B \log_2\!\left(1 + \frac{P}{N_0 B}\right), \)
where $B$ is the channel bandwidth and $P/N_0$ is the signal-to-noise ratio (SNR). Also, in signal processing, the Nyquist sampling theorem states that a band-limited signal with maximum frequency $f_{max}$ can be perfectly reconstructed if sampled at a rate:
\(f_s \ge 2f_{max}.\)
These mathematically grounded principles encode the invariant physical constraints and performance limits, providing a reliable prior knowledge that can be embedded into learning-based models to guide the inference and optimization solutions.
%
%%\vspace{0.2cm}
\subsubsection{World Facts}
World facts refer to empirical observations derived from the natural world. They represent objective data collected from real environments and are used to inform and validate the models and theories. Mathematically, they can be modeled as samples $\{(x_i, y_i)\}_{i=1}^N$ drawn from an unknown real-world data-generating distribution $p_{\text{world}}(x,y)$. For instance, real-world measurements of wireless signal propagation in urban environments capture the path-loss, shadowing, and multipath fading, which are often expressed as
\begin{equation}
P_r(d) = P_t - \big(PL_0 + 10n\log_{10} d + X_\sigma\big),
\end{equation}
where $P_r(d)$ is the received power at distance $d$, $n$ is the path-loss exponent, and $X_{\sigma}$ is a random variable representing shadowing. These empirical observations encode environment-specific characteristics such as building density and material properties.
%, often difficult to capture fully with principle models alone. 
World facts serve as critical inputs for constraining and validating learning-based models. 
%
%%\vspace{0.2cm}
\subsubsection{Expert Knowledge} 
Expert knowledge refers to the specialized understanding acquired through extensive training and practical experience within a specific domain. It is often encoded as rules, heuristics, constrains, or decision policies. Mathematically, expert knowledge can be represented as a set of functions or constrains $\{g_j(\cdot)\}_{j=1}^M$ that restrict the feasible decisions or parameter spaces. For example, in wireless security, access control and intrusion detection policies can be expressed as decision rules of the form:
\(a = \pi_{\text{exp}}(s) \in \mathcal{A}_{\text{safe}},\)
where $s$ denotes the system state, $\pi_{exp}$ is an expert design policy, and $\mathcal{A}_{safe}$ defines feasible actions that mitigate the threats. These policies are typically developed by security professionals with deep insight into the vulnerabilities of wireless networks. 
%Expert knowledge is particularly valuable when designing engineering solutions in limited-data or high-risk settings.
%
\subsection{Scientific Representation}
%In this section, we introduce a set of. 
Scientific knowledge can be represented through a variety of formal structures, including differential equations, logic and algebraic rules, graph-based models, production rules, and geometric or topological properties~\cite{cercone1987knowledge, von2021informed, russell2016artificial}. Such representations enable a systematic integration of domain knowledge into learning architectures by explicitly encoding the physical laws, structural relationships and invariances. A discussion on each structure follows.
%
%%\vspace{0.2cm}
\subsubsection{Algebraic Equations}
Algebra deals with expressions and equations involving elements of sets -- such as the set of real numbers $\mathbb{R}$ -- combined through algebraic operations. A typical algebraic expression is a polynomial of the form 
\(p(x) = a_0 + a_1 x + \cdots + a_n x^n,\)
from which algebraic equations define dependencies between variables and constants through equality or inequality relationships, such as
\(p(x)=0.\)%
%These equations capture static phenomena. 
In wireless communications, for instance, an algebraic expression models the signal-to-interference-plus-noise ratio (SINR), which characterizes link quality in steady-state conditions. For user $k$, the ${\rm SINR}$ is expressed as:
\begin{equation}
\mathrm{SINR}_k = \frac{p_k |h_{kk}|^2}{\sum_{j\neq k} p_j |h_{kj}|^2 + \sigma_k^2},
\end{equation}
where $p_k$ is the transmit power of user $k$, $h_{k_j}$ denotes the channel gain from transmitter $j$ to receiver $k$, and $\sigma_k^2$ is the noise power. This algebraic relationship captures the static interference among users communicating in a wireless environment.
%
%%\vspace{0.2cm}
\subsubsection{Differential Equations}
Differential equations generalize algebraic equations by describing relationships between functional quantities and their rates of change, and are fundamental for modeling dynamic systems. They can capture non-steady physical phenomena such as wave propagation. In a general form, differential equations can be written as
\begin{equation}
F\!\left(x,\,u(x),\,\frac{du}{dx},\,\ldots,\,\frac{d^n u}{dx^n}\right) = 0,
\end{equation}
where $x$ is an independent variable, $u(x)$ is a functional quantity, $\frac{d^k u}{dx^k}$ denote the $k$-th derivate of $u$, for $k=\{1,\ldots,n\}$, and $F(\cdot)$ is a possibly non-linear function. For example, Maxwell's equations, which govern electromagnetic wave propagation, lead to the wave equation for the electric field as follows:
\begin{equation}
\nabla^2 \mathbf{E}(\mathbf{r},t) - \mu\varepsilon\,\frac{\partial^2 \mathbf{E}(\mathbf{r},t)}{\partial t^2} = \mathbf{0}, 
\end{equation}
where $\mathbf{E}(\mathbf{r},t)$ is the electric field, and $\mu$ and $\varepsilon$ denote the magnetic permeability and electric permittivity of the medium, respectively. This partial differential equation characterizes how electromagnetic waves propagate through space, interact with materials, and attenuate with distance. 
%
%%\vspace{0.2cm}
\subsubsection{Logic Rules}
Logic rules provide a formal mechanism for representing knowledge through propositional logic, where statements take binary truth values. Mathematically, a logic rule can be expressed as a Boolean function
\(r(\mathbf{z}) \in \{0,1\},\)
where $\mathbf{z} = (z_1,\dots,z_n)$ are relevant system variables on which the logic rule operates. The rule comprises logical operators such as AND $(\wedge)$, OR $(\vee)$, NOT $(\neg)$, and IMPLIES $(\Rightarrow)$. In wireless communications, an access control rule can encode the protocol constraints and expert knowledge and may be written as
\(({\rm Auth}_u \wedge {\rm SNR}_u \ge \gamma) \;\Rightarrow\; \text{GrantAccess}_u,\)
where ${\rm Auth}_u$ indicates user authentication and $\gamma$ is the minimum ${\rm SNR}$ threshold. 
%
%%\vspace{0.2cm}
\subsubsection{Geometric Properties} These properties describe the characteristics of objects or functions in the system that remain unchanged under mathematical transformations such as translation, rotation, or scaling. Let $G$ be a transformation group acting on an input space $\mathcal{X}$. A function $f:\mathcal{X}\to\mathcal{Y}$ is invariant to $G$ if 
\((f \circ g)(x) = f(x),\)
for all $g \in G$, $x \in \mathcal{X}$. Also, $f$ is equivariant to $G$ if there exists a representation $\rho$ on $\mathcal{Y}$ such that
\((f \circ g)(x) = \rho(g)\, f(x).\)
When an object is unaffected by a specific transformation, it is said to exhibit symmetry. Leveraging symmetry is fundamental in scientific modeling and DL, as these principles enable the models to generalize efficiently by recognizing the patterns that persist across transformed or perturbed representations of the same phenomenon. 
%Similarly, a function is considered invariant if it produces the same output when the input undergoes a symmetric transformation.
%
%%\vspace{0.2cm}
\subsubsection{Graph-based Models}
Graphs provide a structured framework to represent knowledge through interconnected nodes, also known as vertices, and edges. Formally, a graph-based model is defined as a graph
\(\mathcal{G} = (\mathcal{V}, \mathcal{E}),\)
where  $\mathcal{V} = \{v_1,\dots,v_N\}$ is a set of nodes and $\mathcal{E} \subseteq \mathcal{V}\times\mathcal{V}$ is a set of edges encoding relationships among nodes. In this representation, the nodes denote entities or concepts, while the edges capture the relationships or dependencies between them. Common graph-based frameworks include semantic and Bayesian networks. 
%Semantic networks are graph-based representations in which the nodes denote concepts or entities, and the edges represent the relationships among them. Mathematically, semantic relations can be represented as labeled edges $(v_i, r_{ij}, v_j) \in \mathcal{E}$, where $r_{ij}$ denotes the type of relationship between concepts $v_i$ and $v_j$. These networks encode meaning through the structure of interconnected concepts, allowing machines to reason about relationships and hierarchies within the knowledge domains. Semantic networks play a fundamental role in natural language processing (NLP) by capturing the semantics of words, phrases, and their contextual association.
%
%Bayesian networks, also known as belief networks or probabilistic graphical models, are represented as directed acyclic graph (DAG) $\mathcal{G}$ in which each node $X_i$ represents a random variable and the edges encode conditional dependencies. The joint probability distribution of the nodes can be expressed as factors according to the graph structure as follows:
%%
%\begin{equation}
% p(X_1,\dots,X_N) = \prod_{i=1}^{N} p\!\left(X_i \mid \mathrm{Pa}(X_i)\right),   
%\end{equation}
%%
%where $\mathrm{Pa}(X_i)$ are the parent nodes of $X_i$. Bayesian graph models can be used to represent and reason about the knowledge under uncertainty, enabling inference over complex, independent systems.
%
\subsubsection{Production Rules} Production rules represent procedural knowledge and control in role-based reasoning systems. A set of $M$ production rules define a rule base $\mathcal{R} = \{r_1,\dots,r_M\}$ from which the production rules $r_j \in \mathcal{R}$ are typically written as
\begin{equation}
r_j:\quad \text{IF } c_j(\mathbf{z}) = \text{true} \;\; \text{THEN } a_j(\mathbf{z}),
\end{equation}
where $\mathbf{z} \in \mathcal{Z}$ denotes the system state (e.g., observed context variables), $c_j:\mathcal{Z}\rightarrow\{0,1\}$ is a Boolean-valued condition, and $a_j:\mathcal{Z}\rightarrow\mathcal{A}$ is an action. $\mathbf{z}$, the antecedent, specifies the criteria that must be satisfied for the rule to activate, and $c_j$, the consequent, defines the operation to be executed when the condition holds true, implementing the decision functions as follows:
\( a(\mathbf{z}) =
\sum_{j=1}^{M} \mathbf{1}_{\{c_j(\mathbf{z})=1\}} \, a_j(\mathbf{z}),\)
where $\mathbf{1}_{\{c_j(\mathbf{z})=1\}}$ is the indicator function that activates on $c_j(\mathbf{z})=1$, and $a_j{(\mathbf{z})}$ is the action produced by rule $j$ when it fires. Production rules enable the systems to reason through chains of conditional statements, supporting inference, decision-making, and automated control. 
\section{Taxonomy of ScIDL: Embedding Science into DL Models}%Bridging Science and Data for Transparent DL Models
\begin{figure*}
    \centering
    \includegraphics[width=0.75\linewidth]{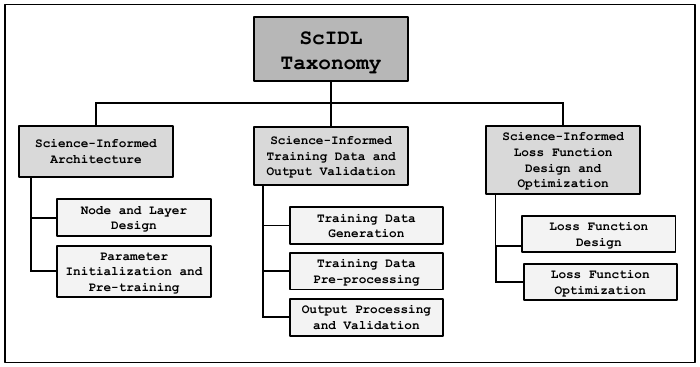}
    \caption{A conceptual overview of the proposed ScIDL taxonomy, organizing science-informed DL methods into three dimensions: (1) Architecture Design, where scientific knowledge is embedded through node/layer design, parameter initialization, and pre-training strategies; (2) Training Data and Output Validation, which integrates scientific principles into data generation, processes and physics-based input and output consistency checks; and (3) Loss Function Formulation and Optimization, which incorporates scientific constraints, governing equations, and priors directly into the loss formulation and the corresponding optimization process.}
    \label{fig:taxonomyMap}
\end{figure*}
%
%\addatefeh{In the previous section, we reviewed the non-transparent challenges of DL and discussed existing analysis methods that provide insight into DL models by examining their interpretability and generalization.}
%In this section, we study how to embed science-based knowledge into DL methods to improve interpretability, generalization, and architecture design. 
%, addressing the challenges that arise from their non-transparent nature. 
As discussed earlier, purely data-driven DL models offer limited insights into how information is represented and processed internally. Also, these models can sometimes produce outcomes that contradict established scientific principles or regulatory standards. For example, DL-based wireless algorithms may infer channel behaviors inconsistent with known laws of electromagnetic propagation, raising concerns about reliability and physical plausibility~\cite{akrout2023domain}.  Another fundamental limitation of purely data-driven DL models is their limited ability to generalize beyond the training conditions, which often leads to degraded performance under unseen scenarios.  Consequently, applying DL methods to non-stationary and highly dynamic wireless environments remains a significant challenge. 

ScIDL is an emerging paradigm to tackle these limitations by embedding scientific knowledge (e.g., laws, rules, models, and physical properties) into different stages of the DL pipeline. This integration not only reduces reliance on massive training datasets but also helps ensure that the learned representations remain consistent with established scientific principles. To organize the diverse integration strategies, we propose a structured taxonomy that defines the conceptual dimensions of ScIDL. This taxonomy serves as a methodological framework for classifying the existing methods and as a guidance for future SciDL design and models. It organizes  integration of scientific knowledge into DL systems  into three broad categories:  
\begin{itemize}
    \item Science-Informed DL Architecture,
    \item Science-Informed Training Data and Output Validation,
    \item Science-Informed Loss Function Design and Optimization.
\end{itemize}
Each dimension captures a distinct way in which domain-specific knowledge can be incorporated into different components of a DL pipeline.
In doing so, it transforms the conventional, non-transparent models into models that are more interpretable, more generalizable, and more physically consistent. A detailed discussion of each category follows, and the organizational map of the proposed ScIDL taxonomy is provided in Figure~\ref{fig:taxonomyMap}.
%
%\vspace{-10}
\subsection{Science-Informed DL Architecture}

The modular design of NNs makes them naturally well-suited for architectural modifications. The architecture of a DNN refers to its nodes, layers, and their interconnections. In a ScIDL architecture, a DL model is augmented with a set of scientific models $\mathcal{M} = \{g_j(\cdot)\}_{j=1}^{M}$ that encode the known domain relationships such as differential equations, conservation laws, symmetries, or other physical principles. Formally, the model in~(\ref{equation:DLmodel}) is updated to:
%$f(\mathbf{x}; \boldsymbol{\theta})$
%
%\begin{equation}
%\mathbf{y} = f_{\text{ScI}}(\mathbf{x}; \boldsymbol{\theta}, \mathcal{M})\\
%\mathbf{y} =  f^{(L)}g \circ f^{(L-1)} g\circ \cdots \circ f^{(1)}g(\mathbf{x})   
%\end{equation}
\begin{eqnarray}
%\scalebox{0.8}{
\mathbf{y}
& = & 
\bigl(f^{(L)} \circ g^{(L)}\bigr)
\circ
\bigl(f^{(L-1)} \circ g^{(L-1)}\bigr)
\circ \cdots \notag \\
 &  & 
\circ
\bigl(f^{(1)} \circ g^{(1)}\bigr)(\mathbf{x},\boldsymbol{\theta}),
\label{eq:ScIDL_comp}
%}
\end{eqnarray}
where $f^{(\ell)}(\cdot)$ is the $\ell$-th learnable DL layer (the transformation performed at the $\ell$-th layer), and $g^{(\ell)}(\cdot)$ is a designed scientific module, chosen from $\mathcal{M}$. Therefore, the modules in $\mathcal{M}$ are incorporated into the architecture by modifying the nodes, layers, and/or their interconnections, as well as by informing parameter initialization (e.g., network weights). This section describing our proposed taxonomy is organized into two categories:
\begin{itemize}
\item Science-Informed Node and Layer Design, 
\item Science-Informed Parameter Initialization and Pre-training.
\end{itemize}
%
%\begin{itemize}
%    \item Nodes and Layers
%    \item Loss Function
%    \item Domain Variables
%    \item Invariance and Symmetry
%\end{itemize}
% 
A summary of the these two categories of ScIDL architectures with examples can be found in Table~\ref{tab:dl-architecture-taxonomy}.

\begin{table*}[htbp]
\centering
\caption{ScIDL architecture: (1) node and layer design with embedded topological priors, and (2) parameter initialization and pre-training strategies with embedded domain knowledge}
\label{tab:dl-architecture-taxonomy}
\resizebox{\linewidth}{!}{
%\small
\begin{tabular}{p{1.5cm} p{4cm} p{4cm} p{4cm}}
%\begin{tabular}{|p{2cm}|p{5cm}|p{5cm}|p{5cm}|p{5cm}|}
\toprule
\textbf{Work} & \textbf{Model in a Few Words} & \textbf{Science-Informed Aspect} & \textbf{Gains} \\ 
\midrule
\multicolumn{4}{l}{\textbf{\textit{Node and Layer Design}}} \\
%\midrule
%\multicolumn{4}{l}{\textit{Graph Neural Networks}} \\
\midrule
\cite{zhang2022topology} & Graph-based DL model with an attention mechanism for resource allocation problems & The GNN architecture is guided by embedding the network's physical topology in the graph. & Near-optimal performance with reduced computational complexity even for unseen topologies\\ 
\addlinespace
\cite{9252917} & MPGNN-based graph convolution network for beamforming optimization problems & The GNN architecture is guided by embedding the network's physical topology in the graph. & Performance surpasses traditional WMMSE optimizers along with strong generalization abilities. \\ 
\addlinespace
\cite{9072356} & Graph-based DL model for optimal resource allocation policies & The GNN architecture is shaped by the network's permutation-invariant topology in the graph. & Performance is comparable to fully-connected NNs with reduced computational complexity. \\
%\midrule
%\textit{Deep Unfolding}
\addlinespace
\cite{10021866,9414561,9667094} & Deep-unfolding model with no matrix inversions for beamforming optimization problems & The deep unfolding architecture is determined by the structure of the WMMSE algorithm. & Near-optimal performance with reduced computational complexity  \\
\addlinespace
\cite{daw2020physics} &
Physics-guided LSTM autoencoder for resource allocation &
The LSTM model embeds physics-informed connections and intermediate processing variables. &
Performance is comparable to baselines, with demonstrated generalization abilities. \\
\addlinespace
\cite{9024538} & Graph-based DL model for power control problems & The graph embeds permutation invariances from interference links in the architecture. & Performance surpasses data-driven MLP, CNN and the model-based WMMSE algorithm. \\
\midrule
\multicolumn{4}{l}{\textbf{\textit{Parameter Initialization and Pre-training}}}\\
\midrule
\cite{10974467} & RNN-based architecture for OFDM symbol detection in MIMO systems & ESN/WESN parameters are deterministically derived from channel statistics and IIR characterizations. & Performance surpasses randomly initialized DL-based detectors with reduced computational complexity. \\
\addlinespace
\cite{8335785} & CNN-based framework for power allocation problems & The model's loss function embeds spectral/energy efficiency terms from Shannon's capacity law. & Performance is comparable to WMMSE with less computation and signaling overhead. \\
\bottomrule
\end{tabular}
}
\end{table*}
\subsubsection{Science-Informed Node and Layer Design}
In the following, we outline the primary directions for embedding scientific knowledge into the design of nodes and layers of DL models. 
\paragraph{Graph Neural Networks}
In wireless communication systems, an increasing number of studies have advocated the use of graph neural networks (GNNs) to embed graph-based knowledge into NN architectures~\cite{he2021overview}. This improves the ability of DL models to represent and learn complex interactions among wireless network elements. %Because network topologies often change dynamically due to channel variation and user mobility, the underlying graph structures are time-varying. 
As an example, Zhang et al.~\cite{zhang2022topology} propose a topology-aware DL (TADL) model that explicitly incorporates both the inter-dependencies among communication devices and the evolving nature of wireless networks. In their work, a wireless network is modeled as a directed graph $\mathcal{G}=\{\mathcal{N},\mathcal{E}\}$, where $\mathcal{N}$, the graph nodes, is the set of wireless devices and $\mathcal{E}\subseteq\mathcal{N}\times\mathcal{N}$, the graph edges, is the set of directed links $(u,v)$, whenever $v$ is within the communication range of $u$.  The problem of optimal link scheduling in order to maximize total network capacity  is formulated as a network flow optimization problem subject to individual link capacity constraint as well as interference constraints. The problem  can be solved with efficient approximation methods such as the delayed column generation (DCG) algorithm~\cite{Patil2024}, however the complexity grows significantly when the number of links grows. The authors replace model-based solvers with a graph neural network (GNN)-based message-passing architecture to embed the wireless graph and predict the optimal traffic flows.  During training, the model processes sample graph instances and learns to compute optimal routing, link scheduling, and resource allocation decisions through an attention mechanism and a weighted loss function that prioritizes more critical links. The authors report near-optimal performance with substantial reductions in computational complexity and strong generalization to unseen network topologies. 

%Each link has a physical capacity $c(u,v)$, representing the supported peak data rate, and scheduling variable $x(u,v) \in \{0,1\}$, 1 if the link is active, 0 otherwise. In the network, there exists a set of commodity flow demands $\mathcal{K}$, where each flow $k$ is defined as a source-destination pair $(s_k,d_k)$, where each of these must be valid nodes in the network. The conflict relations among links can be characterized by a conflict graph, where the edges represent pairs of links that cannot be active at the same time, i.e., if $\ell_1$, $\ell_2 \in \mathcal{E}$ are conflicting links, then $x(\ell_1) + x(\ell_2) \le 1$. An independent set (IS) over the conflict graph indicates a set of links that can be scheduled for transmission simultaneously. The function $f_k(u,v)$ denotes the amount of traffic flow associated with commodity $k$ on the link $(u,v)$. For commodity $k$, the net flow $\lambda_k$ out of the source node is defined as:
%%
%\begin{equation}
%\lambda_k=\sum_vf_k(s_k,v))-f_k(v,s_k).    
%\end{equation}
%%
%The capacity constraint for each link $(u,v)$ is defined as:
%%
%\begin{equation}
%\sum_{k \in \mathcal{K}} f_k(u,v) \le c(u,v) \cdot x(u,v).
%\end{equation}
%%
%The network flow optimization problem maximizes the total flow $z$ delivered by the network. Then, the maximum achievable flow becomes
%%
%\begin{equation}
%z^* = \max_{f_k,x} \sum_{k \in \mathcal{K}} \lambda_k,    
%\end{equation}
%
%subject to capacity and interference constraints. 
%%INCLUDE A FIGURE FOR GNN
%%Explain the learning pipeline and highlight ScIDL aspects
Similarly, Shen et al.~\cite{9252917} propose a wireless channel graph convolution network (WCGCN) based on a message passing GNN (MPGNN) framework for scalable radio resource management. The network is modeled as a graph where the nodes represent transmitters or receivers and  the edges correspond to communication links. The message-passing mechanism enables the nodes to exchange information with their neighboring nodes and iteratively converge toward optimal allocations of beamforming vectors, transmit powers, and scheduling policies, while maintaining  a linear computational complexity. The model is permutation equivariant and can be trained with relatively small datasets. Numerical results demonstrate that the WCGCN solution outperforms the classical weighted minimum mean square error (WMMSE) algorithm in sum-rate maximization. It also identifies near-optimal beamforming solutions. Moreover, it exhibits an acceptable generalization to larger network topologies than those seen during training. 

Eisen and Ribeiro~\cite{9072356} propose a graph-based DL method for wireless resource allocation called random edge GNN (REGNN). The method models a wireless network as a directed graph in which the nodes correspond to transmitters, receivers, or links, and the edges encode interference relationships. REGNN learns the parameterized resource-allocation policies through a model-free primal–dual learning approach, requiring significantly fewer parameters than traditional fully-connected NNs. Each REGNN layer performs graph convolutions that aggregate localized neighborhood information, mirroring how interference propagates through wireless networks. The authors prove that the REGNN policies are permutation equivariant, allowing them to generalize across networks with similar topologies or arbitrary relabelings. Numerical results show that, REGNN achieves performance comparable to fully connected architectures at a fraction of the complexity. Also, the model exhibits generalization: models trained on networks with 50 nodes maintain a near-optimal performance when deployed in substantially larger networks with 75–100 nodes.
%~\cite{10021866}
\paragraph{Deep Unfolding}
Another approach to embed scientific knowledge into DL architectures is deep unfolding (also called ``algorithm unrolling"). 
Deep unfolding constructs a neural network by mapping the iterations of a classical optimization algorithm  (e.g., iterative shrinkage-thresholding algorithm [ISTA], alternating direction method of multipliers [ADMM]) into a finite number of network layers, where each layer corresponds to one iteration. The layer-wise operations follow the computational template of the underlying optimization algorithm (e.g., linear updates, proximal steps, and constraint projections), thereby embedding the problem’s model/constraints and associated domain knowledge. Meanwhile, a subset of algorithmic parameters (e.g., step sizes, thresholds, or penalty weights) is learned from data to improve the solution quality and accelerate the convergence. The authors in~\cite{10021866, 9414561, 9667094} addressed the beamforming problem in multi-user MISO and MIMO wireless networks, respectively. Classical solution approaches based on WMMSE optimization are computationally demanding due to iterative matrix inversions. To overcome this, they propose a matrix-inverse-free deep unfolding (MIFDU) network that decompose the steps of the WMMSE algorithm into a DL architecture. Each layer of the MIFDU corresponds to one iteration of the traditional optimization algorithm. Also, the matrix inversion operations are replaced with learnable linear transformations, allowing the model to learn efficient beamforming mappings through data-driven training. By using deep unfolding, the authors embed the scientific structure and mathematical dependencies of the WMMSE algorithm directly into the DL architecture. This ensures consistency with the underlying physics of the wireless network. In general, deep unfolded networks are much faster (fewer iterations needed), produce faster inference, are often more accurate, and are more interpretable than conventional deep networks. The results show that MIFDU achieves near-optimal beamforming performance comparable to conventional WMMSE while significantly reducing computational complexity and improving scalability across different network sizes and configurations. %Additional models incorporating the physical topology of wireless networks in the configuration of NN nodes and layers can be found in~\cite{} and~\cite{}.
%Zhuang et al.~\cite{8745681} present an alternative to traditional routing protocols such as OSPF and heuristic optimization algorithms, which often struggle with complex and time-varying network conditions. They propose a centralized graph-aware DL architecture that leverages graph topology to capture the structure of the network. The model encodes the network state into node and edge features---including load, latency, throughput, and connectivity---and performs graph-node convolutions that aggregate neighborhood information, mimicking how signals propagate across links. The trained model maps these network states to optimal next-hop routing decisions, enabling adaptive route planning. Training is performed offline via supervised learning using cross-entropy loss with L2 regularization and the Adam optimizer, with datasets derived from both real-world and synthetic topologies. Numerical results show that GADL achieves higher accuracy and significantly faster training times compared to standard DL methods such as DNNs, CNNs, and deep belief architectures (DBA).
\paragraph{Model Decomposition}
%\textbf{Science-Informed Decomposition}: 
%Another approach to integrating domain knowledge into the DL is to %embed science meaning directly into the learning pipeline by define and compute physically relevant variables along the model pathway.  
DL inter-layer connections can be designed by decomposing a model into physically motivated sub-tasks. where each sub-task is implemented by dedicated layers whose inputs and outputs mirror the dependencies of the underlying domain model. As an example, in a THz-enabled BS, resource allocation must jointly satisfy the physical constraints and QoS requirements while maximizing spectral efficiency. In the science-informed framework of~\cite{daw2020physics}, an LSTM-based autoencoder first extracts the temporal features from historical traffic and mobility patterns. These temporal features are combined with environmental parameters (e.g., water-vapor concentration, link distance, and THz carrier frequency) and passed to a second LSTM that predicts the path-loss while enforcing propagation consistency through a science-informed loss function. A final multi-layer perceptron (MLP) combines the temporal and physical features with the user QoS demands to determine resource allocation. This example illustrates how computing intermediate physical variables within the network architecture improves interpretability compared to non-transparent purely data-driven designs.
\paragraph{Invariance and Symmetry Integration}
Symmetries such as translational and rotational invariance serve important roles in physics, influencing the formulation of fundamental laws governing nature's behavior~\cite{gross1996role}. In conventional DL models, state-of-the-art architectures already capture certain types of invariances. For instance, the RRNs capture the temporal invariances, while the CNNs implicitly handle spatial translation, rotation, and scale invariances. Similarly, other types of invariances based on physical laws can be embodied in DL models. 

In many engineering and physical systems, the underlying laws are invariant under certain transformations (e.g., rotation, scaling, permutation, or phase shifts). ScIDL incorporates these invariances directly into the model architecture or preprocessing, allowing the network to learn more efficiently, with fewer samples, and with stronger guarantees of generalization.  For example, in wireless communications, the received signal is phase-rotation invariant: multiplying the complex baseband signal by $e^{j\theta}$
 does not change the underlying channel quality.  Instead of feeding the raw complex input $h$, we input a symmetry-invariant representation such as: magnitude, $|h|$;  power, $|h|^2$; normalized vector, $[\Re(h) /|h|, \Im(h) /|h|]$.  This ensures the network cannot be confused by arbitrary phase rotations, and it learns the underlying physical mapping more efficiently. The authors in \cite{wang2020incorporating} introduced several techniques to enforce various symmetries such as translation, rotation, uniform motion, and scale in CNNs for modeling dynamical systems.
%For example, invariant tensor bases can be used to incorporate rotational invariance into a neural network for improved fluid prediction accuracy in turbulence modeling \cite{ling2016reynolds}. Additionally, 

%A more recent trend involves incorporating spatial invariances as part of the CNN architecture, giving rise to a field known as geometric DL.
%Symmetries such as translational and rotational invariance serve important roles in physics, influencing the formulation of fundamental laws governing nature's behavior. %\cite{gross1996role}. Therefore, incorporating symmetries into DL models is likely to enhance the physical consistency and generalizability of DL. State-of-the-art DL architectures already capture certain types of invariance. For instance, recurrent neural networks (RNNs) capture the temporal invariance, while CNNs implicitly handle the spatial translation, rotation, and scale invariance. A more recent trend involves incorporating spatial invariances as part of the CNN architecture, giving rise to a field known as geometric DL. Similarly, it is required to embody in DL methods other types of invariance based on physical laws. For example, %invariant tensor bases can be used to embed rotational invariance into a neural network for improved fluid prediction accuracy in turbulence modeling \cite{ling2016reynolds}. Additionally, the authors in \cite{wang2020incorporating} introduced several techniques to enforce various symmetries such as translation, rotation, uniform motion, and scale in CNNs for modeling dynamical systems.

Shen et al.~\cite{9024538} introduce IGCNet, a graph-neural architecture for scalable wireless power control that models the $K$-user interference channel as a complete interference graph. IGCNet embeds the domain-specific symmetries, specifically the permutation invariance of interference wireless links, through symmetric graph-aggregation operations. This ensures that the architecture reflects the physical structure of wireless networks. Moreover, the model is trained with an unsupervised loss equal to the negative weighted sum-rate, which directly encodes the Shannon SINR dynamics into the optimization objective. Numerical evaluations show that IGCNet consistently outperforms MLP and CNN baselines and even surpasses the classical WMMSE algorithm, while offering substantial computational reduction and acceptable generalization to larger networks and partial-CSI settings. Although enforcing physical invariances within DL models is not extensively explored in the wireless-systems literature, current evidence suggests that it can improve interpretability and enhance their ability to generalize beyond the training distribution.
%%\vspace{0.2cm}
\subsubsection{Science-Informed Parameter Initialization and Pre-training}
DL models require initial parameter values (weights and biases) to begin the learning process. Inadequate initialization can lead to slow convergence, vanishing/exploding gradients, and suboptimal performance. Unlike conventional DL methods that rely on statistically motivated initialization schemes, 
(e.g., Xavier or He initialization), science-informed initialization embeds the domain knowledge into this early stage, providing a physics-grounded starting point for training. Formally, the initial parameter set $\boldsymbol{\theta}_0$ is defined as
\begin{equation}
\boldsymbol{\theta}_0 = h(\mathcal{M}),
\end{equation}
where $h(\cdot)$ represents an initialization function that maps available scientific knowledge $\mathcal{M}$ (e.g., known physical laws, analytical solutions, or domain constraints) into an informed parameter configuration $\boldsymbol{\theta}_0$. 
%This strategy is particularly valuable in deep networks, where poor initialization can impede gradient flow and slow convergence. 
By grounding initial weights in scientific principles, the training process often converges more rapidly and reaches better local optima ~\cite{jia2021physics}.

Additionally, a practical strategy for science-informed DL is using pre-training  ~\cite{willard2022integrating}. In this approach, a model is first trained on a related task and then fine-tuned on the target task with limited data. The pre-trained weights serve as a safe starting point, placing the parameters much closer to optimal values than random starts. A common implementation involves pre-training the network on simulated data generated by physics-based models. For example, initializing NN weights with simulated domain data can significantly accelerate convergence and improve learning efficiency.

In ~\cite{10974467}, Jere et al. study science-informed DL initialization for MIMO--OFDM receive processing, with a particular focus on OFDM symbol detection using reservoir computing. They propose explainable Echo State Network (ESN) and Windowed ESN (WESN) receivers that are lightweight enough for online adaptation, while explicitly leveraging signal-processing structure.

Lee et al.~\cite{8335785} propose deep power control (DPC), a CNN-based transmitter power-control policy for an $N$-user interference network that maximizes either the sum spectral efficiency (SE) or the sum energy efficiency (EE) with substantially lower runtime than iterative optimizers. The channel power gain from transmitter $i$ to receiver $j$ is modeled as \(
h_{i,j}=|g_{i,j}|^2\,\beta\, d_{i,j}^{-\alpha},\)
where $g_{i,j}$ is the small-scale fading coefficient, $\beta$ is a path-loss constant, $d_{i,j}$ is the distance between transmitter $i$ and receiver $j$, and $\alpha$ is the path-loss exponent. Let $P_i$ denote the transmit power of user $i$, constrained by $0\le P_i\le P_{\max}$, where $P_{\max}$ is the maximum transmit power. With bandwidth $W$ and noise power spectral density $N_0$, the SE of user $i$ is
\begin{equation}
{\rm SE}_i=\log_2\!\left(1+\frac{h_{i,i}P_i}{N_0W+\sum_{k\neq i}h_{k,i}P_k}\right),
\end{equation}
where the denominator includes thermal noise power $N_0W$ and the interference received at user $i$ from all other transmitters $k\neq i$. The EE of user $i$ is defined as \({\rm EE}_i=\frac{{\rm SE}_i}{P_i+P_C},\) 
where $P_C$ is the circuit power consumption. The corresponding design problems maximize $\sum_{i=1}^{N}{\rm SE}_i$ or $\sum_{i=1}^{N}{\rm EE}_i$, which are nonconvex and are typically solved via iterative methods. DPC instead learns a direct mapping from the (normalized) channel matrix $\hat{\mathbf H}\in\mathbb{R}^{N\times N}$ (constructed from $\{h_{i,j}\}$) to normalized powers $\mathbf P_N\in[0,1]^N$, and then forms feasible powers as $\mathbf P=P_{\max}\mathbf P_N$, where $\mathbf P=[P_1,\ldots,P_N]^{\mathsf T}$. Training proceeds in two stages: first, the CNN is pre-trained to imitate a reference optimizer (e.g., WMMSE) by minimizing
\(L_{\mathrm{init}}=\big\|\mathbf P_{\mathrm{WMMSE}}-\mathbf P\big\|_2^2,\)
where $\mathbf P_{\mathrm{WMMSE}}$ is the power vector produced by WMMSE for the same CSI; second, it is fine-tuned with an unsupervised objective that directly maximizes the target utility by minimizing
\begin{equation}
L_{\mathrm{SE}}=-\sum_{i=1}^{N}{\rm SE}_i
\qquad \text{or} \quad
L_{\mathrm{EE}}=-\sum_{i=1}^{N}{\rm EE}_i,
\end{equation}
using the Adam optimizer. The paper also introduces a distributed variant to reduce signaling, where each transmitter uses only locally available CSI to form an approximate $\hat{\mathbf H}$ by setting unknown entries to zero and then applies the same trained CNN. Simulations show that centralized DPC achieves comparable or improved sum-SE/sum-EE relative to WMMSE with much lower computation time, while distributed DPC trades some performance for reduced CSI exchange overhead. In summery, pre-training gives a good initialization (a ``safe starting point") that already reaches at least WMMSE-level performance. Then, the utility-driven (unsupervised) training can improve beyond WMMSE because it optimizes the actual objective (SE or EE).
%To further explain and improve performance over challenging non-minimum-phase (NMP) / mixed-phase channels,~\cite{10974467} connects the WESN architecture to classical stable inversion: a stable inverse of a mixed-phase channel generally requires an IIR component (for the minimum-phase factor) plus an FIR component (feedforward taps) for the NMP factor. Accordingly, WESN augments the readout with explicit skip and delayed input connections (a tap-delay-line), enabling an FIR-like component learned directly from the limited training data, while the reservoir can be configured using the statistics of the minimum-phase factor. Overall, simulations under 5G/5G-Advanced-inspired settings show that such domain-configured ESN/WESN detectors significantly outperform randomly initialized reservoirs, while retaining low-complexity online training.

Although the topic remains largely unexplored, initial evidence suggests that informed parameter initialization can outperform random schemes by giving the model a more intuitive, physically grounded starting point closer to meaningful physical solutions. After all, current evidence indicates that embedding network topologies and channel characteristics into the node, layer design, and parameter initialization of DL models enhances their interpretability and adaptability in dynamic wireless environments. 
%By aligning learned representations with propagation behavior and network geometry, 
%In addition, this ScIDL may also reduce model complexity while maintaining reliable and physically consistent performance. 
%
%\vspace{-5}

\subsection{Science-Informed Training Data and Output Validation}

\begin{table*}[htbp]
\centering
\caption{ScIDL architecture: (1) training data generation, (2) training data pre-processing, and (3) output processing and validation}
\label{tab:scidl-arch-2}
\resizebox{\linewidth}{!}{
%\small
\begin{tabular}{p{1.5cm} p{4cm} p{4cm} p{4cm}}
%\begin{tabular}{|p{2cm}|p{5cm}|p{5cm}|p{5cm}|p{5cm}|}
\toprule
\textbf{Work} & \textbf{Model in a Few Words} & \textbf{Science-Informed Aspect} & \textbf{Gains} \\ 
\midrule
\multicolumn{4}{l}{\textbf{\textit{Training Data Generation}}} \\
%\midrule
%\multicolumn{4}{l}{\textit{Graph Neural Networks}} \\
\midrule
\cite{8444648} & DNN-based framework for power allocation problems in wireless networks & The model is trained with power allocation labels produced by the WMMSE algorithm. & Performance is comparable to WMMSE (90 to 95\%) with reduced computational complexity.\\
\addlinespace
\cite{9285223} & Graph-embedding DL model for link scheduling problems in D2D networks & The model is trained using optimal link-scheduling labels from the model-based FPLinQ algorithm & Performance is comparable to conventional FPLinQ, with reduced computational complexity. \\
\addlinespace
\cite{9175003} & DTL-based framework for downlink CSI prediction in FDD mMIMO systems & The NN model is trained on synthetic data from a geometry-based FDD mMIMO channel model. & Significantly reduces prediction MSE compared with DL trained on non-synthetic data. \\
\addlinespace
\cite{10757719} & Transformer-encoder-based neural receiver for {LLR} estimation in {6G} {V2X} uplink communications & The NN is trained on synthetic data generated using the 3GPP UMa channel model. & Performance surpasses LS+LMMSE-based receivers, with strong generalization abilities.\\
\addlinespace
\cite{KHACHATRIAN2025103696} & UNet-based solution for single-BS NLOS outdoor localization & The NN is trained on the WAIR-D dataset with ray-traced synthetic data. & %Performance achieves 11.3 meters NLOS RMSE with 76.5\% accuracy, 
Outperforms purely data-driven CNN and MLP baselines \\
\addlinespace
\cite{8879693} & MLP-assisted B\&B algorithm for Cloud-RAN power minimization problems & The MLP is trained through DAgger learning to collect features and prune/preserve labels. &  Near-optimal performance comparable to classical B\&B, with significant complexity reduction
\\
%\cite{zhang2022topology} & Graph-based DL model with an attention mechanism for resource allocation problems. & The GNN architecture is guided by embedding the network's physical topology in the graph. & Near-optimal performance with reduced computational complexity even for unseen topologies.\\ 
%\cite{9252917} & MPGNN-based graph convolution network beamforming optimization problems. & The GNN architecture is guided by embedding the network's physical topology in the graph. & Performance surpasses traditional WMMSE optimizers with strong generalization abilities. \\ 
%\cite{9072356} & Graph-based DL model for optimal resource allocation policies. & The GNN architecture is guided by embedding the network's permutation-invariant topology in the graph. & Performance comparable to fully-connected NNs with reduced computational complexity. \\
%\midrule
%\textit{Deep Unfolding}
%\cite{10021866,9414561,9667094} & Deep-unfolding model with no matrix inversions for beamforming optimization problems. & The deep unfolding architecture is determined by the structure of the WMMSE algorithm. & Near-optimal performance with reduced computational complexity.  \\
%\cite{daw2020physics} & & & \\
%\cite{9024538} & Graph-based DL model for power control problems. & The inference graph embeds the permutation invariance condition of interference links in the architecture. & Performance surpasses purely data-driven MLP and CNN learning, and the classical model-based WMMSE algorithm. \\
\midrule
\multicolumn{4}{l}{\textbf{\textit{Training Data Pre-processing}}}\\
\midrule
\cite{224799} & NN-based framework for terrestrial radio-wave field strength prediction & Model-based pre-processing for dominant propagation effects, and NN-based learning for terrain-dependent corrections. & Performance surpasses empirical models and faster inference than deterministic ray-tracing methods. \\
\addlinespace
\cite{9154263} & Hybrid DL architecture for link adaptation under outdated CSI & Wiener filter pre-processing for instantaneous CSI estimation before DNN-based MCS classification. & Near-optimal performance compared to data-driven baselines with perfect CSI \\
\addlinespace
\cite{6755477} & CVNN-based framework for CSI estimation problems & CZT frequency-domain pre-processing followed by the CVNN for amplitude–phase channel prediction & Performance surpasses AR-based CSI predictors with superior BER and strong generalization, especially under fast Doppler conditions. \\
\addlinespace
\cite{10972051,10882105} & Neural-based solution for channel estimation in VLC systems & LS-based channel estimation based on VLC channel equations followed by an NN-based correction stage & Performance surpasses classical LS and MMSE estimators, especially at low-to-medium ${\rm SNR}$.\\
%\cite{10974467} & RNN-based architecture for OFDM symbol detection in MIMO systems. & ESN/WESN parameters are not learned but deterministically configured using channel statistics and IIR filter characterizations.  & Performance surpasses randomly initialized models and matches or exceeds conventional DL-based detectors, all with significant complexity reduction. \\
%\cite{8335785} & CNN-based framework for power allocation problems. & The model's loss function embeds spectral/energy efficiency terms from Shannon's capacity law. & Efficiency performance comparable to WMMSE with less computation and signaling overhead.\\
\midrule
\multicolumn{4}{l}{\textbf{\textit{Output Processing and Validation}}}\\
\midrule
\cite{8979256} & CNN-based TDD framework for CSI estimation for mMIMO with channel aging & The CNN extracts ACF and then fed to a model-based AR module for final CSI prediction. & Performance surpasses pilot-based TDD with faster convergence and reduced pilot-overhead.  \\
%\addlinespace
%\addlinespace
%\cite{termehchisafe} & & & \\
\bottomrule
\end{tabular}
}
\end{table*}
In ScIDL frameworks, the training data and output validation stages ensure that the learning process remains aligned with the established scientific principles from data generation to performance assessment. This section of our taxonomy is organized into three complementary categories:
\begin{itemize}
\item Training Data Generation, 
\item Training Data Pre-processing,
\item Output Processing and Validation.
\end{itemize}
A discussion on these three directions follows and a summary is given in Table~\ref{tab:scidl-arch-2}.
%In the following subsections, we outline these three directions for integrating scientific knowledge into initialization and validation of DL models.
%and substantial computational resources and fast convergence

%\vspace{0.2cm}
\subsubsection{Science-Informed Training Data Generation}
In many research domains, including wireless communications and sensing, acquiring sufficient real-world data is technically challenging, costly, and time-consuming. In wireless systems, physics-based simulation is often used to build virtual datasets; however, such simulations may rely on computationally intensive models. More recently, generative DL has emerged as an alternative for learning data distributions and synthesizing realistic samples. Nevertheless, conventional generative models typically require a large, high-quality labeled training set $\mathcal{D}_{\mathrm{train}}$ to achieve good performance.

Motivated by these challenges, this section reviews two science-informed directions for training data generation: (i) science-informed generative DL frameworks, and (ii) synthetic dataset construction via physics-based simulation.
%In many research domains, including wireless communications, obtaining sufficient real-world data is often technically challenging, costly, and time-consuming. In wireless, specifically, physics-based simulations are frequently used to build virtual datasets, but these methods rely on computationally intensive models or measurements. In addition, Generative DL has emerged as an alternative for learning data distributions that synthesize realistic samples. %In conventional supervised DL, a neural network $f(\mathbf{x};\mathbf{\theta})$ is trained to approximate input-output relationships solely from the labeled training set:
%
%\begin{equation}
%\mathcal{D}_{train}=\{(x^{(i)},y^{(i))}\}.
%\end{equation}
%
%From the labeled data, the network parameters $\theta$ are optimized to minimize a loss $\mathcal{L}$, typically using gradient descent or one of its variants. 
%However, conventional generative models normally require large, high-quality labeled training set $\mathcal{D}_{train}$ to achieve good performance. 

\paragraph{Science-Informed Generative DL Frameworks} 
Recently, science-informed generative DL frameworks have been developed that integrate physical laws and domain knowledge to generate data that are both realistic and scientifically consistent. For instance, Wu et al.~\cite{wu2020enforcing} propose a statistically constrained GAN that enforces the covariance constraints into the training process, ensuring that generated samples remain consistent with the true statistical structure of the underlying physical system. In wireless systems, geometric properties such as antenna-pattern symmetry can be exploited for data augmentation. For example, the symmetric radiation pattern of a dipole antenna enables mirrored samples, expanding the training set and improving DL robustness for tasks like beamforming or antenna selection.

\paragraph{Synthetic Datasets via Physics-Based Simulation}
The authors in~\cite{8444648} propose a DNN-based framework for power allocation in wireless networks. The key idea is to generate labeled training data synthetically by simulating wireless channels and computing near-optimal power allocations using a model-based optimizer. Specifically, a fully connected feed-forward network is trained to approximate the nonlinear mapping from channel realizations to transmit powers, using labels produced by the weighted minimum mean-square error (WMMSE) algorithm.

WMMSE is a classical model-based method derived from wireless interference theory. It reformulates weighted sum-rate maximization via MMSE--SINR relationships and applies block-coordinate descent updates that converge to a Karush--Kuhn--Tucker (KKT) stationary point of the associated non-convex problem. Concretely, for a $K$-user Gaussian interference channel, the $i$-th training sample is generated by drawing a complex channel matrix $\mathbf{H}^{(i)}\in\mathbb{C}^{K\times K}$ with independent Rayleigh fading coefficients $h_{kj}^{(i)}\sim\mathcal{CN}(0,1)$ for all links $j\!\to\!k$. The corresponding DNN input is formed by stacking the channel magnitudes,
\(
x^{(i)}=\operatorname{vec}\!\big(|h_{kj}^{(i)}|\big)\in\mathbb{R}^{K^2}.
\)
For the interfering multiple-access channel scenario, user locations are randomly generated across multiple cells, and each $h_{kj}^{(i)}$ incorporates large-scale fading (path-loss and shadowing computed from user--BS distances) as well as small-scale Rayleigh fading. Given $\mathbf{H}^{(i)}$, a maximum power constraint $P_{\max}$, and noise powers $\{\sigma_k^2\}_{k=1}^K$, the label is obtained by solving the weighted sum-rate maximization problem by using WMMSE:
\begin{equation}
%\scalebox{0.91}{
\max_{0 \le p_k \le P_{\max}} \;\sum_{k=1}^K \alpha_k
\log\!\left(1 + \frac{|h_{kk}^{(i)}|^2 p_k}{\sum_{j\neq k} |h_{kj}^{(i)}|^2 p_j + \sigma_k^2}\right),
%$}
\end{equation}
where $p_k$ is the transmit power of user $k$ and $\alpha_k$ is its weight. In WMMSE, each iteration updates an MMSE receive coefficient $u_k$, a positive weight $w_k$, and a transmit coefficient $v_k$ (initialized, e.g., by $v_k^{(0)}=\sqrt{P_{\max}}$), until convergence (or until a maximum iteration count is reached). The converged powers $p_k^{*(i)}=|v_k^{*(i)}|^2$ define the following label vector:
\begin{equation}
y^{(i)}=(p_1^{*(i)},\ldots,p_K^{*(i)})\in\mathbb{R}_+^K.
\end{equation}
Repeating this procedure for $i=1,\ldots,N_{\text{train}}$ yields the supervised dataset
\(\mathcal{D}_{\mathrm{train}}=\{(x^{(i)},y^{(i)})\}_{i=1}^{N_{\text{train}}}.\)
Because the labels are generated by the analytical WMMSE optimizer, they inherit its structural properties and constraints, making the data-generation process science-informed. Numerical results show that the resulting DNN achieves about $95$--$99\%$ of WMMSE's sum-rate performance while reducing computational cost by orders of magnitude and generalizing well to unseen channel conditions.

Conventional DL-based link-scheduling methods often rely on large labeled datasets and accurate CSI. To reduce this dependency, Lee et al.~\cite{9285223} propose a graph-embedding approach for device-to-device (D2D) link scheduling that uses only topology/geometry information (e.g., distance-based graph features) rather than CSI. The authors generate activation vectors by running a fractional programming-based link-scheduling algorithm FPLinQ, which is a fully model-based optimizer, on many synthetic network realizations. The resulting activation vectors serve as physics/optimization-derived labels for training a lightweight classifier (via cross-entropy) to imitate the FPLinQ decisions. Numerical results indicate that the trained model approaches FPLinQ performance while offering substantially lower computational complexity. 

%Conventional DL-based solutions for link scheduling typically demand large training datasets with accurate CSI. To address this, Lee et al.~\cite{9285223} propose a graph-embedding DL model for link scheduling in device-to-device (D2D) networks that operates without CSI. In this framework. each D2D pair is represented as a graph node and interference relationships as edges. The architecture uses a multi-layer classifier to determine optimal link activation. The network is trained using a standard cross-entropy loss optimized by the fractional programming-based link scheduling (FPLinQ) algorithm---a fully model-based link-scheduling optimizer derived from wireless interference theory. Numerical results show that the DL model achieves performance close to conventional FPLinQ while significantly reducing computational complexity. 
%The authors in~\cite{9175003}
Yan et al.~\cite{9175003} formulate a deep transfer learning (DTL)-based framework for downlink CSI prediction in frequency-division-duplexing (FDD) massive MIMO (mMIMO) systems. The model employs a fully connected feed-forward NN with ${\rm ReLU}$ activations and a standard MSE loss for the real–imaginary CSI representations. The model is trained and evaluated on synthetic datasets generated from a geometry-based FDD mMIMO channel model with environment-specific angular spreads, ensuring that uplink/downlink CSI pairs and labels are grounded in physical propagation behavior. Although the loss function is conventional, the use of science-informed synthetic data embeds channel geometry into the learning process. Numerical results show that DTL significantly reduces the prediction MSE compared with DL models trained directly on non-synthetic data.

%The authors in~\cite{10980283}
Saleem et al.~\cite{10757719} propose TransRx, a transformer-encoder-based neural receiver. It performs end-to-end log-likelihood ratio (LLR) estimation directly from received OFDM resource elements in 6G vehicular-to-everything (V2X) uplink communications. The architecture uses stacked multi-head self-attention blocks and is trained with a standard binary cross-entropy loss. Training relies entirely on synthetic data generated using the 3GPP Urban Macrocell (UMa) channel model---one of the standardized physics-based wireless models used by 3GPP to benchmark 4G/5G/6G link-level algorithms. Because 3GPP UMa captures realistic Doppler, multipath, delay-spread, and fading characteristics, the training distribution is inherently grounded in wireless propagation physics.  Simulation results show that TransRx significantly outperforms classical LS and LMMSE receivers and maintains a strong generalization to previously unseen channel families.

%The authors in~\cite{8291154}
Khachatrian et al.~\cite{KHACHATRIAN2025103696} develop a DL-based solution for single-BS NLOS outdoor localization using a UNet architecture (a type of CNN). The UNet estimates a spatial heatmap that represents the relative likelihood of the user being located at each position on the map. The solution combines map images with link-level features (such angles of arrival and departure drawn as directional rays) via an MLP. The model is trained on the WAIR-D dataset, which provides ray-traced synthetic data, embedding geometric propagation physics and real urban morphology into the learning stage. Experiments on more than 1.3 million samples show that, with the  proposed UNet architecture, the  predicted locations are within 10 meters of real locations in 3 out of 4 cases, outperforming purely data-driven CNN and MLP baselines by 47\%.
 
Many wireless resource-management problems (e.g., power allocation and user association) are modeled as mixed-integer nonlinear programs (MINLPs). These problems are NP-hard and typically solved with classical global optimization methods like branch-and-bound (B\&B) at exponential complexity. To reduce this cost, Shen et al.~\cite{8879693} propose LORM, a learning-to-optimize framework that accelerates B\&B by learning its pruning policy rather than solving the MINLP directly. LORM replaces the standard bound-based pruning rule with an MLP classifier that predicts whether each node should be pruned or preserved using problem-independent and problem-dependent features. The classifier is trained through data aggregation (DAgger) learning, where B\&B is repeatedly run to collect the node features and optimal prune/preserve labels, and the model is updated using weighted cross-entropy to imitate the optimal pruning behavior. Experiments on Cloud-RAN power minimization show that LORM achieves a near-optimal performance with significant speedup over classical B\&B and prior learning-based methods, and it generalizes well across different system configurations.

In summary, existing evidence suggests that science-informed training-data generation encourages DL models to learn representations that are consistent with underlying wireless physics. %, rather than relying on spurious statistical correlations. 

%This integration may lead to a performance comparable to classical optimization-based methods, while keeping transparent predictions. 
%
%%\vspace{0.2cm}
\subsubsection{Science-Informed Pre-processing of Training Data}
Science-informed feature pre-processing can improve NN performance. As an example, consider the problem of predicting signal quality (or equivalently  ${\rm SNR}$ or achievable rate) from raw channel measurements (e.g., complex channel coefficients $h(t)$). For this, let us do the physics-based pre-processing by computing the path-loss corrected channel gain:
\begin{equation}
g_{\text {norm }}(t)=\frac{|h(t)|^2}{d^{-\alpha}},
\end{equation}
where $d$ is the distance between transmitter and receiver, $\alpha$ is the path-loss exponent. Then, we compute angular spread or delay spread to represent the multipath effects. 
We can now use $g_{\text {norm }}(t)$, angular spread, and delay spread as input to the NN instead of raw $h(t)$ so that the network sees just the small-scale fading, which is more uniform and easier to learn (and generalization improves drastically).  Without the science-informed pre-processing, the network must learn the $1/d^\alpha$
 relationship from data, and small datasets would make it hard to generalize to new distances.

Another example of wireless channel prediction is proposed by Gschwendtner and Landstorfer~\cite{224799}, who present one of the earliest NN-based approaches for terrestrial radio-wave propagation modeling. Their method uses an MLP, trained with a standard squared-error loss, to map topographical features (e.g., terrain profiles, land-use classes, and antenna height) to the received field strength. Although the learning stage is purely data-driven, the framework becomes partially science-informed through a physics-based pre-processing step: analytical components such as free-space path-loss and antenna radiation patterns are computed in advance, allowing the MLP to focus on learning the residual, terrain-dependent effects. Numerical results show that the proposed MLP achieves low prediction error and generalizes well to unseen locations, outperforming semi-empirical models in soft line-of-sight (LOS) and non-LOS (NLOS) transitions, while enabling substantially faster inference than deterministic ray-tracing methods.
%Gschwendtner and Landstorfer~\cite{224799} present one of the earliest NN-based approaches to terrestrial radio-wave propagation prediction. Their method uses an MLP trained with a standard squared-error loss to map topographical features (such as terrain profiles, land-use classes, and antenna height) to received field strength. Although the training process is purely data-driven, the framework becomes partially science-informed through a physics-based pre-processing stage: analytical components such as free-space path loss and antenna radiation patterns are computed beforehand, allowing the NN to focus solely on learning residual terrain-dependent effects. Numerical results show that the MLP achieves low prediction error and generalizes well to unseen locations, outperforming semi-empirical models in soft line-of-sight (LOS) and non-LOS (NLOS) transitions and offering substantially faster inference than deterministic ray-tracing methods.

Pellaco et al.~\cite{9154263} address the link adaption problem under outdated CSI by proposing a hybrid DL architecture that incorporates a Wiener MMSE channel estimator (i.e., a model-based predictor grounded in Jakes’ Doppler correlation model) into the learning pipeline preceding a conventional NN trained with a standard cross-entropy loss. The Wiener filter first predicts the instantaneous CSI, which are then passed to the NN responsible for selecting the modulation and coding scheme (MCS). Although the model training is conventional, the inclusion of the Wiener MMSE estimator guides the learning process toward physically consistent outcomes. Simulation results using LTE channel models show that the hybrid approach achieves near-optimal spectral efficiency with reduced computational complexity, faster convergence, and improved interpretability compared with purely data-driven baselines that assume a perfect CSI. 

Ding and Hirose~\cite{6755477} propose a hybrid physics–data-driven channel prediction model that integrates the Chirp Z-Transform (CZT) with a complex-valued neural network (CVNN) for accurate fading-channel prediction. The CZT is a deterministic signal processing algorithm based on the Discrete Fourier Transform (DFT) that enables frequency-domain analysis by extracting the channel parameters from the observed complex-valued channel coefficients. In the proposed CZT-CVNN architecture, these frequency-domain parameters are used as inputs to a multilayer or recurrent CVNN that predicts the complex-valued channel coefficients directly in the amplitude–phase domain. Although the CVNN is trained on a conventional complex-domain mean-squared loss, the inclusion of the CZT pre-processing module embeds frequency-domain physics into the learning pipeline, guiding the model toward physically consistent representations. Both simulation and over-the-air experiments show that the CZT–CVNN achieves superior bit-error-rate (BER) performance, enhanced prediction stability, and lower generalization error compared with conventional AR-based predictors, especially in scenarios with fast Doppler dynamics. 

In~\cite{10972051,10882105}, the authors present an NN-based network architecture for channel estimation in Visible Light Communication (VLC) systems. A key distinction between VLC and classical wireless channels lies in their noise characteristics. In classical systems, signal and noise are independent, whereas in VLC, the noise (shot and thermal) is signal-dependent because the shot noise increases with signal intensity. Under these conditions, an accurate channel estimation is crucial for optimized transmission. The authors propose a modified NN architecture consisting of two steps. First, a classical least-squares (LS) estimator based on VLC channel equations that produces preliminary channel estimates from pilot measurements. This stage incorporates the physics of the channel, including signal–noise dependency. As second step, an NN is used that corrects the estimation errors from the first step, guiding the network to converge to the true channel matrix. Numerical results show that the ``LS and NN" architecture outperforms the classical LS and MMSE estimators, with notable accuracy gains in low-to-medium ${\rm SNR}$ regimes where signal-dependent noise dominates. 

In summary, evidence suggests that enforcing scientific principles in training-data pre-processing promotes interpretability, improves generalization, and enhances physical consistency.

%\vspace{0.2cm}
\subsubsection{Science-Informed Output Processing and Validation}
The DL model outputs can be evaluated using metrics derived from established scientific principles (e.g., physical laws and system constraints). This enables verification of not only predictive accuracy but also physical consistency. A representative approach is to post-process the learned outputs using a science-based module: the DL model produces an initial estimate, and a physics- or model-based routine then refines, projects, or validates it to enforce scientific consistency. In wireless applications, such validation leverages known wireless laws, constraints, invariances, and theoretical bounds, rather than relying solely on statistical error metrics.

%For example, assume that a DL model predicts the precoding matrices for a MIMO system. For science-informed validation, the resulting Shannon capacity $C=\log _2 \operatorname{det}\left(\mathbf{I}+\mathrm{SNR} \cdot \mathbf{H} \mathbf{H}^H\right)$, where $\mathbf{H}$ is the channel matrix and $\mathbf{I}$ is an identity matrix, can be computed using predicted precoders. Then, it can be checked if the predicted capacity is physically achievable (e.g., does not exceed the MIMO bounds).  

As an example,  DL models can be tested on scenarios with known theoretical outcomes. Consider that the  model predicts BER vs. SNR in a wireless system. A science-informed test can be done by comparing the DL-predicted BER curve with the analytical $Q$-function for a given modulation scheme (e.g., for $\mathrm{BPSK}$, $\mathrm{BER}=Q\left(\sqrt{2 \mathrm{SNR}}\right)$). This will help  to ensure that the model generalizes physically, not just statistically. 

Yan et al.~\cite{8979256} address the channel estimation overhead problem in massive MIMO (mMIMO) system under channel aging by proposing an ML-based TDD framework for efficient CSI prediction. The method first uses a CNN to extract the channel’s auto-correlation function (ACF), whose features are then passed to a model-based autoregressive (AR) module for CSI prediction. Importantly, the AR coefficients are not learned from data but are pre-computed from a physics-based formulation and selected according to the CNN output. The approach is science-informed in the sense that data-driven learning captures the temporal patterns while the AR module enforces 
physically-consistent channel dynamics. Simulation results show that the CNN–AR model achieves superior prediction accuracy, faster convergence, and significant pilot-overhead reduction compared with conventional pilot-based TDD channel estimation schemes.
%Recheck before adding
%Science-informed validation of DL outputs can also be used to assess symmetry or invariance in DL models. For instance, assume that a DL model predicts beamforming weights for a uniform circular array (UCA). If the array is rotated, predicted beam patterns should rotate correspondingly. Any violations will indicate the DL model has not captured physics-informed invariances. In~\cite{termehchisafe}, the authors apply a DRL output-refinement strategy that enforces speed constraints in a THz-enabled UAVs network. They further use a clip($\cdot$) function to ensure UAV trajectories remain within predefined boundaries. Rather than encoding these physical constraints as additional reward terms, which increases complexity and hyperparameter count, the authors refine the outputs directly, simplifying training while preserving physical feasibility and model reliability.

To summarize, physics-based consistency checks and output refinement helps ensure that DL predictions remain physically valid and reliable. 

%Such science-informed mechanisms reduce the risk of non-physical outputs and enhance interpretability compared with purely data-driven models.
%
\subsection{Science-Informed Loss Function Design and Optimization}

The third major direction for integrating scientific knowledge into DL is through the learning objective and the optimization procedure. Instead of training solely with data-fitting losses (e.g., mean-squared error or cross-entropy), science-informed learning augments the loss with terms that penalize violations of known physical laws, system constraints, invariances, or model-based priors. These additional terms can act as regularizers that steer the learned representations toward physically meaningful solutions, improve sample efficiency, and enhance generalization. Moreover, because such composite objectives may introduce competing gradients and mismatched scales across loss components, the resulting optimization procedure often requires a careful weighting and normalization. The optimization procedure may also need explicit constraint-handling strategies (e.g., penalty or Lagrangian methods, projection steps, or alternating updates) to achieve stable and efficient training.

The different approaches used for science-informed loss function design and optimization in the literature are discussed below and a summary of those approaches is given in Table~\ref{tab:loss-function-opt}.

%\vspace{0.2cm}
\subsubsection{Science-Informed Loss Function Design}
\begin{table*}[htbp]
\centering
\caption{Science-informed loss function design and optimization: A Summary}
\label{tab:loss-function-opt}
\resizebox{\linewidth}{!}{
%\small
\begin{tabular}{p{1.5cm} p{4cm} p{4cm} p{4cm}}
%\begin{tabular}{|p{2cm}|p{5cm}|p{5cm}|p{5cm}|p{5cm}|}
\toprule
\textbf{Work} & \textbf{Model in a Few Words} & \textbf{Science-Informed Aspect} & \textbf{Gains} \\ 
\midrule
\multicolumn{4}{l}{\textbf{\textit{Loss Function Design}}} \\
\midrule
\cite{xu2018semantic} & Semantic-guided DL-based model for structured output learning & The model employs a semantic loss with symbolic knowledge embedded in the formulation. & Improved generalization in semi-supervised learning under complex output constraints settings \\
\addlinespace
\cite{li2024physics} & PINO architecture for physics-constrained estimation of solution operators & The model embeds physics constraints in the loss function through partial differential equations. & Performance surpasses data-driven DL and conventional PINNs, with strong generalization abilities.\\
\addlinespace
\cite{10719669} & HV-DL architecture for complex-valued electromagnetic field prediction & PINN-style loss function that embeds the electric field integral equation (EFIE) & Performance comparable to full-wave MoM solver and superior than data-driven DL baseline.\\
\addlinespace
\cite{8935405} & DL-based framework for MISO downlink beamforming optimization problems & The model's loss function is derived from Shannon's sum-rate capacity criterion. & Near-optimal performance comparable to WMMSE wih reduced computational complexity  \\
\addlinespace
\cite{11016226} & DL-based DT framework for RIS-assisted beamforming optimization problems & The model's loss function is built from Maxwell partial differential equations, boundary-conditions, and data-fitting terms. & Performance surpasses data-driven DL baseline solutions. \\
\addlinespace
\cite{10720822} & DL-based frequency detection model for high-resolution sonar applications & The loss function combines the data-fitting and physical linear models. & Performance surpasses FFT and SBL methods, especially under low SNR conditions.\\
\addlinespace
\cite{8922744} & DL-based solution for multi-user power allocation problems & The model's loss function is derived from Shannon's sum-rate capacity criterion. & Performance is comparable to WMMSE, especially in high interference scenarios.\\
\midrule
\multicolumn{4}{l}{\textbf{\textit{Loss Function Optimization}}} \\
\midrule
\cite{li2023physics} & PINN model based on an adaptive gradient descent method for solving partial differential equations & The model adjusts gradient directions based on the interaction of the loss function terms. & Performance is comparable to conventional Adam and PCGrad loss optimizers, with reduced training complexity.
\\
\addlinespace
\cite{he2023generalizing} & DL model based on a generalized projected gradient descent method for massive MIMO detection & The model evaluates multiple gradient descent steps involving two physics-informed loss functions, before defining a projection. & Detection performance surpasses conventional MMSE-based baselines.  \\
\bottomrule
\end{tabular}
}
\end{table*}
Conventional DL models rely on loss functions to guide the parameter updates through optimization methods such as gradient descent. In traditional behavior-prediction settings, these losses simply measure the difference between the predicted and the true target values. In such a setting, a purely data-driven standard DL method struggles to generalize to unseen scenarios~\cite{willard2022integrating}.
%However, in physics-based domains, physical relationships evolve across space and time and span multiple scales under complex governing equations. As a result, a standard DL method often struggles to capture these dependencies purely from data, limiting its ability to generalize to unseen scenarios~\cite{willard2022integrating}. 
To address this limitation, recent research proposes augmenting the loss function with scientific constraints~\cite{karniadakis2021physics}, integrating the physical laws, invariances, or governing equations directly into the optimization objective. Formally, an augmented loss function $\mathcal{L}$ can be written as:
\begin{equation}
\mathcal{L} = w_{data}\mathcal{L}_{data}(\mathbf{\hat{y}},\mathbf{y}) + w_{ScI}\mathcal{L}_{ScI}(\mathbf{\hat{y}}),     
\label{eq:augmentedLoss}
\end{equation}
where $\mathcal{L}_{data}$ measures the supervised error between the predicted labels $\hat{\mathbf{y}}$ and the ground-truth labels $\mathbf{y}$, and $\mathcal{L}_{ScI}$ denotes the unsupervised loss associated with the scientific constraints. The weights $w_{data}$ and $w_{ScI}$ balance the contribution of the two terms and may be user-defined or automatically tuned. This strategy (often referred to as soft aggregation) allows the DL model to incorporate scientific structure during training. It promotes representations that generalize better and remain consistent with the underlying physical principles. However, during training, the DL model parameters are learned by minimizing the expected loss over the dataset. Thus, the resulting solution is not guaranteed to satisfy the scientific constraints exactly; rather, the knowledge-based term acts as a regularizer that biases learning toward constraint-consistent solutions \cite{moseley2022physics}. Consequently, choosing $w_{data}$ and $w_{ScI}$ is critical. Instead of treating them as fixed hyperparameters, these weights can be determined using theory-grounded methods (e.g., adaptive Lagrangian or penalty approaches) to balance data fidelity and constraint satisfaction \cite{han2021reinforcement}. %Instead of treating them as fixed hyperparameters, they can be determined using theory-grounded methods (e.g., Lagrangian or penalty methods with adaptive/data-driven updates) that adjust the weights to better balance data fidelity and constraint satisfaction, as in \cite{han2021reinforcement}.
Consider a learning-based beamforming in a MIMO system, where the goal is to obtain a beamforming vector $\mathbf{w}_\theta$ for transmission that maximizes the received signal strength or minimizes the bit error rate. This is done under a total transmit power constraint:
\(
\left\|\mathbf{w}_\theta\right\|^2 \leq P_{\max }.
\)
The  data/performance loss can be defined in terms of mean squared error for the received signal:
\begin{equation}
\mathcal{L}_{\text {data }}=\frac{1}{N} \sum_{i=1}^N\left\|\mathbf{y}_i-\mathbf{H}_i \mathbf{w}_\theta\left(\mathbf{H}_i\right) \mathbf{x}_i\right\|^2,
\end{equation}
where $\mathbf{y}_i=$ desired received signal,  $\mathbf{H}_i=$ channel matrix, and $\mathbf{x}_i=$ transmitted symbol. 
Let us now define an unsupervised loss function associated with the  power constraint as follows:
\begin{equation}
\mathcal{L}_{\text {power }}=\operatorname{ReLU}\left(\left\|\mathbf{w}_\theta\right\|^2-P_{\max }\right)^2,
\end{equation}
which penalizes the network if it predicts a beamforming vector exceeding the allowed transmit power. To this end,  the augmented loss function can be defined as:
\begin{equation}
\mathcal{L} =\mathcal{L}_{\text {data }}+ w \mathcal{L}_{\text {power }}, 
\end{equation}
where $w$ balances fitting the signal versus respecting the power limit.

Xu et al.~\cite{xu2018semantic} incorporate logic rules as constraints into a DL model by formulating a semantic loss function. This loss is defined as the negative log-likelihood that the network’s probabilistic outputs satisfy a Boolean logic formula, thereby aligning the predictions with predefined logical structures. Similarly, Stewart and Ermon~\cite{stewart2017label} propose a constraint-based supervised learning framework, replacing traditional labeled training. The authors incorporate algebraic equations into the loss function to enforce domain consistency. They minimize the constraint loss by penalizing the deviations from known scientific laws, showing that NNs can be guided entirely by domain-knowledge replacing the traditional labeled training. 

Li et al.~\cite{li2024physics} introduce physics-informed neural operators (PINOs), a hybrid framework that embeds physics-based constraints into the loss function of Fourier neural operators (FNOs). The FNOs learn the mappings between function spaces by applying stacked Fourier layers that map the inputs into the frequency domain and truncate them to a fixed number of modes. This enables the model to capture invariant functional relationships and remain robust under varying input discretizations. The authors report that PINOs outperform purely data-driven approaches and traditional physics-informed neural networks (PINNs), achieving higher accuracy, faster inference than numerical solvers, and strong generalization to unseen scenarios.

Similarly, Wan and Pan~\cite{10719669} develop a hybrid-value DL (HV-DL) architecture for predicting complex-valued electromagnetic fields directly from real-valued spatial coordinates. HV-DL integrates real-valued layers with shared weights for phase representation followed by complex-valued layers that explicitly model the amplitude–phase interactions. The model is trained using a PINN-style loss function that embeds the electric field integral equation (EFIE), allowing the network to learn the Maxwell-consistent electromagnetic behavior without requiring labeled data. Numerical experiments demonstrate that HV-DL achieves an accuracy comparable to that of full-wave solvers such as MoM while outperforming both purely real-valued and purely complex-valued neural architectures with significantly fewer parameters.

%The authors in~\cite{10750167}
%Narang and Lingam~\cite{10750167} propose a DL framework for electromagnetic modeling of functional materials, introducing an autoencoder–GAN–assisted PINN (XA-PINN) that predicts reflection coefficients and absorption bandwidths from dielectric parameters. Their PINN model incorporates the residual of the Riccati differential equation into the loss function, to ensure the outputs are consistent with Maxwell-based scattering physics. To address data scarcity, the architecture integrates an autoencoder–GAN module that generates synthetic EM samples derived from analytical Riccati solutions, providing physics-informed data that further guides the optimization process. Numerical experiments over 0.5–18 GHz show that XA-PINN achieves $\sim$ 92.7\% accuracy, outperforming conventional DL architectures such as LSTMs, while requiring significantly less training data.

Xia et al.~\cite{8935405} propose a DL-based framework for MISO downlink beamforming using science-informed beamforming neural networks (BNNs). Their approach exploits the uplink–downlink duality to structure the learning process, ensuring physically-consistent beam patterns. The loss function incorporates Shannon sum-rate expressions, directly maximizing downlink throughput while grounding DL optimization in established wireless theory. Numerical results show that BNNs achieve a near-optimal performance comparable to well-established iterative solvers such as WMMSE, with substantially lower computational complexity, reduced latency, and acceptable generalization when tested on unseen channel conditions.  

Han et al.~\cite{11016226} propose a physics-informed digital twin framework for RIS-assisted wireless systems that combines a PINN-based electromagnetic solver with a real-time virtual replica of the environment. The architecture integrates sensing, 3D scene modeling, and physics-constrained prediction to dynamically optimize the RIS phase shifts. The PINN is trained using a science-informed loss that embeds the  partial differential equation residuals, boundary-condition constraints, and data-fitting terms, enforcing physically consistent electromagnetic field predictions. This hybrid approach enables accurate and efficient RIS optimization while substantially reducing the computational overhead compared to that for finite element methods (FEM). Numerical results show that PINN–DT improves reflection efficiency, transmission rates, and inference speed, highlighting the benefits of integrating the physical laws with data-driven learning for real-time wireless optimization.

Liang et al.~\cite{8922744} propose DL-based power control network (PCNet) and its ensemble variant (ePCNet) for power allocation in multi-user wireless interference channels. The models are trained in an unsupervised manner using a science-informed loss function derived from Shannon’s capacity law. This loss directly maximizes the system sum-rate and enforces physical consistency in the learned power-control policy. Numerical results show that PCNet and ePCNet achieve a performance comparable to (and in some cases surpassing) that of classical optimization-based solvers such as WMMSE, particularly in high-interference scenarios. Both models also generalize well to unseen channel distributions, demonstrating a strong robustness and adaptability across diverse wireless environments.
%Ko et al.~\cite{10720822} introduce a DL-based frequency detection model for high-resolution sonar applications. Their system-informed NN (SINN) builds on the adaptive learned iterative shrinkage thresholding algorithm (Ada-LISTA), embedding iterative optimization steps into neural layers to approximate sparse signal recovery. The loss function incorporates a data-fitting term alongside the physical linear measurement model $y = Ax + n$ as a soft term, guiding the network toward solutions that remain consistent with the underlying frequency-domain physics. This system-informed loss acts as a physical prior, enabling stable and physically meaningful signal reconstruction with lower computational complexity. Numerical experiments show that SINN and its multi-measurement extension (MM-SINN) outperform classical approaches such as fast fourier transform (FFT) and sparse bayesian learning (SBL), especially under low signal-to-noise ratio (SNR) conditions.

In DRL, where an agent seeks to maximize cumulative rewards, embedding scientific knowledge into the reward function plays a role analogous to incorporating domain knowlege into a supervised-learning loss function. In~\cite{zhuang2020adaptive}, Zhuang et al. embed queuing dynamics into the reward by constructing a Lyapunov candidate function and deriving its Lyapunov drift from stability theory. To promote a stable queue behavior in mobile edge computing networks, the Lyapunov drift is included in the reward alongside a penalty for end-to-end latency. Consequently, the DRL agent implicitly minimizes the Lyapunov drift, improving stability while reducing the delay. This illustrates how physics or model-based priors can be injected into DRL through reward shaping, mirroring the role of science-informed loss functions in supervised learning.
%with a tunable hyperparameter balancing these objectives.

%similar to inserting priors in Bayesian modeling \cite{moseley2022physics}
In many cases, incorporating scientific terms into the reward structure can improve the convergence, generalization, and data efficiency of DRL models. However, as discussed earlier, these terms typically act only as regularizers, loosely constraining the feasible policy space. The authors in~\cite{han2021reinforcement} modify the DRL optimization objective itself to enforce the scientific constraints more strictly. Instead of adding knowledge-based terms directly to the reward, the constraints are embedded into the proximal policy optimization (PPO) update through a Lagrangian formulation, turning the problem into a constrained policy optimization task. Specifically, each DRL agent $i$ learns a parameterized policy $\pi_i^{\theta_i}$ subject to the scientific constraint (stability constraint), i.e.,
\begin{equation}
\text{Find } \pi_i^{\theta_i}\quad \text{s.t.}\quad \text{ (stability constraint),}
\end{equation}
where $\theta_i$ is the parameter of each DRL agent $i$ policy. 
In the paper, the DRL objective consists of (i) the standard PPO clipped surrogate term and (ii) an additional Lagrange term weighted by a multiplier $\lambda_i$:
\begin{equation}
L_i(\theta_i)= L_i^{\text{PPO}}(\theta_i)\;+\;\lambda_i\,\mathbb{E}\big[g_i(\cdot)\big],
\end{equation}
where $L_i^{\text{PPO}}$ is the clipped PPO loss and $g_i(\cdot)$ denotes the constraint-related expression that encodes the requirement in the constraint (e.g., stability/safety).

Yang et al.~\cite{9322615} study the secure beamforming problem in reconfigurable intelligent surface (RIS)-aided multi-user wireless networks. They propose a DRL framework based on post-decision state (PDS) learning to jointly optimize the base station (BS) beamforming and RIS phase-shift configuration. The DRL agent observes the CSI, secrecy rate, and QoS indicators, and selects the beamforming actions to maximize a physics-based reward derived from the Shannon-capacity secrecy expressions and QoS constraints. Although the model employs a conventional TD-error loss, the science-informed aspect appears in the reward function and MDP formulation, which embed the wireless propagation model and secrecy-rate equations. Simulation results show that the proposed deep PDS-learning method converges faster and achieves higher QoS satisfaction than classical DQN and alternating optimization (AO) baselines, with performance gains becoming more pronounced as the number of reflecting elements increases.

%In summary, incorporating scientific knowledge into the loss function serves as a regularization mechanism, analogous to imposing priors in Bayesian modeling. The approach guides DL models toward solution spaces in which their predictions align with established scientific principles. %guiding DL models to physically consistent solution spaces making their predictions aligned with established scientific principles.%Further studies on science-informed loss functions in DL can be found in~\cite{}.
%
%\begin{table*}[]
%\centering
%\caption{Caption}
%\resizebox{\linewidth}{!}{
%\begin{tabular}{|p{4.3cm}|p{0.7cm}|p{6cm}|p{6cm}|p{6cm}|}
%\hline
%\textbf{Sub-category} & \textbf{Work} & \textbf{Model Description} & \textbf{Science-Informed Aspect} & \textbf{Gains} \\
%\hline
%& \cite{li2023physics} & & & \\
%& \cite{8943940} & & & \\
%\hline
%\end{tabular}    
%}
%\label{tab:SocialApplications}
%\end{table*}
%
%%\vspace{0.2cm}
\subsubsection{Science-Informed Parameter Optimization}
Gradient-based optimization is widely used in DL for parameter adjusting due to its computational efficiency. However, when science-informed losses combine data-based and scientific-constraint terms, their gradients inevitably interact. If not handled carefully, these interactions can drive optimization toward suboptimal solutions or cause convergence difficulties. To address this issue, Li et al.~\cite{li2023physics} propose an adaptive gradient descent algorithm (AGDA) that analyzes and compensates for these interaction effects. AGDA dynamically adjusts the gradient direction based on how different loss terms influence one another, enabling the optimizer to better navigate complex loss shapes and improving the model stability, performance, and generalization compared to the conventional gradient descent (GD) algorithm.
%it is crucial to modify the standard gradient-based optimization to effectively manage the complex interactions between multiple terms in science-informed loss functions. Specifically, different loss components---such as those enforcing the physical constraints and those fitting data---can have gradients that interact in complex ways. These interactions can lead to suboptimal performance if not managed correctly. Considering this challenge, an adaptive gradient descent algorithm (AGDA) is introduced in \cite{li2023physics} based on an analysis of the interaction mechanisms analysis. Indeed, AGDA dynamically adjusts the \st{learning rates} gradient direction based on the interactions between various loss gradients. Consequently, AGDA can overcome the limitations of traditional gradient descent (GD) methods, ensuring more effective and stable convergence thereby enhancing the model performance, stability, and generalization.
%and enhance the model performance, stability, and generalization. %This modification can ensure more effective and stable convergence. 
%

Advanced optimization methods such as projected GD (PGD) help maintain parameter feasibility and enforce the science-informed constraints. In PGD, after computing the gradient of the loss function, the parameters are updated in the negative gradient direction and then projected back onto the feasible region defined by the physical constraints. This projection step is applied after back propagation and before parameter updates and ensures that the model parameters are kept within physically valid regions.  As an example, consider the MIMO beamforming problem under power constraint: 
\begin{equation}
\|\mathbf{w}\|^2 \leq P_{\max}.
\end{equation}
To solve this problem, the standard update will be as:
\(\mathbf{w} \leftarrow \mathbf{w}-\eta \nabla L.\)
With the projected gradient step, the update will be given by:
\begin{equation}
\mathbf{w} \leftarrow P_{\max }^{1 / 2} \frac{\mathbf{w}}{\max \left(\|\mathbf{w}\|, P_{\max }^{1 / 2}\right)}. 
\end{equation}
Consequently, the DL training never produces non-physical beamforming vectors.

To reduce the computational cost of the standard PGD (which performs one gradient step and one projection per iteration), He et al.~\cite{he2023generalizing} propose a generalized PGD framework that executes multiple gradient steps before each projection, improving efficiency and scalability.
%Advanced optimization methods such as projected GD (PGD) help maintain parameter feasibility and enforce science-informed constraints. In PGD, the model first updates its parameters along the negative gradient of the loss function, after which the updated parameters are projected back onto the feasible region defined by the underlying physical constraints. 
%the model first updates its parameters along the negative gradient of the loss function, after which the updated parameters are projected back onto the feasible region defined by the underlying physical constraints. This projection step, applied after backpropagation and before parameter updates, ensures that the model remains within physically valid regions throughout training. To reduce the computational burden of standard PGD—which performs one gradient step followed by one projection per iteration—He et al.~\cite{he2023generalizing} propose a generalized PGD framework that performs multiple gradient updates before each projection, improving efficiency and scalability.

Although evidence is also limited in this category, Li's AGDA in~\cite{li2023physics} and He's generalized PGD in~\cite{he2023generalizing} suggest that science-informed optimizers achieve a more stable convergence and a near-model performance at lower complexity than the standard gradient descent method.

\section{ScIDL Case Studies: Addressing Complex Wireless System Problems}

\subsection{Case Study 1: Beamforming for MU-MISO}
In this case study, we investigate science-informed node and layer design for downlink MU-MISO beamforming. The objective is to improve the generalization  of the learned model under changes in the number of active users and the resulting interference conditions.

\subsubsection{System Model}
We consider a downlink multi-user multiple-input single-output (MU-MISO) channel where a base station (BS) equipped with $M$ transmit antennas serves $K$ single-antenna users indexed by $\mathcal{K}=\{1,\dots,K\}$. Let $\mathbf{h}_k\in\mathbb{C}^{M}$ denote the channel vector from the BS to user $k$, and let $s_k\sim\mathcal{CN}(0,1)$ be the mutually independent transmitted data symbol intended for user $k$. The BS employs linear precoding with beamforming vectors $\{\mathbf{v}_k\in\mathbb{C}^{M}\}$ and transmits
\(
\mathbf{x}=\sum_{k=1}^{K}\mathbf{v}_k s_k,
\) such that $ \sum_{k=1}^{K}\|\mathbf{v}_k\|_2^2 \le P$, where $P$ is the total transmit power.
The received signal at user $k$ is
\(
y_k=\mathbf{h}_k^{H}\mathbf{v}_k s_k+\sum_{j\neq k}\mathbf{h}_k^{H}\mathbf{v}_j s_j+n_k,
\)
where $n_k\sim\mathcal{CN}(0,\sigma^2)$ denotes additive white Gaussian noise and the corresponding {\rm SINR} is:
\begin{equation}
\mathrm{SINR}_k = 
\frac{\big|\mathbf{h}_k^{H}\mathbf{v}_k\big|^2}
{\sum_{j\in\mathcal{K}\setminus\{k\}}\big|\mathbf{h}_k^{H}\mathbf{v}_j\big|^2+\sigma^2}.
\end{equation}
Then the achievable rate of user $k$ is given as
\(
R_k=\log_2\!\left(1+\mathrm{SINR}_k\right).
\)

%---------------------------------------------------
\subsubsection{Problem Definition}
We aim to compute the transmit beamformers that maximize the weighted sum-rate (WSR):
\begin{subequations}\label{eq:wsr_problem}
\begin{align}
\max_{\mathbf{V}} \quad & \sum_{k=1}^{K}\alpha_k \log_2\!\left(1+\mathrm{SINR}_k\right) \label{eq:wsr_problem_a}\\
\text{s.t.}\quad & \mathrm{tr}\!\left(\mathbf{V}\mathbf{V}^H\right) \le P, \label{eq:wsr_problem_b}
\end{align}
\end{subequations}
where $\mathbf{V} = [\mathbf{v}_1,\mathbf{v}_2,\dots,\mathbf{v}_K]\in\mathbb{C}^{M\times K}$, and $\alpha_k>0$ is the priority of the $k$-th user. 

%-------------------------------------------------
\subsubsection{Classical Model-Based Solution}
Problem~\eqref{eq:wsr_problem} is non-convex, and a standard approach to obtain a stationary point is the weighted minimum mean-square error (WMMSE) algorithm~\cite{liu2012achieving}. For a given transmit precoder $\mathbf{V}$, user $k$ applies a scalar equalizer $u_k\in\mathbb{C}$ to form $\hat{s}_k=u_k y_k$, leading to the MSE:
\begin{equation}
\begin{aligned}
e_k &= \mathbb{E}\!\left[|\hat{s}_k-s_k|^2\right] \\
&= |u_k|^2\!\left(\sum_{j=1}^{K}|\mathbf{h}_k^H\mathbf{v}_j|^2+\sigma^2\right)
-2\Re\!\left\{u_k\,\mathbf{h}_k^H\mathbf{v}_k\right\}+1 .
\end{aligned}
\end{equation}
By introducing auxiliary MSE weights $w_k>0$, the WSR maximization can be converted into an equivalent weighted-MSE minimization, which is then solved by alternating updates over $\mathbf{u}$, $\mathbf{w}$, and $\mathbf{V}$~\cite{liu2012achieving}. The equalizer and weight variables use closed-form updates, whereas the transmit update requires solving a constrained quadratic problem and typically involves a matrix inversion together with a power enforcing Lagrange multiplier. Although WMMSE provides a strong model-based benchmark, its iterative nature and the cost of the transmit update can be prohibitive in fast-varying scenarios and for large antenna dimensions. This motivates learning based alternatives that aim to retain a near optimal performance while enabling low latency inference.

%---------------------------------------------
\subsubsection{DL-Based Solutions: Purely Data-Driven versus Science-Informed Alternatives}
We now describe the learning-based baselines used in this case study, focusing on the network architectures and how they interface with the beamforming variables and constraints.
%---------------------------------------------------
\paragraph{DNN-Based Solution}
We consider a one-shot fully-connected DNN that maps the channel realization to the transmit beamforming matrix. The input is a real-valued representation of the MU-MISO channel $\{\mathbf{h}_k\}_{k=1}^{K}$ obtained by stacking the real and imaginary parts of each channel vector. In our implementation, this produces a tensor of size $K\times 2M\times 2$, which is flattened into a single feature vector per sample.
The network outputs a real-valued tensor $\tilde{\mathbf{V}}\in\mathbb{R}^{K\times 2M\times 1}$, which represents the stacked real and imaginary components of the complex beamformers $\mathbf{V}=[\mathbf{v}_1,\dots,\mathbf{v}_K]\in\mathbb{C}^{M\times K}$. The last layer is linear and produces $K\times 2M$ real values, which are reshaped to form $\tilde{\mathbf{V}}$. We use a fully connected multilayer perceptron with three hidden layers and ${\rm ReLU}$ activations. In our experiments, the hidden-layer widths are set to $(512,1024,512)$, and the final linear layer outputs $K\cdot 2M$ real values that are reshaped into $\tilde{\mathbf{V}}\in\mathbb{R}^{K\times 2M}$.
After converting $\tilde{\mathbf{V}}$ to its complex form $\mathbf{V}$, we normalize the precoder to satisfy the power constraint:
\begin{equation}
\mathbf{V} = \mathbf{V}\sqrt{\frac{P}{\mathrm{tr}(\mathbf{V}\mathbf{V}^H)}}.
\label{eq:powerconstraint}
\end{equation}
%---------------------------------------------------
\paragraph{Mask-Aware DNN-Based Solution - Adaptive Solution}
The baseline DNN assumes a fixed number of users $K$ and always produces beamformers for all $K$ user indices. To explicitly handle dynamic scheduling and improve generalization across different active-user sets, we adopt a mask-aware architecture based on DeepSets \cite{zaheer2017deep}. The key idea is to process each user through a shared encoder, aggregate the information across users through a permutation-invariant pooling operator, and then decode per-user beamformers using both the local and global features. This yields a permutation-equivariant mapping from the set of user channels to the set of beamformers. We refer to this approach as an adaptive solution because the DNN is explicitly trained to handle specific variations in the number of active users. This notion of adaptivity differs from domain generalization (DG), where the objective is to achieve robust performance on previously unseen domains without explicit exposure during training. In contrast, this adaptive solution (Mask-Aware DNN) is trained on scenarios with varying active users and learns to adjust its behavior accordingly, rather than generalizing to entirely unseen operating conditions.

For each sample, user $k$ is represented by a feature vector formed by stacking the real and imaginary parts of its channel and appending a binary activity indicator $m_k\in\{0,1\}$. More specifically, we flatten the channel representation of user $k$ into a vector and construct
\(\mathbf{x}_k = \big[\mathrm{vec}(\Re\{\mathbf{h}_k\}),\ \mathrm{vec}(\Im\{\mathbf{h}_k\}),\ m_k\big],\)
so that inactive users are explicitly marked in the input. Each $\mathbf{x}_k$ is passed through a shared multilayer perceptron (encoder) $\phi(\cdot)$ to produce an embedding $\mathbf{e}_k\in\mathbb{R}^{D}$. The embeddings are then gated by the activity mask and pooled via sum to form a global context vector
\(\mathbf{g} = \sum_{k=1}^{K} m_k\,\mathbf{e}_k,\)
which is permutation-invariant and depends only on the set of active users. For each user, the local embedding and the global context are concatenated and passed through a shared decoder $\rho(\cdot)$ to generate the corresponding beamformer parameters:
\(\tilde{\mathbf{v}}_k = \rho\big([\mathbf{e}_k,\mathbf{g}]\big)\in\mathbb{R}^{2M}.\)
Stacking the outputs over all users yields $\tilde{\mathbf{V}}\in\mathbb{R}^{K\times 2M\times 1}$, which is converted to the complex precoder $\mathbf{V}\in\mathbb{C}^{M\times K}$.
Finally, the predicted beamformers are gated to enforce zero transmission toward the inactive users,
\(\mathbf{v}_k \leftarrow m_k\,\mathbf{v}_k,\qquad k\in\mathcal{K},\)
and the resulting precoder is normalized to satisfy the total power constraint using (\ref{eq:powerconstraint}).
%\vspace{-1}

\paragraph{Deep Unfolding-Based Solution}
We follow the deep-unfolding framework proposed in~\cite{pellaco2021matrix}, which turns the WMMSE beamforming iterations~\cite{liu2012achieving} into a trainable architecture by removing the matrix inversion in the transmit update and learning a small set of algorithm parameters. In standard WMMSE~\cite{liu2012achieving}, the receiver equalizers and MSE weights are updated in closed form, whereas the transmit precoder update is obtained by solving a constrained quadratic problem that typically requires a matrix inversion together with a power-enforcing Lagrange multiplier search. The inverse-free formulation in~\cite{pellaco2021matrix} preserves the closed-form auxiliary updates, but replaces the transmit update with a few projected-gradient steps followed by an explicit projection onto the total-power constraint. As a result, each unrolled iteration is lightweight, relies only on matrix multiplications, and enforces feasibility by construction.

For fixed $\mathbf{u}=[u_1,\dots,u_K]^T$ and $\mathbf{w}=[w_1,\dots,w_K]^T$, the transmit subproblem is
\begin{subequations}\label{eq:V_subproblem}
\begin{align}
\min_{\mathbf{V}} \quad & f(\mathbf{V}) = \sum_{k=1}^{K}\alpha_k\!\left(w_k e_k - \log w_k\right) \label{eq:V_subproblem_a}\\
\text{s.t.}\quad & \mathbf{V}\in\mathcal{C}=\left\{\mathbf{V}\;|\;\mathrm{tr}(\mathbf{V}\mathbf{V}^H)\le P\right\}. \label{eq:V_subproblem_b}
\end{align}
\end{subequations}
Starting from $\mathbf{V}^{(0)}$, the inverse-free transmit update is implemented via $N_{\mathrm{pgd}}$ projected-gradient steps~\cite{pellaco2021matrix}:
\begin{subequations}\label{eq:pgd_update}
\begin{align}
\tilde{\mathbf{V}}^{(n)} &= \mathbf{V}^{(n-1)} - \gamma^{(n)}\,\nabla f\!\left(\mathbf{V}^{(n-1)}\right), \label{eq:pgd_update_a}\\
\mathbf{V}^{(n)} &= \Pi_{\mathcal{C}}\!\left(\tilde{\mathbf{V}}^{(n)}\right), \label{eq:pgd_update_b}
\end{align}
\end{subequations}
where $\gamma^{(n)}>0$ is the step size and $\Pi_{\mathcal{C}}(\cdot)$ denotes the projection performed using~\eqref{eq:powerconstraint}.

An unfolded layer consists of: (i) updating $\mathbf{u}$ and $\mathbf{w}$ using the WMMSE closed-form expressions~\cite{liu2012achieving}, followed by (ii) updating $\mathbf{V}$ via the $N_{\mathrm{pgd}}$ projected-gradient steps in~\eqref{eq:pgd_update} and the projection~\eqref{eq:powerconstraint}. Stacking $L$ such layers yields a deep unfolded network with output $\mathbf{V}^{(L)}$. The step sizes are treated as learnable parameters and may vary across layers and inner steps.

We also consider an accelerated variant of the projected-gradient descent, as in~\cite{pellaco2021matrix}, which introduces a momentum through the difference between consecutive iterates. Let $\mathbf{V}^{(n)}$ denote the precoder iterate and define $\bar{\mathbf{V}}^{(n)}=\mathbf{V}^{(n)}-\mathbf{V}^{(n-1)}$. The accelerated projected-gradient step is
\begin{subequations}\label{eq:acc_update}
\begin{align}
\tilde{\mathbf{V}}^{(n)} &=
\begin{aligned}[t]
&\mathbf{V}^{(n-1)} + \theta^{(n)}\bar{\mathbf{V}}^{(n-1)} \\
&\quad - \gamma^{(n)} \nabla f\!\Big(\mathbf{V}^{(n-1)}+\xi^{(n)}\bar{\mathbf{V}}^{(n-1)}\Big),
\end{aligned}
\label{eq:acc_update_a}\\
\mathbf{V}^{(n)} &= \Pi_{\mathcal{C}}\!\left(\tilde{\mathbf{V}}^{(n)}\right),
\label{eq:acc_update_b}
\end{align}
\label{ANees1}
\end{subequations}
where $\theta^{(n)}$ controls the momentum contribution and $\xi^{(n)}$ specifies the gradient evaluation point. In the unfolded implementation, $\gamma^{(n)}$, $\theta^{(n)}$, and $\xi^{(n)}$ are treated as learnable parameters and can vary across the layers and the inner steps.
\subsubsection{Simulation Results and Discussion}
\paragraph{Simulation Setup}
We consider a downlink MU-MISO system with $M=8$ BS antennas and $K=8$ single-antenna users. In each channel realization, $K_{\mathrm{a}}=4$ users are active (during training). The transmit power budget and noise power are set to $P = 10\,\mathrm{W}$ and $\sigma^2=1$, respectively.  The channels are generated as i.i.d.\ Rayleigh fading realizations.

For the model-based reference, WMMSE is run for a single outer iteration. For the unfolded architectures, we use $L=1$ unfolded outer iteration and perform $N_{\mathrm{pgd}}=4$ PGD steps within the beamforming update. The accelerated variant uses the same setting with a learnable step size and momentum parameters.

All learning-based models are trained using Adam optimizer with learning rate $10^{-3}$ and mini-batches of size $B=100$ for $3\times 10^{4}$ training batches. The performance is evaluated over $1000$ independently generated test batches. For the one-shot DNN baseline, we use a fully connected network with three hidden layers of widths $(512,1024,512)$ and ${\rm ReLU}$ activations. For the mask-aware DNN model, the encoder and decoder each use two hidden layers of widths $(256,256)$ with an embedding dimension of $256$.
\begin{figure}
    \centering
    \includegraphics[width=0.99\linewidth]{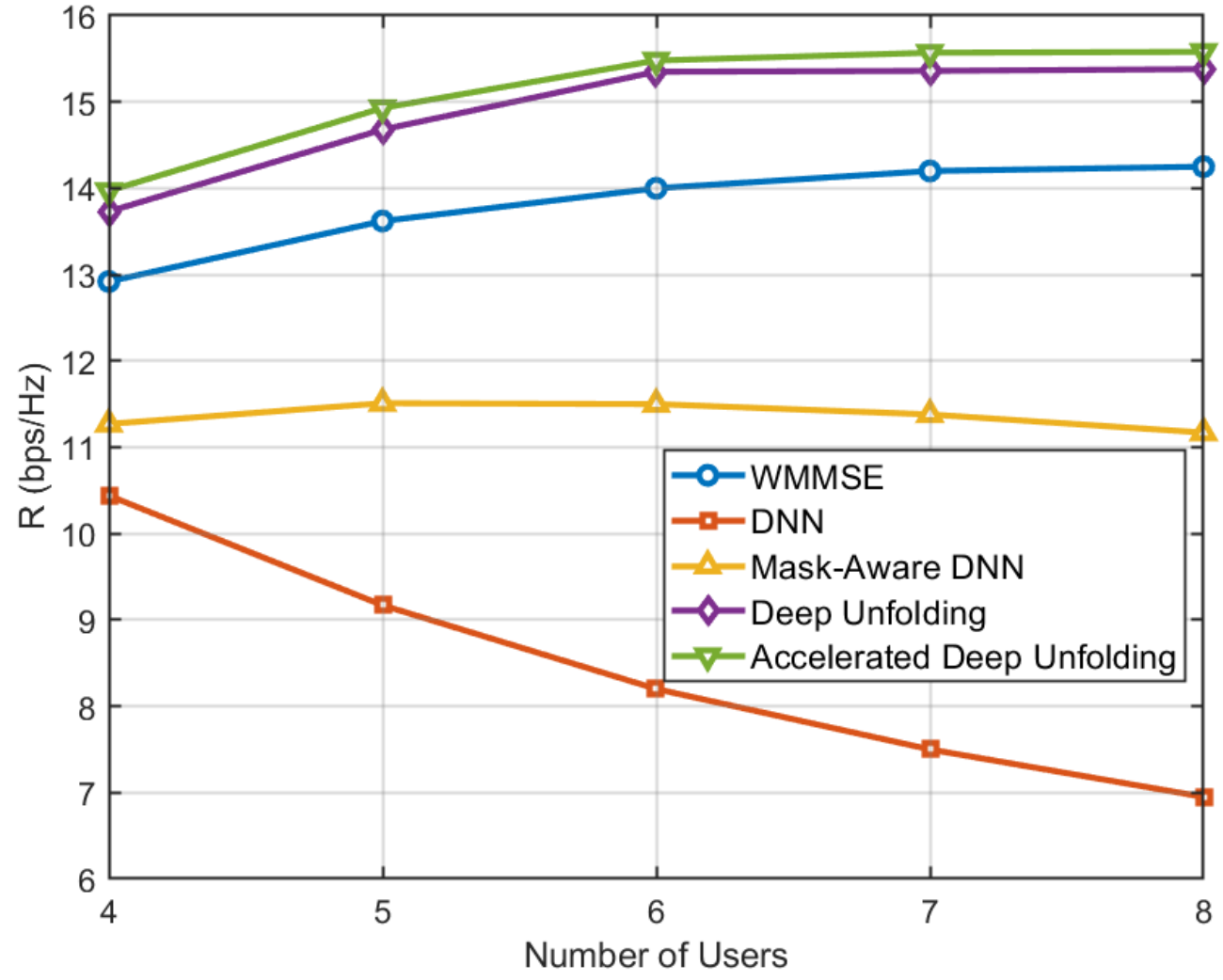}
    \caption{Sum-rate $R$ versus the number of users, comparing WMMSE, DNN, mask-aware DNN, deep unfolding, and accelerated deep unfolding.}
    \label{fig:wsr_vs_users}
\end{figure}

\paragraph{Simulation Results}
Fig.~\ref{fig:wsr_vs_users} shows the achieved sum rate $R$ as a function of the number of users. All learning-based models are trained with $K_{\mathrm{a}}=4$ active users and are then tested by varying the number of active users, which changes the interference conditions relative to those seen during training.

The purely data-driven DNN shows the weakest generalization. It achieves a reasonable sum rate at the training point ($K=4$), but its performance degrades as $K$ increases, indicating limited robustness to the shift in operating conditions.

The mask-aware DNN (adaptive solution) improves over the vanilla DNN across the full range of 
$K$. Its performance remains comparatively stable as 
$K$ increases, suggesting that providing activity information helps the network adapt its output to different active-user configurations. Moreover, this adaptive behavior is achieved under known and expected scenarios that are already considered during training. However, a consistent performance gap relative to unfolding-based methods remains.

Deep unfolding and accelerated deep unfolding achieve the best performance across all values of $K$. Both unfolded models maintain a high sum rate even when the number of active users at test time differs from the training configuration and the scenario that is not considered during training. The accelerated variant consistently attains the highest sum rate, indicating that the additional learnable momentum parameters improve the effectiveness of the truncated updates.

Overall, the results show that purely data-driven learning is sensitive to changes in the number of active users, whereas ScIDL approaches, particularly unfolding and its accelerated variant, generalize more reliably across different user configurations.

%---------------------------------------------------
%
%\begin{figure*}[!t]
%  \centering
 % \subfloat[Similar domain to training]{%
    %\includegraphics[width=0.32\textwidth]%{./Figures/confusion_matrix.pdf}%
 %   \label{fig:confusion-source}}
 % \hfil
 % \subfloat[Target domain]{%
    %\includegraphics[width=0.32\textwidth]%{./Figures/confusion_matrix_target_domain.pdf}%
%    \label{fig:confusion-target}}
%  \hfil
 % \subfloat[Fourier-informed (time–frequency) LSTM on target domain]{%
    %\includegraphics[width=0.32\textwidth]%{./Figures/confusion_matrix_dualpath_target.pdf}%
   % \label{fig:confusion-fourier}}
%  \caption{Confusion matrices of LSTM-based HAR models evaluated on: (a) a test set drawn from a domain similar to the training data, (b) a different target domain, and (c) a Fourier-informed (time–frequency) LSTM evaluated on the target domain.}
 % \label{fig:domain-shift-confusion}
%\end{figure*}

\subsection{Case Study 2: Sensing Channel State Information for Human Activity Recognition}

In this case study, we investigate science-informed pre-processing of training data methods (belonging to the science-informed training data and output validation category) with the objective of improving the domain generalization performance of DL models.
\paragraph{Problem Definition}
Human Activity Recognition (HAR) concerns the problem of inferring and classifying human motions and behaviors from observed signals. The goal of HAR is to identify the physical activities performed by one or multiple individuals. Here, we consider Wi-Fi sensing–based HAR, where variations in wireless channel state information (CSI) induced by human movement are leveraged to recognize activity patterns without requiring wearable sensors or cameras. The main challenge is generalizing the mapping from high-dimensional, time-varying wireless channel measurements to activity labels across diverse environments, user behaviors, and system configurations.%Such a sensing modality enables a wide range of human-centric and privacy-preserving applications, particularly in intelligent environments.
%Typical use cases include monitoring daily activities in smart homes, detecting abnormal events such as falls in healthcare and assisted-living facilities, enabling adaptive environmental control (e.g., lighting or temperature adjustment). 
%The core challenge lies in accurately mapping the high-dimensional, time-varying wireless channel measurements to semantic activity labels under diverse environmental conditions and user behaviors. Therefore, addressing the generalization challenge (arising from distribution shifts in propagation environments, user motion patterns, and hardware configurations) is critical for the reliable deployment of Wi-Fi-based HAR systems.
\paragraph{Dataset and DL model}
To carry out our simulation experiments, we use the CSI Human Activity dataset \cite{-21}. The dataset consists of real CSI measurements collected using commodity Wi-Fi devices, where human movements induce characteristic fluctuations in the wireless channel response. These time-varying CSI sequences are labeled with corresponding activity classes and are commonly used to evaluate DL models for Wi-Fi-based HAR. %We use an LSTM network to model the temporal dependencies in the CSI sequences and to learn a mapping from the observed channel dynamics to activity labels. 
To model the temporal dynamics of Wi-Fi CSI measurements, we employ a stacked LSTM network designed for sequence-to-label classification. The input to the network is a sequence of CSI feature vectors with dimensionality 104 per time step. The architecture consists of two LSTM layers with 128 and 64 hidden units, respectively, followed by dropout with probability 0.3 to mitigate overfitting. The final LSTM output is passed to a fully connected layer that maps the learned temporal representation to four activity classes (i.e., empty, sit, stand, walk). The network is trained end-to-end using cross-entropy loss. In addition, to construct the target domain, the original CSI sequences are deliberately perturbed at the signal level to emulate the deployment conditions that differ from those observed during training. Specifically, a lower signal-to-noise ratio (SNR) is simulated by injecting additive noise into the CSI, corresponding to an approximately 20\% reduction in SNR, which represents noisier wireless environments or lower-quality hardware. Phase noise is introduced by applying random phase perturbations to the complex CSI, with the phase-noise level increased by approximately 20\%, modeling oscillator instability and carrier frequency offsets commonly encountered in practical systems. In addition, amplitude scaling and drift are applied with an overall 15\% increase in scaling range, capturing larger variations in path loss, user position, and automatic gain control effects across different environments. Together, these perturbations modify the statistical properties of the CSI dataset, yielding a target domain that reflects realistic distribution shifts.% in Wi-Fi sensing scenarios.
%the target domain is generated by introducing lower SNR, higher phase noise, wider amplitude scaling, stronger temporal warping, and increased subcarrier dropout, thereby inducing a distribution shift in the wireless channel measurements.
\begin{figure}[!ht]
  \centering
  \subfloat[Similar domain to training]{%
    \includegraphics[width=0.4\textwidth]{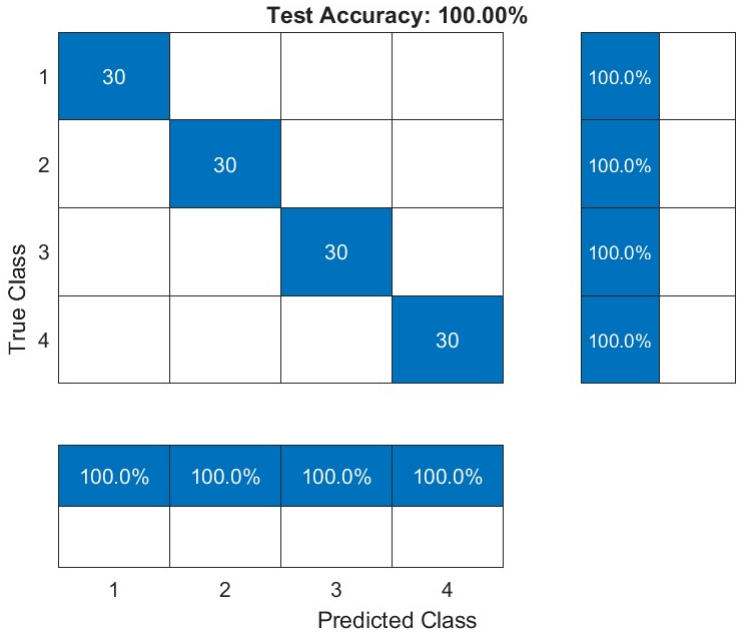}%
    \label{fig:confusion-source}}
  \par%\vspace{0.2em}
  \subfloat[Target domain]{%
    \includegraphics[width=0.4\textwidth]{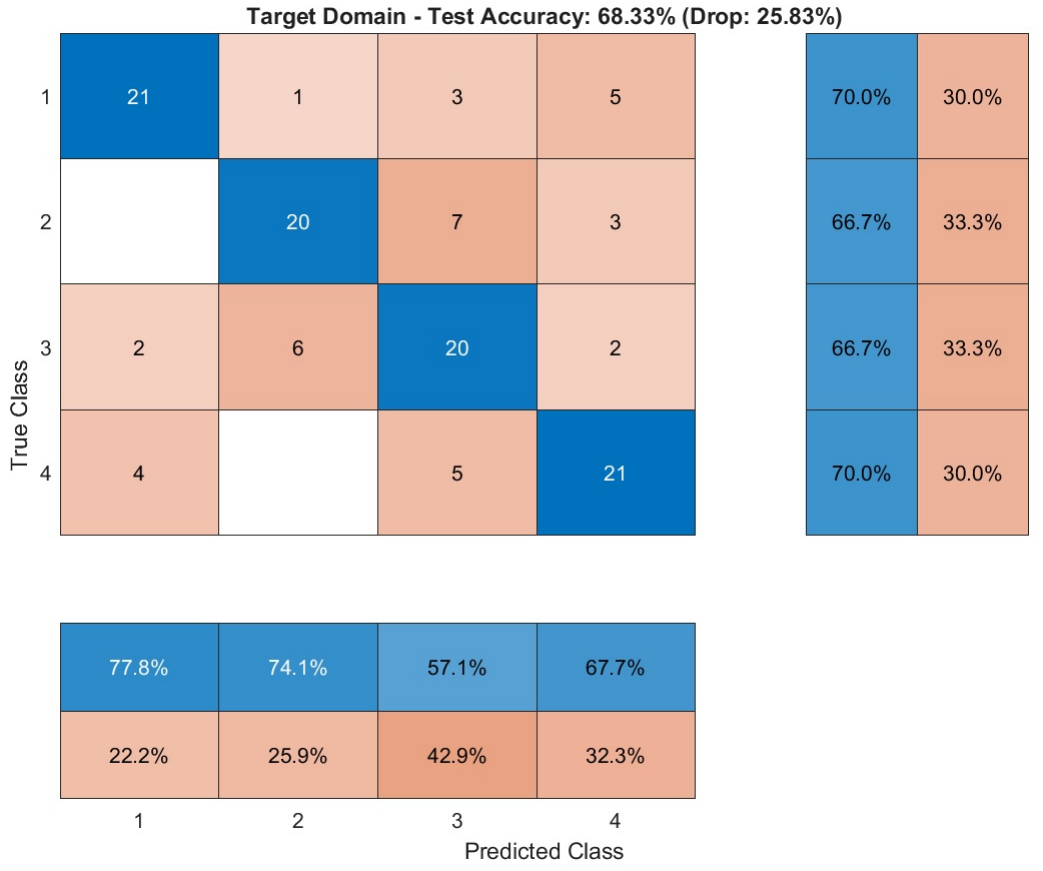}%
    \label{fig:confusion-target}}
  \caption{Confusion matrices of LSTM-based HAR models evaluated on (a) a distribution-like test domain and (b) a target domain.}%Confusion matrices of LSTM-based HAR models evaluated on: (a) a test set drawn from a domain similar to the training data, and (b) a different target domain.}
  \label{fig:domain-shift-confusion}
\end{figure}
\paragraph{Generalization Challenge and Solutions}
The results of testing the trained LSTM on data drawn from a distribution similar to the training dataset, as well as on data from a different distribution (target domain), are shown in Figs.~\ref{fig:confusion-source} and~\ref{fig:confusion-target}. These results illustrate a significant performance degradation under domain shift. To address this challenge, we first incorporate frequency-domain information by applying a Fourier transform to the CSI measurements and augmenting the training dataset with the resulting spectral features. As discussed earlier, augmenting the training data with its Fourier-based representations can be viewed as a form of science-informed preprocessing.

Specifically, we jointly leverage the time-domain and frequency-domain CSI features by concatenating them into a unified input sequence for the stacked LSTM classifier. However, the performance under domain shift is similar to that shown in Fig.~\ref{fig:confusion-target}, indicating that this ScIDL approach does not yield a noticeable improvement.
%However, as illustrated in Fig.~\ref{fig:confusion-fourier}, this ScIDL approach does not lead to a noticeable improvement in performance under domain shift. 
The underlying reason is that, in Wi-Fi CSI–based HAR, applying a Fourier transform to CSI sequences does not introduce additional task-relevant knowledge. This is because CSI already captures the spectral information, while human activity is primarily reflected in temporal channel dynamics. %As a result, frequency-domain features do not provide invariance to environmental changes and therefore fail to mitigate domain shift.
This observation indicates that the effectiveness of ScIDL methods is inherently application-dependent.

In the next step, we evaluate the separability of the training dataset. The results show that the real training dataset is easily separable. Specifically, we analyze the class separability of CSI samples in a reduced feature space. Let $\mathbf{x}_i \in \mathbb{R}^d$ denote a CSI sample with class label $y_i \in \{1,\ldots,C\}$. Class separability can be characterized by the ratio between the between-class scatter 
$\mathbf{S}_B=\sum_{c=1}^C n_c(\boldsymbol{\mu}_c-\boldsymbol{\mu})(\boldsymbol{\mu}_c-\boldsymbol{\mu})^\top$ 
and the within-class scatter 
$\mathbf{S}_W=\sum_{c=1}^C\sum_{i\in c}(\mathbf{x}_i-\boldsymbol{\mu}_c)(\mathbf{x}_i-\boldsymbol{\mu}_c)^\top$, 
where $\boldsymbol{\mu}_c$ and $\boldsymbol{\mu}$ denote the class-wise and global means, respectively. Our analysis shows that $\mathrm{tr}(\mathbf{S}_B)\gg \mathrm{tr}(\mathbf{S}_W)$, indicating that samples from different activity classes form well-separated clusters in the training domain. This confirms that the dataset exhibits a strong intrinsic separability, and the observed performance degradation under domain shift is primarily due to weak training dataset rather than learning model.

Now, we train the LSTM model on an augmented dataset. Specifically, the augmented data are created by injecting additive noise to reduce the effective ${\rm SNR}$, introducing random phase perturbations to model the oscillator and synchronization imperfections, applying amplitude scaling to capture the variations in path-loss and user distance, and performing temporal warping and subcarrier dropout to emulate the differences in motion dynamics and channel availability. These transformations modify the statistical properties of the CSI while preserving the underlying activity labels, thereby exposing the model to realistic domain variations during training and encouraging the learning of more robust representations. 

To test this LSTM model trained on the augmented dataset, we create a new target domain with random perturbation added to the augmented training dataset.
 %that  follows a uniform and Gaussian distribution. 
 To generate the target domain, the CSI samples $\mathbf{x}$ are perturbed using bounded and stochastic disturbances that emulate domain shift. Specifically, uniform disturbances are applied as bounded amplitude perturbations, modeled by additive disturbance $\mathbf{x}'=\mathbf{x} + \mathbf{w}_u$, where $\odot$ denotes the Hadamard product and $\mathbf{w}_u\sim\mathcal{U}(-\delta\mathbf{I},\delta\mathbf{I})$. In our implementation, the range of uniform perturbation is widened in the target domain, corresponding to an approximately $33\%$ increase in amplitude-scaling variability relative to the training domain, thereby modeling larger fluctuations in signal strength due to changes in path-loss, user location, and automatic gain control. In addition, additive Gaussian noise $\mathbf{w}_g\sim\mathcal{N}(\mathbf{0},\sigma^2\mathbf{I})$ is injected to model random channel and hardware noise with an increased power to reflect an effective $20\%$ reduction in {\rm SNR} compared to the training setting. Thus, the resulting target-domain samples are given by $\mathbf{x}'=\mathbf{x}+\mathbf{w}_i$, where $i \in \{u,g\}$, which preserves the underlying activity labels. 
%Together, these perturbations preserve the underlying activity labels while inducing a controlled and physically plausible distribution shift in the CSI statistics.
%To generate the target domain, the CSI samples $\mathbf{x}$ are perturbed using bounded and stochastic disturbances that emulate domain shift. Specifically, uniform disturbances are applied as bounded amplitude perturbations, modeled by multiplicative noise $\mathbf{x}'=\mathbf{x}\odot(1+\mathbf{w}_u)$, where $\odot$ is the Hadamard product with $\mathbf{w}_u\sim\mathcal{U}(-\delta,\delta)$, capturing unpredictable variations in signal strength. In addition, additive Gaussian noise $\mathbf{w}_g\sim\mathcal{N}(\mathbf{0},\sigma^2\mathbf{I})$ is injected to model random channel and hardware noise, yielding $\mathbf{x}'=\mathbf{x}+\mathbf{w}_g$. 
%These perturbations preserve the underlying activity labels while inducing a distribution shift in the CSI.
%To generate the target domain, the CSI samples $\mathbf{x}$ are perturbed by additive disturbances. Specifically, we consider disturbance terms $\mathbf{w}_u$ and $\mathbf{w}_g$, where $\mathbf{w}_u\sim\mathcal{U}(-\delta,\delta)$ represents uniformly distributed, bounded perturbations. In addition, $\mathbf{w}_g\sim\mathcal{N}(\mathbf{0},\sigma^2\mathbf{I})$ denotes additive Gaussian disturbance. The resulting target-domain samples are given by $\mathbf{x}'=\mathbf{x}+\mathbf{w}_{i}$, where $i \in \{u,g\}$, which preserves the underlying activity labels. 
Moreover, we employ a domain generalization method based on domain-shift rejection, as presented in~\cite{termehchi2026generalization}, which is derived by using an $H_{\infty}$ control framework. 

\begin{figure}[!ht]
    \centering
    \includegraphics[width=0.99\linewidth]{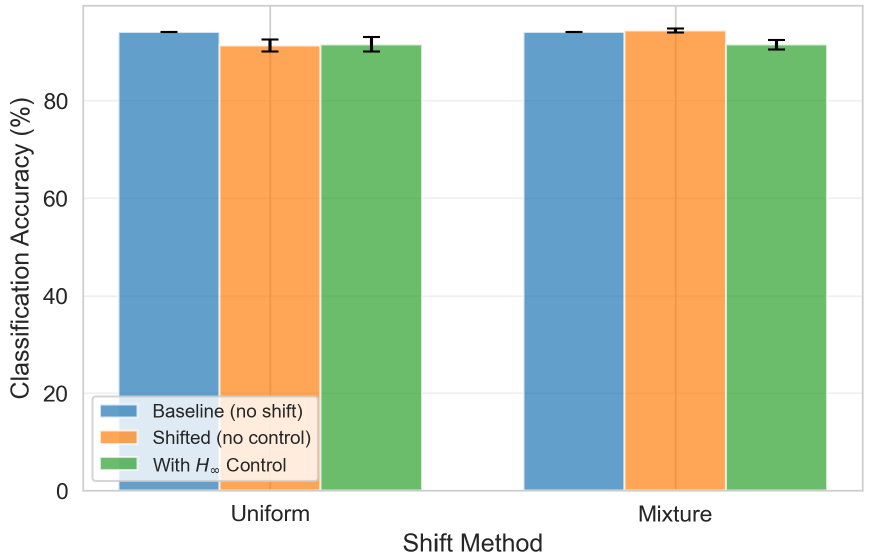}
  \caption{Performance comparison of the baseline LSTM (trained on the augmented dataset), the baseline LSTM under domain-shift, and an baseline LSTM with $H_{\infty}$ control.}
  \label{Improvement}
\end{figure}

Fig.~\ref{Improvement} compares: (i) the performance of the LSTM trained on the augmented dataset and evaluated without domain shift (baseline), (ii) the performance degradation of the LSTM trained on the augmented dataset and evaluated under domain shift, and (iii) the performance degradation of the LSTM trained on the augmented dataset augmented with the proposed $H_{\infty}$-based domain-shift rejection controller.
\begin{table}[!ht]
\centering
\caption{Performance comparison under uniform and mixture domain shifts}
\label{tab:hinf_results}
\begin{tabular}{lcccc}
\hline
\textbf{Method} & \textbf{Baseline} & \textbf{Shifted-no control} & \textbf{With $H_\infty$} \\
\hline
Uniform  & 94.17\% & $91.33 \pm 1.19$\% & $91.58 \pm 1.49$\% \\
Mixture  & 94.17\% & $94.42 \pm 0.40$\% & $91.50 \pm 0.95$\% \\
\hline
\end{tabular}
\end{table}
As shown in Fig.~\ref{Improvement} and Table~\ref{tab:hinf_results}, the baseline LSTM evaluated without domain shift achieves a lower accuracy ($94.17\%$) than the LSTM trained on real dataset and tested on a similar data distribution (Fig.~\ref{fig:confusion-source}). However, its performance does not degrade significantly when it is evaluated under domain shift. Moreover, since the baseline LSTM already exhibits a good robustness to input perturbations, the $H_{\infty}$-based controller provides limited additional benefit ($0.25\%$) under uniform-based shift and does not lead to an improved performance under gaussian-based shift. Indeed, the 
$H_{\infty}$-based controller seeks to reduce the effect of disturbances by limiting the deviations between the nominal and the perturbed states. In this case, the deviation is already small due to the robustness of the model, and the controller yields negligible additional benefits.
%Moreover, since the baseline LSTM already exhibits a relatively low $H_{\infty}$ gain from input disturbances to the output, the proposed $H_{\infty}$-based controller provides limited additional benefit and does not lead to a noticeable improvement in performance under domain shift.

\section{Toward the Future of ScIDL: Challenges and Research Directions}

As discussed earlier, ScIDL provides a promising framework for addressing key limitations of conventional deep learning, particularly limited theoretical interpretability and weak generalization. This is achieved by integrating well-established scientific principles and domain knowledge with data-driven learning. However, ScIDL also introduces a number of challenges that must be resolved to enable effective and scalable deployment across scientific and engineering domains, especially in wireless systems design. 

%The following subsections outline some open challenges and future research directions that can guide the continued development of the ScIDL field.
%
\subsection{Challenges}

%While ScIDL provides clear advantages over purely data-driven, non-transparent DL, it also introduces several challenges across the development and deployment process. 

The challenges include limited availability of high-quality real-world data and test benchmark, the difficulty of balancing data-fitting objectives with science-based terms in composite loss functions, the computational overhead of science-augmented DL models, and the limitation of theoretical methods for interpretability and
generalization analysis. %Moreover, ScIDL methods must remain compatible with conventional DL frameworks despite differing assumptions and training dynamics. 

%The following sections discuss each challenge in detail.
%

%\vspace{0.2cm}
\subsubsection{Lack of Real-World Datasets and Benchmarks}
In mature ML domains such as computer vision (CV) and natural language processing (NLP), widely adopted benchmarks like ImageNet~\cite{russakovsky2015imagenet} and GLUE~\cite{wang2019glue} provide public datasets and standardized evaluations that enable reproducible experiments and fair model comparison.
%In mature ML application domains such as computer vision (CV) and natural language processing (NLP), widely adopted benchmark suites, such as ImageNet~\cite{russakovsky2015imagenet}, COCO~\cite{lin2014coco}, and GLUE~\cite{wang2019glue}, provide publicly available datasets and standardized evaluation protocols that enable reproducible experimentation and fair comparison of model performance and computational cost.
In contrast, for DL in wireless systems, datasets and benchmarks derived from real-world scenarios remain scarce, a limitation that is particularly pronounced for ScIDL. Existing studies often rely on domain-specific or proprietary datasets, custom simulation environments, and case-dependent evaluation protocols, which significantly hinders reproducibility and cross-study comparison. Even when similar physical systems are considered, differences in problem formulations, system models, and performance metrics make it difficult to compare the results meaningfully.

As ScIDL methods continue to mature, evaluating their scalability and generalization to real-world wireless problems becomes increasingly important. While proof-of-concept studies on simplified or toy problems provide useful initial insights, their conclusions do not necessarily extend to realistic deployments. In such settings, computational complexity, system scale, and physical constraints can pose significant challenges. In particular, methods that outperform conventional DL on small-scale examples may still face prohibitive computational or implementation barriers in practical settings. Moreover, the lack of established real-world benchmarks limits the ability to assess the generalization performance under realistic operating conditions. This limitation is especially critical in wireless systems (e.g., communication and sensing), where DL model errors may lead to violations of physical constraints. Consequently, developing shared datasets and unified benchmarks grounded in real-world wireless implementations is essential for rigorous evaluation and for accelerating the adoption of ScIDL methods. An instructive precedent is the CERTS Microgrid benchmark at the University of California, Irvine, which provides a realistic and standardized testbed for smart grid research; analogous benchmark datasets would be required for wireless systems.
\subsubsection{Balancing Data and Science in the Loss Function}
As discussed before, soft aggregation (e.g., see  equation~\eqref{eq:augmentedLoss}) is a common strategy for aggregating data-fitting and science-based terms in the loss function. Properly scaling these terms is crucial, since random weighting can cause gradient-based training to converge to meaningless solutions or fail to converge entirely~\cite{karpatne2017theory}. For instance, when multi-term losses are highly imbalanced, some terms are weighted far more heavily than others, the resulting backpropagation gradients become dominated by stronger directions, causing the optimizer to ignore other components. If data terms dominate, the model may fit the observations while violating science (e.g., physical laws); if science terms dominate, the model may under-fit the noisy real data. Also, the gradients of loss terms may be misaligned, leading to destructive gradient directions that can partially cancel each other and destabilizing convergence. Additionally, multi-scale loss components with different magnitudes can create severe conditioning issues that guide training toward inaccurate or non-convergent solutions. For these reasons, understanding the gradient interactions, adaptive loss-balancing (like data-driven Lagrangian in \cite{termehchisafe}) and normalization mechanisms are essential to mitigate these issues and prevent the naive gradient descent from oscillating between objectives or converging to poor local minima.
%

%\vspace{0.2cm}
\subsubsection{New Computational Techniques and Tools}
The deployment of ScIDL methods require close integration of DL, optimization theory, numerical analysis, and domain-specific scientific knowledge~\cite{karniadakis2021physics}. While conventional DL is typically trained using generic gradient-based optimization algorithms with relatively low computational cost, such methods often lack transparency in their training dynamics. As a result, they are not well suited for ScIDL settings, where interpretability and generalization are critical. ScIDL approaches frequently rely on customized DL architectures and science-guided loss functions with multiple interacting terms. In addition, they often require specialized optimization strategies, which increase the algorithmic complexity and computational burden.

As a result, training ScIDL models is generally more demanding than standard DL, requiring additional computation and memory to evaluate embedded physics constraints, enforce prior knowledge, and manage complex optimization landscapes~\cite{Grossman1093}. %Although classical model-based solvers remain the most computationally intensive option, they offer high physical fidelity, whereas ScIDL occupies a middle ground—trading increased computational cost relative to standard deep learning for improved interpretability and physical consistency, while remaining more scalable than fully model-based approaches. Nevertheless, 
These elevated computational requirements pose a significant barrier to adoption. They also highlight the need for new computational algorithms that are both more interpretable and more efficient, as well as software frameworks that natively support the structure of ScIDL problems.

Addressing these challenges necessitates the development of dedicated computational frameworks, including scalable software tools, libraries, and infrastructure tailored to ScIDL workflows. Such frameworks can facilitate constrained optimization, and physics-informed learning in a unified environment, thereby lowering the barrier to practical implementation. In this context, Neural Modules with Adaptive Nonlinear Constraints and Efficient Regularizations (NeuroMANCER)~\cite{Neuromancer2023} represents a promising open-source PyTorch-based differentiable programming framework for parametric constrained optimization and physics-informed system identification, enabling systematic integration of DL with scientific computing.
\subsubsection{New Theoretical Methods for Interpretability and Generalization Analysis}
The primary objective of ScIDL is to integrate scientific knowledge with DL in order to address limitations in interpretability and generalization. It is therefore essential to assess how effectively ScIDL approaches can overcome these challenges. Several existing works have developed theoretical analysis methods that go beyond purely data-driven approaches by incorporating science-based principles to better characterize and explain DL behavior. We have  these methods in Section~IV.
Nevertheless, many open challenges remain, which include: (i) development of interpretable metrics and generalization bounds that are applicable across different families of DL models, (ii) analysis of generalization for learning methods with non-independent training data (e.g., sequential or time-series data) \cite{hellstrom2025generalization,rodriguez2024information}, and (iii) design of scalable interpretability and generalization analyses that can be applied to large-scale DL models.

%\vspace{0.2cm}
\subsection{Research Directions}
Future xG use cases, such as cyber–physical immersive environments, joint communication–sensing–control systems, and brain-to-brain communications, are expected to require massive-scale deployments with near-real-time latency, ultra-high data rates, and stringent reliability guarantees, exceeding those of mobile broadband predecessors~\cite{11195786,hossain2025hyperdimensionalconnectivity6gconceptual}. Meeting these demands involve solving complex, high-dimensional problems across the wireless protocol stack. Traditional model-based optimization methods often become impractical in this setting, as they rely on large-scale analytical models and iterative solvers whose computational complexity grows rapidly with system dimensions such as the number of users, antennas, and carriers. Moreover, the highly dynamic and non-stationary nature of xG environments requires frequent re-optimization, further limiting the feasibility of globally optimal solvers for real-time operation.

Although conventional DL have been used to address some of these problems, their limited interpretability and generalization restrict their applicability in critical wireless applications. These limitations motivate research on ScIDL as an alternative to integrates physical models and domain knowledge into learning pipelines. This section outlines six research directions for applying ScIDL to xG wireless systems: (1) ScIDL-based optimization methods, (2) ScIDL-based environment and channel estimation, (3) ScIDL-based goal-oriented and semantic communication, (4) ScIDL-based secure and safe system design, (5) ScIDL-based scientific knowledge extraction, and (6) foundations and frontiers of ScIDL.  
%

%\vspace{0.2cm}
\subsubsection{ScIDL-Based Optimization Methods}
Many xG wireless problems are characterized by high dimensionality, strong non-convexity, known optimization objectives, strict constraints, and the need for frequent re-solving under dynamic conditions. Model-based solvers for tasks such as beamforming, power control, interference management, scheduling, handover, load balancing, rate-splitting, joint communication and sensing, and trust modeling are often prohibitive for their high computational complexity and poor scalability. By embedding optimization structure, physical constraints, or algorithmic steps into learning pipelines, ScIDL-based optimization methods can achieve substantial reductions in computational complexity, enabling near-real-time decision-making in large-scale, dynamic wireless networks where classical model-based solvers are impractical.  

%\vspace{0.2cm}
\subsubsection{ScIDL-Based Environment and Channel Estimation}
Wireless communication environments are inherently dynamic due to user mobility, time-varying traffic demands, and constantly changing channel conditions. To account for this variability, modern wireless systems employ adaptive mechanisms such as dynamic link adaption, network load balancing, adaptive MAC design, and flexible resource allocation, all which are essential for maintaining high performance and quality user experience. For example, dynamic link adaption and beamforming rely heavily on accurate CSI. In many practical systems, CSI estimation depends on over-the-air pilot and reference signaling, such as CSI reference signals (CSI-RS) in the downlink and sounding reference signals (SRS) in the uplink for LTE and 5G-NR, which are used by the receivers to infer the channel conditions~\cite{tse2005fundamentals}. However, extensive pilot signaling is problematic, as it consumes limited radio resources, increases signaling overhead and measurement costs, and becomes unreliable in large-scale, fast-varying, and high-interference xG environments. These limitations motivate the reduced-pilot or pilot-free CSI estimation approaches based on synthetic data, where physical models and digital twins can complement or partially replace real measurements. In this context, ScIDL-powered digital twins offer a promising solution by enabling physics-consistent virtual channel replicas at scale, significantly reducing the dependence on real-world pilot signaling while preserving physical fidelity. 

%\vspace{0.2cm}
\subsubsection{ScIDL-Based Goal-oriented and Semantic Communication} 
Conventional Shannon-style communication systems are designed to maximize throughput and minimize bit error probability, without explicitly accounting for the meaning of the transmitted information. In contrast, goal-oriented and semantic communication focus on transmitting only task-relevant information needed to achieve a specific objective, such as sensing, control, or decision-making. Rather than optimizing the bit-level performance, semantic communication targets task-level metrics, such as semantic distortion or control errors. The paradigm shift can significantly reduce the communication overhead and it has been identified as a key enabler for  6G systems~\cite{liu2024survey_semantic6G}. However, semantic communication introduces substantial computational and learning challenges, particularly for encoding and decoding semantic representations under strict latency, energy, and hardware constraints. These systems typically rely on complex DNN architectures and a shared knowledge base between transmitter and receiver, making them difficult to train and deploy on edge or IoT devices. An ScIDL approach, such as symbolic neural networks with loss functions that incorporate domain knowledge, can substantially reduce training complexity while preserving semantic consistency. ScIDL offers a promising alternative to embed scientific structure into learning pipelines for potential simplification of goal-oriented and semantic communication architectures. 
%
%\vspace{0.2cm}
\subsubsection{ScIDL-Based Secure and Safe System Design}
Safety and security are critical in wireless systems, particularly in safety-critical applications such as autonomous vehicles and remote healthcare. In these settings, compromises in security or safety can lead to data breaches, increased latency, or system failures that adversely affect system responsiveness and reliability. A major limitation of most of the conventional DL-based solutions for these applications is their vulnerability to adversarial training data designed to produce incorrect outputs~\cite{pellaco2022machine}. Such vulnerabilities can be exploited by attackers to degrade the performance or compromise the system operation, making safety and security in DL-enabled wireless systems a challenging open problem. A promising research direction in xG wireless system design is the explicit incorporation of safety and security rules into the learning process. For instance, the authors in~\cite{termehchisafe} employ a science-informed loss function formulation to enforce cooperative safety constraints, specifically collision avoidance among UAVs, while optimizing energy efficiency in THz-enabled UAV-aided wireless networks. More broadly, ScIDL can also enhance anomaly detection and system resilience by incorporating domain knowledge, physical laws, and expected operational behaviors directly into loss functions or optimization algorithms, thereby improving robustness against adversarial and unseen conditions.
%
%\vspace{0.2cm}
\subsubsection{ScIDL-Based Scientific Knowledge Extraction}
Next generation (xG) wireless networks are evolving into highly nonlinear, multiscale, and multiphysics systems due to the integration of emerging technologies such as aerial and space-based platforms (e.g., UAVs and satellites), integrated sensing and communications, multi-band massive MIMO, cell-free massive MIMO, and adaptive beamforming. The growing complexity and heterogeneity of these systems exceed the scope of current analytical models, making it essential to advance our scientific understanding of their behaviors. The ScIDL framework could play a big role in this scientific discovery process, as learning-based methods can be used not only for optimization or prediction, but also for improving the interpretability and understanding of system dynamics. In ScIDL, the interaction between DL and scientific knowledge is bidirectional: domain knowledge is inserted into DL pipelines to improve the learning efficiency and prediction accuracy, while purely data-driven DL models can exploit large volumes of wireless sensory data to improve and validate the partially known system models. This strategy may uncover hidden knowledge such as effective channel models in complex environments, low-dimensional structures in massive MIMO architectures or interference dynamics in wireless environments. The resulting understanding can be fed back into DL model architectures, training strategies, and optimization processes to refine the existing analytical models.
%
%\vspace{0.2cm}
\subsubsection{Expanding the ScIDL Frontier}
From a training-scale and parameter-count perspective, DL models span a wide range of architectures. At one end are small task-specific models, such as CNNs, RNNs, PINN, and lightweight GNNs, which typically contain on the order of $10^3$--$10^5$ parameters and are trained from scratch on limited datasets. At the other end are medium- and large-scale domain-specialized models, including neural operators and equivariant networks, which can contain $10^5$--$10^9$ parameters and are trained on substantially larger datasets. At the top of this spectrum lie Large Foundation Models (LFMs) and Large Reasoning Models (LRMs), which typically contain $10^9$ to $10^{12+}$ parameters, are trained on massive heterogeneous datasets, and enable broad cross-task generalization through parameter fine-tuning. To the best of our knowledge, the majority of ScIDL applications in wireless networks focus on small- and medium-scale DL architectures. Consequently, the applicability of foundation-scale ScIDL models to wireless networks remains largely unexplored, representing a promising research direction for expanding the impact of AI in xG wireless systems.
\section{Conclusion}

%Conventional scientific methods, including physics-based models and traditional optimization algorithms, have effectively driven the evolution of wireless systems from 1G to 5G.
%However, these approaches are inadequate for the next generation of wireless systems because of the continuously ever-increasing computational complexity and scaling challenges. On the other hand, while DL methods have been suggested to tackle the challenges in 5G and 6G wireless systems, they encounter significant limitations. These limitations include the lack of interpretability, physics consistency, theoretical guidance in designing learning algorithms, and limited generalization. 

We envision that the emerging field of ScIDL will play a pivotal role in tackling the current and future challenges in multi-functional wireless systems (e.g., for communication, sensing). 
The primary objective of ScIDL is to integrate the existing scientific knowledge with DL to overcome the limitations of traditional DL models. In addition, ScIDL seeks to assess how effectively these approaches address the challenges related to lack of interpretability, limited generalization, and physical inconsistency. As xG wireless systems increase in complexity, this evaluation becomes even more critical, as these networks must integrate and optimize across diverse functional layers and physical phenomena to advance their capabilities. 
%A rigorous theoretical analysis of generalization and interpretability is, therefore, imperative to align the development of future ScIDL models to advance the field in the same philosophy horizon. In summary, ScIDL aims to leverage centuries of scientific knowledge to overcome the limitations of conventional DL.  In ScIDL, the scientific knowledge acts as a teacher for DL, improving the performance of DL algorithms. 
Moreover, DL and data-driven methods can provide valuable tools for enhancing our scientific knowledge and increasing the accuracy and applicability of physics-based models in wireless communications. In this paper, we have provided a concise tutorial on ScIDL. Furthermore, a roadmap for researchers has been provided, which outlines the challenges, implementation guidelines, and future research directions for applying ScIDL in wireless systems.
\bibliography{ref}

@Inbook{Patil2024,
author="Patil, Adithya
and Sohoni, Milind G.",
editor="Hamid, Faiz",
title="Delayed Column Generation: Solving Large-Scale Optimization Models From the Airline Industry",
bookTitle="Optimization Essentials: Theory, Tools, and Applications",
year="2024",
publisher="Springer Nature Singapore",
address="Singapore",
pages="279--296",
isbn="978-981-99-5491-9",
doi="10.1007/978-981-99-5491-9_9",
url="https://doi.org/10.1007/978-981-99-5491-9_9"
}

@article{karniadakis2021physics,
  title={Physics-informed machine learning},
  author={Karniadakis, George Em and Kevrekidis, Ioannis G and Lu, Lu and Perdikaris, Paris and Wang, Sifan and Yang, Liu},
  journal={Nature Reviews Physics},
  volume={3},
  number={6},
  pages={422--440},
  year={2021},
  publisher={Nature Publishing Group}
}

@phdthesis{moseley2022physics,
  title={Physics-informed machine learning: from concepts to real-world applications},
  author={Moseley, Benjamin},
  year={2022},
  school={University of Oxford}
}

@article{akrout2023domain,
  title={Domain Generalization in Machine Learning Models for Wireless Communications: Concepts, State-of-the-Art, and Open Issues},
  author={Akrout, Mohamed and Feriani, Amal and Bellili, Faouzi and Mezghani, Amine and Hossain, Ekram},
  journal={IEEE Communications Surveys \& Tutorials},
  year={2023},
  publisher={IEEE}
}

@article{karpatne2017theory,
  title={Theory-guided data science: A new paradigm for scientific discovery from data},
  author={Karpatne, Anuj and Atluri, Gowtham and Faghmous, James H and Steinbach, Michael and Banerjee, Arindam and Ganguly, Auroop and Shekhar, Shashi and Samatova, Nagiza and Kumar, Vipin},
  journal={IEEE Transactions on \uppercase{k}nowledge and \uppercase{d}ata \uppercase{e}ngineering},
  volume={29},
  number={10},
  pages={2318--2331},
  year={2017},
  publisher={IEEE}
}

@techreport{baker2019workshop,
  title={Workshop report on basic research needs for scientific machine learning: Core technologies for artificial intelligence},
  author={Baker, Nathan and Alexander, Frank and Bremer, Timo and Hagberg, Aric and Kevrekidis, Yannis and Najm, Habib and Parashar, Manish and Patra, Abani and Sethian, James and Wild, Stefan and others},
  year={2019},
  institution={USDOE Office of Science (SC), Washington, DC, US)}
}

@article{willard2022integrating,
  title={Integrating scientific knowledge with machine learning for engineering and environmental systems},
  author={Willard, Jared and Jia, Xiaowei and Xu, Shaoming and Steinbach, Michael and Kumar, Vipin},
  journal={ACM Computing Surveys},
  volume={55},
  number={4},
  pages={1--37},
  year={2022},
  publisher={ACM New York, NY}
}

@article{von2021informed,
  title={Informed machine learning--a taxonomy and survey of integrating prior knowledge into learning systems},
  author={Von Rueden, Laura and Mayer, Sebastian and Beckh, Katharina and Georgiev, Bogdan and Giesselbach, Sven and Heese, Raoul and Kirsch, Birgit and Pfrommer, Julius and Pick, Annika and Ramamurthy, Rajkumar and others},
  journal={IEEE Transactions on Knowledge and Data Engineering},
  volume={35},
  number={1},
  pages={614--633},
  year={2021},
  publisher={IEEE}
}

@article{karpatne2017physics,
  title={Physics-guided neural networks ({PGNN}): An application in lake temperature modeling},
  author={Karpatne, Anuj and Watkins, William and Read, Jordan and Kumar, Vipin},
  journal={arXiv preprint arXiv:1710.11431},
  volume={2},
  year={2017}
}

@article{kharazmi2019variational,
  title={Variational physics-informed neural networks for solving partial differential equations},
  author={Kharazmi, Ehsan and Zhang, Zhongqiang and Karniadakis, George Em},
  journal={arXiv preprint arXiv:1912.00873},
  year={2019}
}

@book{russell2016artificial,
  title={Artificial intelligence: a modern approach},
  author={Russell, Stuart J and Norvig, Peter},
  year={2016},
  publisher={Pearson}
}

@incollection{cercone1987knowledge,
  title={What is knowledge representation?},
  author={Cercone, Nick and McCalla, Gordon},
  booktitle={The Knowledge Frontier: Essays in the Representation of Knowledge},
  pages={1--43},
  year={1987},
  publisher={Springer}
}

@article{wu2020enforcing,
  title={Enforcing statistical constraints in generative adversarial networks for modeling chaotic dynamical systems},
  author={Wu, Jin-Long and Kashinath, Karthik and Albert, Adrian and Chirila, Dragos and Xiao, Heng and others},
  journal={Journal of Computational Physics},
  volume={406},
  pages={109209},
  year={2020},
  publisher={Elsevier}
}

@article{jia2021physics,
  title={Physics-guided machine learning for scientific discovery: An application in simulating lake temperature profiles},
  author={Jia, Xiaowei and Willard, Jared and Karpatne, Anuj and Read, Jordan S and Zwart, Jacob A and Steinbach, Michael and Kumar, Vipin},
  journal={ACM/IMS Transactions on Data Science},
  volume={2},
  number={3},
  pages={1--26},
  year={2021},
  publisher={ACM New York, NY}
}

@article{zhang2022topology,
  title={Topology aware deep learning for wireless network optimization},
  author={Zhang, Shuai and Yin, Bo and Zhang, Weiyi and Cheng, Yu},
  journal={IEEE Transactions on Wireless Communications},
  volume={21},
  number={11},
  pages={9791--9805},
  year={2022},
  publisher={IEEE}
}

@article{he2021overview,
  title={An overview on the application of graph neural networks in wireless networks},
  author={He, Shiwen and Xiong, Shaowen and Ou, Yeyu and Zhang, Jian and Wang, Jiaheng and Huang, Yongming and Zhang, Yaoxue},
  journal={IEEE Open Journal of the Communications Society},
  volume={2},
  pages={2547--2565},
  year={2021},
  publisher={IEEE}
}

@article{gross1996role,
  title={The role of symmetry in fundamental physics},
  author={Gross, David J},
  journal={Proceedings of the National Academy of Sciences},
  volume={93},
  number={25},
  pages={14256--14259},
  year={1996},
  publisher={National Acad Sciences}
}

@article{wang2020incorporating,
  title={Incorporating symmetry into deep dynamics models for improved generalization},
  author={Wang, Rui and Walters, Robin and Yu, Rose},
  journal={International Conference on Learning Representations (ICLR)},
  year={2021}
}

@inproceedings{xu2018semantic,
  title={A semantic loss function for deep learning with symbolic knowledge},
  author={Xu, Jingyi and Zhang, Zilu and Friedman, Tal and Liang, Yitao and Broeck, Guy},
  booktitle={International \uppercase{c}onference on \uppercase{m}achine \uppercase{l}earning},
  pages={5502--5511},
  year={2018},
  organization={PMLR}
}

@inproceedings{stewart2017label,
  title={Label-free supervision of neural networks with physics and domain knowledge},
  author={Stewart, Russell and Ermon, Stefano},
  booktitle={Proceedings of the {AAAI} Conference on Artificial Intelligence},
  volume={31},
  number={1},
  year={2017}
}

@article{zhuang2020adaptive,
  title={Adaptive and robust routing with Lyapunov-based deep {RL} in {MEC} networks enabled by blockchains},
  author={Zhuang, Zirui and Wang, Jingyu and Qi, Qi and Liao, Jianxin and Han, Zhu},
  journal={IEEE Internet of Things Journal},
  volume={8},
  number={4},
  pages={2208--2225},
  year={2020},
  publisher={IEEE}
}

@article{li2024physics,
  title={Physics-informed neural operator for learning partial differential equations},
  author={Li, Zongyi and Zheng, Hongkai and Kovachki, Nikola and Jin, David and Chen, Haoxuan and Liu, Burigede and Azizzadenesheli, Kamyar and Anandkumar, Anima},
  journal={ACM/JMS Journal of Data Science},
  volume={1},
  number={3},
  pages={1--27},
  year={2024},
  publisher={ACM New York, NY}
}

@article{han2021reinforcement,
  title={Reinforcement learning control of constrained dynamic systems with uniformly ultimate boundedness stability guarantee},
  author={Han, Minghao and Tian, Yuan and Zhang, Lixian and Wang, Jun and Pan, Wei},
  journal={Automatica},
  volume={129},
  pages={109689},
  year={2021},
  publisher={Elsevier}
}

@article{termehchisafe,
  author = {Termehchi, Atefeh and Syed, Aisha and Kennedy, William Sean and Erol-Kantarci, Melike},
  title = {\uppercase{D}istributed Safe Multi-Agent Reinforcement Learning: Joint Design of {TH}z-enabled UAV Trajectory and Channel Allocation},
  journal={IEEE Transactions on Vehicular Technology},
  year = {2024}
}

@article{li2023physics,
  title={Physics-informed neural network based on a new adaptive gradient descent algorithm for solving partial differential equations of flow problems},
  author={Li, Xiaojian and Liu, Yuhao and Liu, Zhengxian},
  journal={Physics of Fluids},
  volume={35},
  number={6},
  year={2023},
  publisher={AIP Publishing}
}

@article{pellaco2021matrix,
  title={Matrix-inverse-free deep unfolding of the weighted {MMSE} beamforming algorithm},
  author={Pellaco, Lissy and Bengtsson, Mats and Jald{\'e}n, Joakim},
  journal={IEEE Open Journal of the Communications Society},
  volume={3},
  pages={65--81},
  year={2021},
  publisher={IEEE}
}

@phdthesis{pellaco2022machine,
  title={Machine learning for wireless communications: Hybrid data-driven and model-based approaches},
  author={Pellaco, Lissy},
  year={2022},
  school={KTH Royal Institute of Technology}
}

@article{he2023generalizing,
  title={Generalizing Projected Gradient Descent for Deep-Learning-Aided Massive {MIMO} Detection},
  author={He, Lanxin and Wang, Zheng and Yang, Shaoshi and Liu, Tao and Huang, Yongming},
  journal={IEEE Transactions on Wireless Communications},
  year={2023},
  publisher={IEEE}
}

@inproceedings{daw2020physics,
  title={Physics-guided architecture ({PGA}) of neural networks for quantifying uncertainty in lake temperature modeling},
  author={Daw, Arka and Thomas, R Quinn and Carey, Cayelan C and Read, Jordan S and Appling, Alison P and Karpatne, Anuj},
  booktitle={Proceedings of the 2020 \uppercase{siam} \uppercase{i}nternational \uppercase{c}onference on \uppercase{d}ata \uppercase{m}ining},
  pages={532--540},
  year={2020},
  organization={SIAM}
}

@ARTICLE{10719669,
  author={Wang, Ji-Yuan and Pan, Xiao-Min},
  journal={IEEE Transactions on Antennas and Propagation}, 
  title={Universal Approximation Theorem and Deep Learning for the Solution of Frequency-Domain Electromagnetic Scattering Problems}, 
  year={2024},
  volume={72},
  number={12},
  pages={9274-9285},
  doi={10.1109/TAP.2024.3476915}}

@ARTICLE{9252917,
  author={Shen, Yifei and Shi, Yuanming and Zhang, Jun and Letaief, Khaled B.},
  journal={IEEE Journal on Selected Areas in Communications}, 
  title={Graph Neural Networks for Scalable Radio Resource Management: Architecture Design and Theoretical Analysis}, 
  year={2021},
  volume={39},
  number={1},
  pages={101-115},
  doi={10.1109/JSAC.2020.3036965}}

@ARTICLE{9072356,
  author={Eisen, Mark and Ribeiro, Alejandro},
  journal={IEEE Transactions on Signal Processing}, 
  title={Optimal Wireless Resource Allocation With Random Edge Graph Neural Networks}, 
  year={2020},
  volume={68},
  number={},
  pages={2977-2991},
  doi={10.1109/TSP.2020.2988255}}

@ARTICLE{9667094,
  author={Pellaco, Lissy and Bengtsson, Mats and Jaldén, Joakim},
  journal={IEEE Open Journal of the Communications Society}, 
  title={Matrix-Inverse-Free Deep Unfolding of the Weighted {MMSE} Beamforming Algorithm}, 
  year={2022},
  volume={3},
  number={},
  pages={65-81},
  doi={10.1109/OJCOMS.2021.3139858}}

@ARTICLE{10021866,
  author={Pellaco, Lissy and Jaldén, Joakim},
  journal={IEEE Transactions on Signal Processing}, 
  title={A Matrix-Inverse-Free Implementation of the {MU-MIMO} {WMMSE} Beamforming Algorithm}, 
  year={2022},
  volume={70},
  number={},
  pages={6360-6375},
  doi={10.1109/TSP.2023.3238275}}

@ARTICLE{10972051,
  author={ElFar, Sara H. and Yaseen, Maysa and Ikki, Salama},
  journal={IEEE Wireless Communications Letters}, 
  title={A Novel Machine Learning Algorithm With Mathematical Modeling for Channel Estimation in VLC Systems}, 
  year={2025},
  volume={14},
  number={7},
  pages={2084-2088},
  doi={10.1109/LWC.2025.3563156}}

@ARTICLE{10720822,
  author={Ko, Sunyoung and Shin, Myoungin and Kim, Geunhwan and Choo, Youngmin},
  journal={IEEE Signal Processing Letters}, 
  title={System-Informed Neural Network for Frequency Detection}, 
  year={2024},
  volume={31},
  number={},
  pages={2980-2984},
  doi={10.1109/LSP.2024.3483036}}

@INPROCEEDINGS{9024538,
  author={Shen, Yifei and Shi, Yuanming and Zhang, Jun and Letaief, Khaled B.},
  booktitle={2019 IEEE Globecom Workshops (GC Wkshps)}, 
  title={A Graph Neural Network Approach for Scalable Wireless Power Control}, 
  year={2019},
  volume={},
  number={},
  pages={1-6},
  doi={10.1109/GCWkshps45667.2019.9024538}}

@ARTICLE{8335785,
  author={Lee, Woongsup and Kim, Minhoe and Cho, Dong-Ho},
  journal={IEEE Communications Letters}, 
  title={Deep Power Control: Transmit Power Control Scheme Based on Convolutional Neural Network}, 
  year={2018},
  volume={22},
  number={6},
  pages={1276-1279},
  doi={10.1109/LCOMM.2018.2825444}}

@ARTICLE{8922744,
  author={Liang, Fei and Shen, Cong and Yu, Wei and Wu, Feng},
  journal={IEEE Transactions on Communications}, 
  title={Towards Optimal Power Control via Ensembling Deep Neural Networks}, 
  year={2020},
  volume={68},
  number={3},
  pages={1760-1776},
  doi={10.1109/TCOMM.2019.2957482}}

@ARTICLE{8935405,
  author={Xia, Wenchao and Zheng, Gan and Zhu, Yongxu and Zhang, Jun and Wang, Jiangzhou and Petropulu, Athina P.},
  journal={IEEE Transactions on Communications}, 
  title={A Deep Learning Framework for Optimization of MISO Downlink Beamforming}, 
  year={2020},
  volume={68},
  number={3},
  pages={1866-1880},
  doi={10.1109/TCOMM.2019.2960361}}

@ARTICLE{9285223,
  author={Lee, Mengyuan and Yu, Guanding and Li, Geoffrey Ye},
  journal={IEEE Transactions on Wireless Communications}, 
  title={Graph Embedding-Based Wireless Link Scheduling With Few Training Samples}, 
  year={2021},
  volume={20},
  number={4},
  pages={2282-2294},
  doi={10.1109/TWC.2020.3040983}}

@INPROCEEDINGS{9154263,
  author={Pellaco, Lissy and Saxena, Vidit and Bengtsson, Mats and Jaldén, Joakim},
  booktitle={2020 IEEE 21st International Workshop on Signal Processing Advances in Wireless Communications (SPAWC)}, 
  title={Wireless link adaptation with outdated CSI — a hybrid data-driven and model-based approach}, 
  year={2020},
  volume={},
  number={},
  pages={1-5},
  doi={10.1109/SPAWC48557.2020.9154263}}

@INPROCEEDINGS{9414561,
  author={Pellaco, Lissy and Bengtsson, Mats and Jaldén, Joakim},
  booktitle={ICASSP 2021 - 2021 IEEE International Conference on Acoustics, Speech and Signal Processing (ICASSP)}, 
  title={Deep Weighted {MMSE} Downlink Beamforming}, 
  year={2021},
  volume={},
  number={},
  pages={4915-4919},
  doi={10.1109/ICASSP39728.2021.9414561}}

@ARTICLE{8979256,
  author={Yuan, Jide and Ngo, Hien Quoc and Matthaiou, Michail},
  journal={IEEE Transactions on Wireless Communications}, 
  title={Machine Learning-Based Channel Prediction in Massive {MIMO} With Channel Aging}, 
  year={2020},
  volume={19},
  number={5},
  pages={2960-2973},
  doi={10.1109/TWC.2020.2969627}}

@ARTICLE{6755477,
  author={Ding, Tianben and Hirose, Akira},
  journal={IEEE Transactions on Neural Networks and Learning Systems}, 
  title={Fading Channel Prediction Based on Combination of Complex-Valued Neural Networks and Chirp Z-Transform}, 
  year={2014},
  volume={25},
  number={9},
  pages={1686-1695},
  doi={10.1109/TNNLS.2014.2306420}}

@ARTICLE{11016226,
  author={Han, Qiaojian and Zhang, Haixia and Li, Yueheng and Yuan, Dongfeng and Ma, Xiaoyan},
  journal={IEEE Wireless Communications}, 
  title={Physics Informed Digital Twin for {RIS}-Assisted Wireless Communication System}, 
  year={2025},
  volume={32},
  number={3},
  pages={106-112},
  doi={10.1109/MWC.003.2400418}}

@INPROCEEDINGS{9322615,
  author={Yang, Helin and Zhao, Yang and Xiong, Zehui and Zhao, Jun and Niyato, Dusit and Lam, Kwok-Yan and Wu, Qingqing},
  booktitle={GLOBECOM 2020 - 2020 IEEE Global Communications Conference}, 
  title={Deep Reinforcement Learning Based Intelligent Reflecting Surface for Secure Wireless Communications}, 
  year={2020},
  volume={},
  number={},
  pages={1-6},
  doi={10.1109/GLOBECOM42002.2020.9322615}}

@INPROCEEDINGS{10882105,
  author={Yaseen, Maysa and El-Far, Sara and Ikki, Salama},
  booktitle={2024 IEEE Middle East Conference on Communications and Networking (MECOM)}, 
  title={Machine Learning-Based Channel Estimation in Visible Light Communication with Signal-Dependent Noise}, 
  year={2024},
  volume={},
  number={},
  pages={263-267},
  doi={10.1109/MECOM61498.2024.10882105}}

@INPROCEEDINGS{224799,
  author={Gschwendtner, B.E. and Landstorfer, F.M.},
  booktitle={1993 Eighth International Conference on Antennas and Propagation}, 
  title={An application of neural networks to the prediction of terrestrial wave propagation}, 
  year={1993},
  volume={},
  number={},
  pages={804-807 vol.2},
  doi={}}

@ARTICLE{8444648,
  author={Sun, Haoran and Chen, Xiangyi and Shi, Qingjiang and Hong, Mingyi and Fu, Xiao and Sidiropoulos, Nicholas D.},
  journal={IEEE Transactions on Signal Processing}, 
  title={Learning to Optimize: Training Deep Neural Networks for Interference Management}, 
  year={2018},
  volume={66},
  number={20},
  pages={5438-5453},
  doi={10.1109/TSP.2018.2866382}}

@article{KHACHATRIAN2025103696,
title = {Deep learning with synthetic data for wireless NLOS positioning with a single base station},
journal = {Ad Hoc Networks},
volume = {167},
pages = {103696},
year = {2025},
issn = {1570-8705},
doi = {https://doi.org/10.1016/j.adhoc.2024.103696},
url = {https://www.sciencedirect.com/science/article/pii/S157087052400307X},
author = {Hrant Khachatrian and Rafayel Mkrtchyan and Theofanis P. Raptis}
}

@ARTICLE{9175003,
  author={Yang, Yuwen and Gao, Feifei and Zhong, Zhimeng and Ai, Bo and Alkhateeb, Ahmed},
  journal={IEEE Transactions on Communications}, 
  title={Deep Transfer Learning-Based Downlink Channel Prediction for {FDD} Massive {MIMO} Systems}, 
  year={2020},
  volume={68},
  number={12},
  pages={7485-7497},
  doi={10.1109/TCOMM.2020.3019077}}

@INPROCEEDINGS{10757719,
  author={Saleem, Osama and Ribouh, Soheyb and Alfaqawi, Mohammed and Bensrhair, Abdelaziz and Merdrignac, Pierre},
  booktitle={2024 IEEE 100th Vehicular Technology Conference (VTC2024-Fall)}, 
  title={TransRx-{6G-V2X }: Transformer Encoder-Based Deep Neural Receiver For Next Generation of Cellular Vehicular Communications}, 
  year={2024},
  volume={},
  number={},
  pages={1-7},
  doi={10.1109/VTC2024-Fall63153.2024.10757719}}

@ARTICLE{8879693,
  author={Shen, Yifei and Shi, Yuanming and Zhang, Jun and Letaief, Khaled B.},
  journal={IEEE Transactions on Wireless Communications}, 
  title={{LORM}: Learning to Optimize for Resource Management in Wireless Networks With Few Training Samples}, 
  year={2020},
  volume={19},
  number={1},
  pages={665-679},
  doi={10.1109/TWC.2019.2947591}}

@ARTICLE{10974467,
  author={Jere, Shashank and Zheng, Lizhong and Said, Karim and Liu, Lingjia},
  journal={IEEE Transactions on Wireless Communications}, 
  title={Toward x{AI}: Configuring RNN Weights Using Domain Knowledge for {MIMO} Receive Processing}, 
  year={2025},
  volume={24},
  number={9},
  pages={7581-7597},
  doi={10.1109/TWC.2025.3561242}}

@article{Grossman1093,
    author = {Grossmann, Tamara G and Komorowska, Urszula Julia and Latz, Jonas and Schönlieb, Carola-Bibiane},
    title = {Can physics-informed neural networks beat the finite element method?},
    journal = {IMA Journal of Applied Mathematics},
    volume = {89},
    number = {1},
    pages = {143-174},
    year = {2024},
    month = {05},
    issn = {0272-4960},
    doi = {10.1093/imamat/hxae011},
    url = {https://doi.org/10.1093/imamat/hxae011},
    eprint = {https://academic.oup.com/imamat/article-pdf/89/1/143/58325885/hxae011.pdf},
}

@article{russakovsky2015imagenet,
  title={ImageNet Large Scale Visual Recognition Challenge},
  author={Russakovsky, Olga and Deng, Jia and Su, Hao and Krause, Jonathan and Satheesh, Sanjeev and Ma, Sean and Huang, Zhiheng and Karpathy, Andrej and Khosla, Aditya and Bernstein, Michael and Berg, Alexander C and Fei-Fei, Li},
  journal={International Journal of Computer Vision},
  volume={115},
  number={3},
  pages={211--252},
  year={2015},
  publisher={Springer}
}

@inproceedings{wang2019glue,
  title={{GLUE}: A Multi-Task Benchmark and Analysis Platform for Natural Language Understanding},
  author={Wang, Alex and Singh, Amanpreet and Michael, Julian and Hill, Felix and Levy, Omer and Bowman, Samuel R},
  booktitle={International Conference on Learning Representations (ICLR)},
  year={2019}
}

@article{xu2025interpretability,
  title={Interpretability research of deep learning: A literature survey},
  author={Xu, Biao and Yang, Guanci},
  journal={Information Fusion},
  volume={115},
  pages={102721},
  year={2025},
  publisher={Elsevier}
}

@article{zhang2021survey,
  title={A survey on neural network interpretability},
  author={Zhang, Yu and Ti{\v{n}}o, Peter and Leonardis, Ale{\v{s}} and Tang, Ke},
  journal={IEEE transactions on emerging topics in computational intelligence},
  volume={5},
  number={5},
  pages={726--742},
  year={2021},
  publisher={IEEE}
}

@article{bunge1963general,
  title={A general black box theory},
  author={Bunge, Mario},
  journal={Philosophy of Science},
  volume={30},
  number={4},
  pages={346--358},
  year={1963},
  publisher={Cambridge University Press}
}

@article{li2022interpretable,
  title={Interpretable deep learning: Interpretation, interpretability, trustworthiness, and beyond},
  author={Li, Xuhong and Xiong, Haoyi and Li, Xingjian and Wu, Xuanyu and Zhang, Xiao and Liu, Ji and Bian, Jiang and Dou, Dejing},
  journal={Knowledge and Information Systems},
  volume={64},
  number={12},
  pages={3197--3234},
  year={2022},
  publisher={Springer}
}

@article{csahin2025unlocking,
  title={Unlocking the black box: an in-depth review on interpretability, explainability, and reliability in deep learning},
  author={{\c{S}}AHiN, Emrullah and Arslan, Naciye Nur and {\"O}zdemir, Durmu{\c{s}}},
  journal={Neural Computing and Applications},
  volume={37},
  number={2},
  pages={859--965},
  year={2025},
  publisher={Springer}
}

@article{rodriguez2024information,
  title={An information-theoretic approach to generalization theory},
  author={Rodr{\'\i}guez-G{\'a}lvez, Borja and Thobaben, Ragnar and Skoglund, Mikael},
  journal={arXiv preprint arXiv:2408.13275},
  year={2024}
}

@article{hellstrom2025generalization,
  title={Generalization bounds: Perspectives from information theory and {PAC}-{B}ayes},
  author={Hellstr{\"o}m, Fredrik and Durisi, Giuseppe and Guedj, Benjamin and Raginsky, Maxim and others},
  journal={Foundations and Trends{\textregistered} in Machine Learning},
  volume={18},
  number={1},
  pages={1--223},
  year={2025},
  publisher={Now Publishers, Inc.}
}

@article{suh2025survey,
  title={A survey on statistical theory of deep learning: Approximation, training dynamics, and generative models},
  author={Suh, Namjoon and Cheng, Guang},
  journal={Annual Review of Statistics and Its Application},
  volume={12},
  number={1},
  pages={177--207},
  year={2025},
  publisher={Annual Reviews}
}

@article{lecun1998gradient,
  title={Gradient-based learning applied to document recognition},
  author={LeCun, Yann and Bottou, L{\'e}on and Bengio, Yoshua and Haffner, Patrick},
  journal={Proceedings of the IEEE},
  volume={86},
  number={11},
  pages={2278--2324},
  year={1998}
}

@article{Elman1990Finding,
  author  = {Elman, Jeffrey L.},
  title   = {Finding Structure in Time},
  journal = {Cognitive Science},
  volume  = {14},
  number  = {2},
  pages   = {179--211},
  year    = {1990},
  publisher = {Wiley}
}

@article{Rumelhart1986Learning,
  author  = {Rumelhart, David E. and Hinton, Geoffrey E. and Williams, Ronald J.},
  title   = {Learning Representations by Back-Propagating Errors},
  journal = {Nature},
  volume  = {323},
  number  = {6088},
  pages   = {533--536},
  year    = {1986},
  publisher = {Nature Publishing Group}
}

@article{IEEEComSocTutorials,
  author  = {IEEE Communications Society},
  title   = {IEEE Communications Surveys \& Tutorials: Scope and Editorial Policy},
  journal = {IEEE Communications Surveys \& Tutorials},
  year    = {2020},
  note    = {Editorial scope statement describing tutorial and pedagogical survey articles}
}

@article{mnih2015human,
  title={Human-level control through deep reinforcement learning},
  author={Mnih, Volodymyr and Kavukcuoglu, Koray and Silver, David and others},
  journal={Nature},
  volume={518},
  number={7540},
  pages={529--533},
  year={2015}
}

@inproceedings{goodfellow2014generative,
  title={Generative adversarial nets},
  author={Goodfellow, Ian and Pouget-Abadie, Jean and Mirza, Mehdi and Xu, Bing and Warde-Farley, David and Ozair, Sherjil and Courville, Aaron and Bengio, Yoshua},
  booktitle={Advances in Neural Information Processing Systems},
  year={2014}
}

@inproceedings{vaswani2017attention,
  title={Attention Is All You Need},
  author={Vaswani, Ashish and Shazeer, Noam and Parmar, Niki and Uszkoreit, Jakob and Jones, Llion and Gomez, Aidan N. and Kaiser, {\L}ukasz and Polosukhin, Illia},
  booktitle={Advances in Neural Information Processing Systems},
  volume={30},
  year={2017}
}

@article{bengio2003neural,
  title={A Neural Probabilistic Language Model},
  author={Bengio, Yoshua and Ducharme, R{\'e}jean and Vincent, Pascal and Jauvin, Christian},
  journal={Journal of Machine Learning Research},
  volume={3},
  pages={1137--1155},
  year={2003}
}

@ARTICLE{11195786,
  author={Hossain, Ekram and Vera-Rivera, Angelo},
  journal={IEEE Transactions on Technology and Society}, 
  title={{6G} Cellular Networks: Mapping the Landscape for the IMT-2030 Framework}, 
  year={2025},
  volume={},
  number={},
  pages={1-16},
  doi={10.1109/TTS.2025.3611364}}

@misc{hossain2025hyperdimensionalconnectivity6gconceptual,
      title={Toward Hyper-Dimensional Connectivity in Beyond {6G}: A Conceptual Framework}, 
      author={Ekram Hossain and Angelo Vera-Rivera},
      year={2025},
      eprint={2510.12896},
      archivePrefix={arXiv},
      primaryClass={cs.NI},
      url={https://arxiv.org/abs/2510.12896}, 
}

@article{tse2005fundamentals,
  title   = {Fundamentals of Wireless Communication},
  author  = {Tse, David and Viswanath, Pramod},
  year    = {2005},
  publisher = {Cambridge University Press}
}

@article{liu2024survey_semantic6G,
  title   = {A Survey on Semantic Communications: Technologies, Challenges, and {6G} Applications},
  author  = {Liu, Ying and others},
  journal = {Computer Networks},
  year    = {2024}
}

@data{-21,
doi = {https://doi.org/10.3390/app11198860},
url = {https://dx.doi.org/https://doi.org/10.3390/app11198860},
author = {Jörg Schäfer},
publisher = {IEEE Dataport},
title = {CSI Human Activity},
year = {2021} }

@article{termehchi2026generalization,
  title={Generalization Analysis and Method for Domain Generalization for a Family of Recurrent Neural Networks},
  author={Termehchi, Atefeh and Hossain, Ekram and Woungang, Isaac},
  journal={arXiv preprint arXiv:2601.08122},
  year={2026}
}

@article{Neuromancer2023,
  title={{NeuroMANCER: Neural Modules with Adaptive Nonlinear Constraints and Efficient Regularizations}},
  author={Drgona, Jan and Tuor, Aaron and Koch, James and Shapiro, Madelyn and Jacob, Bruno and Vrabie, Draguna},
  Url= {https://github.com/pnnl/neuromancer}, 
  year={2023}
}

@article{liu2012achieving,
  title={Achieving global optimality for weighted sum-rate maximization in the K-user Gaussian interference channel with multiple antennas},
  author={Liu, Liang and Zhang, Rui and Chua, Kee-Chaing},
  journal={IEEE Transactions on Wireless Communications},
  volume={11},
  number={5},
  pages={1933--1945},
  year={2012},
  publisher={IEEE}
}

@article{zaheer2017deep,
  title={Deep sets},
  author={Zaheer, Manzil and Kottur, Satwik and Ravanbakhsh, Siamak and Poczos, Barnabas and Salakhutdinov, Russ R and Smola, Alexander J},
  journal={Advances in neural information processing systems},
  volume={30},
  year={2017}
}
\bibliographystyle{IEEEtran}

%\bibliographystyle{ieeetr}
%\bibliography{ref}
\end{document}